\documentclass[longauth]{aa}

\usepackage{graphicx}
\usepackage{txfonts}
\usepackage{ae,aecompl}
\usepackage[dvipsnames]{xcolor}
\usepackage{amsmath}
\usepackage{amssymb}
\usepackage{savesym}
\savesymbol{tablenum}
\usepackage{makecell}
\usepackage{siunitx}
\usepackage{xspace}
\usepackage{listings}
\usepackage{soul}
\usepackage{float}
\restoresymbol{SIX}{tablenum}
\usepackage{hyperref}
\usepackage{comment}
\usepackage{placeins}

\usepackage[T1]{fontenc}
\usepackage{fontawesome}
\definecolor{linkcolor}{HTML}{2AAD2E}

\usepackage{newtxtext,newtxmath}

\usepackage{multirow}
\usepackage{caption}
\usepackage{subcaption}
\usepackage{graphicx}

\newcommand{\casa}{\texttt{CASA}\xspace} 
\newcommand{\clean}{\texttt{CLEAN}\xspace} 
\newcommand{\frank}{\texttt{frank}\xspace}
\newcommand{\rave}{\texttt{rave}\xspace}
\newcommand{\mpol}{\texttt{MPoL}\xspace}
\newcommand{\arksia}{\texttt{arksia}\xspace}

\newcommand{\efm}[1]{\textcolor{black}{#1}}

\begin{document}

    \title{The ALMA survey to Resolve exoKuiper belt Substructures (ARKS)}

    \subtitle{II. The radial structure of debris discs}

    \author{Yinuo~Han\textsuperscript{1}\fnmsep\thanks{E-mail: yinuo@caltech.edu} \and Elias~Mansell\textsuperscript{2} \and Jeff~Jennings\textsuperscript{3} \and Sebastian~Marino\textsuperscript{4} \and A.~Meredith~Hughes\textsuperscript{2} \and Brianna~Zawadzki\textsuperscript{2} \and Anna~Fehr\textsuperscript{5} \and Jamar~Kittling\textsuperscript{2} \and Catherine~Hou\textsuperscript{2} \and Aliya~Nurmohamed\textsuperscript{2} \and Junu~Lee\textsuperscript{2} \and Allan~Cheruiyot\textsuperscript{2} \and Yamani~Mpofu\textsuperscript{2} \and Mark~Booth\textsuperscript{6} \and Richard~Booth\textsuperscript{7} \and Myriam~Bonduelle\textsuperscript{8} \and Aoife~Brennan\textsuperscript{9} \and Carlos~del~Burgo\textsuperscript{10,11} \and John~M.~Carpenter\textsuperscript{12} \and Gianni~Cataldi\textsuperscript{13,14} \and Eugene~Chiang\textsuperscript{15} \and Steve~Ertel\textsuperscript{16,17} \and Thomas~Henning\textsuperscript{18} \and Marija~R.~Jankovic\textsuperscript{19} \and \'Agnes~K\'osp\'al\textsuperscript{20,21,18} \and Alexander~V.~Krivov\textsuperscript{22} \and Joshua~B.~Lovell\textsuperscript{5} \and Patricia~Luppe\textsuperscript{9} \and Meredith~A.~MacGregor\textsuperscript{23} \and Sorcha~Mac~Manamon\textsuperscript{9} \and Jonathan~P.~Marshall\textsuperscript{24} \and Luca~Matr\`a\textsuperscript{9} \and Julien~Milli\textsuperscript{8} \and Attila~Mo\'or\textsuperscript{20} \and Johan~Olofsson\textsuperscript{25} \and Tim~Pearce\textsuperscript{26} \and Sebasti\'an~P\'erez\textsuperscript{27,28,29} \and Antranik~A.~Sefilian\textsuperscript{16} \and Philipp~Weber\textsuperscript{27,28,29} \and David~J.~Wilner\textsuperscript{5} \and Mark~C.~Wyatt\textsuperscript{30}}

    \institute{
    Division of Geological and Planetary Sciences, California Institute of Technology, 1200 E. California Blvd., Pasadena, CA 91125, USA \and
    Department of Astronomy, Van Vleck Observatory, Wesleyan University, 96 Foss Hill Dr., Middletown, CT, 06459, USA \and
    Center for Computational Astrophysics, Flatiron Institute, 162 Fifth Ave., New York, NY 10010, USA \and
    Department of Physics and Astronomy, University of Exeter, Stocker Road, Exeter EX4 4QL, UK \and
    Center for Astrophysics | Harvard \& Smithsonian, 60 Garden St, Cambridge, MA 02138, USA \and
    UK Astronomy Technology Centre, Royal Observatory Edinburgh, Blackford Hill, Edinburgh EH9 3HJ, UK \and
    School of Physics and Astronomy, University of Leeds, Leeds LS2 9JT, UK \and
    Univ. Grenoble Alpes, CNRS, IPAG, F-38000 Grenoble, France \and
    School of Physics, Trinity College Dublin, the University of Dublin, College Green, Dublin 2, Ireland \and
    Instituto de Astrof\'isica de Canarias, Vía L\'actea S/N, La Laguna, E-38200, Tenerife, Spain \and
    Departamento de Astrof\'isica, Universidad de La Laguna, La Laguna, E-38200, Tenerife, Spain \and
    Joint ALMA Observatory, Avenida Alonso de C\'ordova 3107, Vitacura 7630355, Santiago, Chile \and
    National Astronomical Observatory of Japan, Osawa 2-21-1, Mitaka, Tokyo 181-8588, Japan \and
    Department of Astronomy, Graduate School of Science, The University of Tokyo, Tokyo 113-0033, Japan \and
    Department of Astronomy, University of California, Berkeley, Berkeley, CA 94720-3411, USA \and
    Department of Astronomy and Steward Observatory, The University of Arizona, 933 North Cherry Ave, Tucson, AZ, 85721, USA \and
    Large Binocular Telescope Observatory, The University of Arizona, 933 North Cherry Ave, Tucson, AZ, 85721, USA \and
    Max-Planck-Insitut f\"ur Astronomie, K\"onigstuhl 17, 69117 Heidelberg, Germany \and
    Institute of Physics Belgrade, University of Belgrade, Pregrevica 118, 11080 Belgrade, Serbia \and
    Konkoly Observatory, HUN-REN Research Centre for Astronomy and Earth Sciences, MTA Centre of Excellence, Konkoly-Thege Mikl\'os \'ut 15-17, 1121 Budapest, Hungary \and
    Institute of Physics and Astronomy, ELTE E\"otv\"os Lor\'and University, P\'azm\'any P\'eter s\'et\'any 1/A, 1117 Budapest, Hungary \and
    Astrophysikalisches Institut und Universit\"atssternwarte, Friedrich-Schiller-Universit\"at Jena, Schillerg\"a{\ss}chen 2-3, 07745 Jena, Germany \and
    Department of Physics and Astronomy, Johns Hopkins University, 3400 N Charles Street, Baltimore, MD 21218, USA \and
    Academia Sinica Institute of Astronomy and Astrophysics, 11F of AS/NTU Astronomy-Mathematics Building, No.1, Sect. 4, Roosevelt Rd, Taipei 106319, Taiwan. \and
    European Southern Observatory, Karl-Schwarzschild-Strasse 2, 85748 Garching bei M\"unchen, Germany \and
    Department of Physics, University of Warwick, Gibbet Hill Road, Coventry CV4 7AL, UK \and
    Departamento de Física, Universidad de Santiago de Chile, Av. V\'ictor Jara 3493, Santiago, Chile \and
    Millennium Nucleus on Young Exoplanets and their Moons (YEMS), Chile \and
    Center for Interdisciplinary Research in Astrophysics Space Exploration (CIRAS), Universidad de Santiago, Chile \and
    Institute of Astronomy, University of Cambridge, Madingley Road, Cambridge CB3 0HA, UK
    }

    \date{Received xxx; accepted xxx}
 
  \abstract
   {Debris discs are populated by belts of planetesimals, whose structure  carries dynamical imprints of planets and the formation and evolutionary history of the planetary system. The relatively faint emission of debris discs has previously made it challenging to obtain a large sample of  high-resolution ALMA images to characterise their substructures. }
   {The ALMA survey to Resolve exoKuiper belt Substructures (ARKS) was recently completed to cover the lack of high-resolution observations and to investigate the prevalence of substructures such as radial gaps and rings in a sample of 24 debris discs. This study characterises the radial structure of debris discs in the ARKS programme. }
   {We modelled all discs with a range of non-parametric and parametric approaches, including those that deconvolve and deproject the image or fit the visibilities directly, in order to identify and quantify the disc substructures. }
   {Across the sample we find that of the 24 discs, 5 host multiple rings, 7 are single rings that display halos or additional low-amplitude rings, and 12 are single rings with at most tentative evidence of additional substructures. The fractional ring widths that we measured are significantly narrower than previously derived values, and they follow a distribution similar to the fractional widths of individual rings resolved in protoplanetary discs. However, there exists a population of rings in debris discs that are significantly wider than those in protoplanetary discs.
   We also find that discs with steep inner edges consistent with planet sculpting tend to be found at smaller ($<$100\,au) radii, while more radially extended discs tend to have shallower edges more consistent with collisional evolution. 
   An overwhelming majority of discs have radial profiles that are well-described by either a double power law or double-Gaussian parametrisation. 
   }
   {While our findings suggest that it may be possible for some debris discs to inherit their structures directly from protoplanetary discs, there exists a sizeable population of broad debris discs that cannot be explained in this way. Assuming that the distribution of millimetre dust reflects the distribution of planetesimals, mechanisms that cause rings in protoplanetary discs to migrate or debris discs to broaden soon after formation may be at play, possibly mediated by planetary migration or scattering. }

   \keywords{Planetary systems -- Circumstellar matter -- Planet-disc interactions -- Submillimeter: planetary systems -- Methods: data analysis -- Planets and satellites: dynamical evolution and stability}

   \maketitle

\section{Introduction} \label{sec:introduction}
Planetesimal belts analogous to the Solar System's Edgeworth-Kuiper belt provided some of the earliest hints of the presence of extrasolar planetary systems, which manifest as an infrared excess in the spectral energy distribution (SED) of main-sequence stars \citep{Aumann1984, Harper1984}. Subsequent decades of imaging campaigns have spatially resolved these belts from optical (e.g. GPIES, \citealp{Crotts2024}) to far-infrared (e.g. DEBRIS, \citealp{Matthews2010}; DUNES, \citealp{Eiroa2013, Montesinos2016}; \citealp{Moor2015, Marshall2021}) and (sub)millimetre (e.g. SONS, \citealp{Holland2017}; REASONS, \citealp{Matra2025}) wavelengths, progressively achieving higher resolution that has revealed their morphology. In a handful of systems observed with sufficiently high resolution and sensitivity, substructures such as radial gaps \citep{Marino2018, Marino2019, Marino2020, MacGregor2019, Nederlander2021}, azimuthal asymmetries \citep{Dent2014, MacGregor2017Fom}, and various vertical profiles \citep{Kennedy2018, Matra2019, Daley2019, Vizgan2022, Terrill2023} have been identified. These substructures could be caused by planets possibly residing in these discs, interacting with the planetesimals and the dust, and/or they could be an outcome of planetesimal formation processes, possibly linked with substructures commonly observed in protoplanetary discs \citep{Bae2023}.

The ALMA survey to Resolve exoKuiper belt Substructures (ARKS) is the most recent campaign in this effort to obtain a sample of high-resolution thermal emission images of debris discs and investigate the prevalence of substructures \citep{Marino2025}. This is the only ALMA large programme targeting debris discs, with targets selected from the sample of all ALMA- and SMA-resolved debris discs known as the REASONS sample \citep{Matra2025}. 

The term debris discs loosely refers to the collection of Kuiper belt, asteroid belt, and zodiacal dust analogues in a planetary system \citep{Wyatt2008, Hughes2018}. In the context of this study, we use the term to refer specifically to analogues of the Kuiper belt at 10s to 100s of au, which extend over a large enough angular size for their detailed structure to be imaged. 
While substructures have been identified in debris discs at other observing wavelengths, ALMA observations are unique in imaging the distribution of millimetre-sized dust, which is thought to most closely trace the distribution of planetesimals, as smaller dust grains observed at shorter wavelengths are more significantly subject to radiative forces \citep{Wyatt2008, Krivov2010, Hughes2018, Pawellek2019}. This high-resolution ALMA survey is, therefore, well-suited for studying the detailed substructures of planetesimal belts, their dynamical interactions with planets in the system, and any structural similarities and differences with protoplanetary discs during the earlier phases of planet formation. This paper focuses on the radial structure of debris discs in the ARKS programme. 

Perhaps the Solar System provides the most compelling examples of how the radial disc structure reflects the architecture and dynamical history of the planetary system. The Kuiper belt is known to host an extended scattered disc component that  has been linked to the past migration of \citep{Brown2001, Morbidelli2004}. Furthermore, a wide radial gap lies between the Kuiper belt and asteroid belt, within which the four giant planets of the Solar System reside. Even within the asteroid belt are a series of Kirkwood gaps, which are cleared by orbital resonances with Jupiter and Saturn \citep{Wisdom1983, Morbidelli1991}. 

Similar disc features have also been observed and modelled in extrasolar systems. The location and sharpness of the belt's edges could constrain the presence and migration history of planets, as an inner edge with a shallow slope is expected for debris discs evolving under steady-state collisions alone \citep{Wyatt2007, Loehne2008, Geiler2017, ImazBlanco2023}, whereas as a steeper slope is expected if a planet truncates the inner edge \citep{Quillen2006, Chiang2009, Mustill2012, Rodigas2014, Marino2021, ImazBlanco2023, Pearce2024}. 
Similarly, the outer edge of debris discs could record dynamical events such as the scattering of planetesimals possibly due to migrating planets (e.g. HR\,8799, \citealp{Geiler2019}, and q$^1$~Eri, \citealp{Lovell2021}), whereas any azimuthal asymmetries could yield further constraints on planet migration \citep{Wyatt2003, Booth2023}. 
For discs with sufficiently well-constrained radial profiles, it may be possible to retrieve the semi-major axis and eccentricity distributions of planetesimals within the belt \citep{Rafikov2023}, which closely reflect the dynamical state of the system and any potential perturbing planets \citep{Quillen2006, Chiang2009, Faramaz2014, Pearce2014, Kennedy2020}, as well as the disc's total mass \citep{Sefilian2024}. 
Furthermore, radial gaps have been observed in debris discs \citep{Marino2018, Marino2019, Marino2020} and used to infer the presence of $<$1 Jupiter mass planets at tens of au. These gaps could be carved by planets embedded within the disc \citep{Faber2007, Shannon2016, Morrison2018, Friebe2022} or orbiting interior to its inner edge \citep{Pearce2015, Yelverton2019, Sefilian2021, Sefilian2023}. 

In addition to the link between disc structure and planetary architecture, we also consider the evolution of planetary systems from the few million-year-old protoplanetary discs to the more evolved debris disc phase ($>$10\,Myr). The higher dust continuum luminosity of protoplanetary discs has enabled a number of ALMA programmes to map their detailed structure in dozens of systems \citep{Bae2023}. These observations have found that protoplanetary discs with multiple concentric rings appear to be common \citep{Andrews2018, Huang2018, Long2018}. Out of 479 young stellar objects surveyed in nearby star-forming regions, 355 discs have been detected and 62 have clearly identified ring-like structures, of which approximately half display multiple rings \citep{Bae2023}. However, debris discs known to unambiguously host multiple rings are comparatively few in number, and typical fractional widths (i.e. FWHM of a debris disc ring divided by its radius, or $\Delta R/R_c$) measured from available ALMA observations appear to be larger than in protoplanetary discs \citep{Matra2025}. Prior to the ARKS programme, there were no debris disc surveys with comparable resolution and sample size to those targeting protoplanetary discs, so it was unclear whether the difference is real or due to limitations in the resolution and sensitivity of debris disc observations. 
Such a comparison of disc fractional widths is nonetheless important. If debris discs directly inherit the structure of protoplanetary rings, their radial distribution should be consistent if we are detecting systems of the same population and offset only in age \citep{Najita2022, Orcajo2025}. However, there could also exist mechanisms that broaden debris disc rings, such as   debris self-scattering \citep{Ida1993} or the migration of planetesimal-forming dust traps in protoplanetary discs, which could form wider planetesimal belts than would be formed outside a non-migrating planet \citep{Miller2021, Jiang2023}. 

The ARKS programme attempts to shed light on these questions. 
In this ARKS paper, we describe our modelling approach in Sect.~\ref{sec:modelling}, and show the deconvolved and deprojected radial profiles of the dust continuum emission in Sect.~\ref{sec:results}. The implications of these findings on our general understanding of debris disc structure are discussed in Sect.~\ref{sec:discussion}. Our findings are summarised in Sect.~\ref{sec:conclusions}. Some further technical details are provided in Appendices~\ref{app:emissivity}--\ref{app:app_mmr_l}.

\section{Extracting radial profiles} \label{sec:modelling}
The primary quantity that we aim to determine in this section is the radial profile of the dust continuum emission for all discs in the ARKS programme. The reduced and corrected dataset and the \clean images of all discs are described and displayed in \citet{Marino2025}, which include 18 discs with new observations and 6 with archival observations that meet the sensitivity and resolution requirements of the programme. The targets were selected to cover discs over a range of inclinations, and observations were designed to resolve gaps carved by Neptune-sized planets for face-on discs and to resolve the vertical height for edge-on discs \citep{Marino2025}. 

Radial surface brightness profiles can be directly extracted from clean images by deprojecting and azimuthally averaging the emission (e.g. \citealp{Marino2016}). However, these beam-convolved profiles are limited by the beam size and do not perform well for highly inclined belts. Fortunately, there are parametric and non-parametric methods that can circumvent these issues and robustly infer the deprojected and deconvolved radial profiles. 
This radial profile recovery can be performed either by modelling the visibilities measured by ALMA directly, or by going to image space and modelling reconstructed images. Both are non-trivial, as the source distribution is convolved with the beam (with the dirty beam taking a non-Gaussian form), and the radial distribution is partly degenerate with the vertical structure when the disc is not perfectly face-on. 
To obtain a radial brightness profile, non-parametric and parametric methods each have pros and cons. Non-parametric methods such as those in \citet{Jennings2020} and \citet{Han2022} avoid the risk of biasing a fitted profile towards specific shapes parametrised by certain functions, allowing them to more easily fit the observed visibilities accurately and identify features not obvious by eye from a \clean image. However, their lack of enforcement of a functional form (such as a Gaussian or power law radial profile) can give rise to erroneous features (such as oscillations in the fitted profile) in the absence of sufficient regularisation. To mitigate this, the balance between resolution and noise in the fitted profile can be tuned by varying hyperparameters specific to each method. Parametric methods provide a more direct quantification of specific features in the radial profile by allowing us to comparing different functional forms to theoretical models of debris disc evolution. Parametric models also typically allow for easier marginalisation over physical parameters like the system orientation and stellar flux. 

In addition to deriving radial brightness profiles from \clean images, we applied both non-parametric and parametric approaches to recover deconvolved and deprojected radial profiles across the ARKS sample. We fit each source with: 

(1) \frank \citep{Jennings2020, Terrill2023}, a non-parametric method which operates in visibility space;

(2) \rave \citep{Han2022, Han2025}, a non-parametric method which operates in image space; and 

(3) parametric models in visibility space, motivated by the shape of the non-parametric brightness profiles as well as theoretical predictions of debris disc structure. 

Each of these methods enables radial profile recovery under the assumption of azimuthal symmetry, even if the disc is perfectly edge-on (due to the low optical depth of the discs in the sample). Together, they provide an informal type of verification to more reliably identify and quantify disc substructures. The assumption of azimuthal symmetry here is appropriate for the purpose of extracting the radial profile as any asymmetries of discs in the sample are generally subtle. Details on any asymmetries across the sample are analysed in an accompanying paper \citep{Lovell2025}. We describe the \clean profile extraction and the modelling approaches used in this study in the following sections. The resulting profiles are available for download from the Harvard Dataverse repository at \href{https://dataverse.harvard.edu/dataverse/arkslp}{dataverse.harvard.edu/dataverse/arkslp} and the ARKS website at \href{https://arkslp.org}{arkslp.org}.

\subsection{Extracting \clean profiles}

We extracted deprojected radial profiles directly from the primary beam corrected \clean images by azimuthally averaging the emission \citep[as in][]{Marino2016}. This was performed by drawing a set of ellipses with the same orientation as the discs and different semi-major axes \citep{Marino2025}. For sufficiently inclined belts, we additionally masked sectors of the disc near their minor axes where the deprojected beam is $\geq30$\% the original beam major axis, as including these regions would artificially smooth the azimuthally averaged profile. The uncertainty was estimated as the square root of the average variance divided by the number of independent points being averaged. The latter was estimated from the length of the arcs being averaged divided by the beam major axis. We note that this method does not work for edge-on discs where the minor axis is not well resolved.

\subsection{Non-parametric modelling: \frank} \label{sec:frank_methods}
The \frank code recovers the deprojected and deconvolved radial profile of discs from ALMA visibilities under the assumption of axisymmetry by modelling the real component of the visibilities in 1D with a Fourier-Bessel series regularised by a fast Gaussian process \citep{Jennings2020}. Originally developed to model protoplanetary disc observations \citep{Jennings2022a, Jennings2022b}, the method has been adapted to model optically thin discs, such as debris discs, while also constraining the vertical aspect ratio \citep{Terrill2023}. Further adaptations of the method to fit the surface brightness in logarithmic space reduce oscillatory artefacts and prevent negative surface brightnesses from arising in the fitted profile\footnote{\href{https://github.com/discsim/frank}{https://github.com/discsim/frank}} \citep{Miley2024}, which improves the quality of the fit over previous versions. 

We adopted the updated \frank method to fit all discs, which can be run specifically for the ARKS sample via the pipeline, \texttt{arksia} \footnote{\href{https://github.com/jeffjennings/arksia}{https://github.com/jeffjennings/arksia}} \citep{arksia}. The position angle, inclination and centering offset of the discs were assumed to be those fitted in \citet{Marino2025} for consistency. Before the radial profiles were fitted, stellar flux values (or, to be precise, emission of any unresolved central component) were subtracted from the real component of the visibilities as a constant offset in amplitude; these values were assumed to be those fitted in \citet{Marino2025} if they were well-constrained, or equal to the stellar flux fitted from the SED otherwise. We note that while the stellar flux value for HD\,32297 was fitted in \citet{Marino2025}, it is potentially overestimated. The fitted stellar flux for HD\,32297 is consistent with 0 at 2$\sigma$, so we reverted to the SED flux value instead.  
For HD\,109573, we found that the uncertainties of the radial profile better converged near the star when assuming no stellar emission detected, so we conservatively did not subtract any stellar flux before fitting the radial profile. 

To fit the radial profile of each source, two hyperparameters were varied: $\alpha$ and $w_{\rm smooth}$. $\alpha$ sets the minimum signal-to-noise ratio (S/N) down to which \frank attempts to fit the visibilities. $w_\text{smooth}$ determines the amount of smoothing applied to the reconstructed power spectrum. We performed fits over a few sets of $\alpha$ and $w_\text{smooth}$ values for each system, choosing the best values according to two criteria: the radial brightness profile should capture structural detail in the disc without introducing obvious oscillatory artefacts (a sign of over-fitting), and the reconstructed power spectrum should reasonably reproduce that from the observations without over-fitting to baselines that are undersampled and carry high uncertainty. Discriminating between oscillatory artefacts and real, narrow structures is not always trivial because the observed visibility distributions do not constrain structure in the radial brightness profile below some minimum spatial scale; across sources, we chose more conservative values of $\alpha$ and $w_{\rm smooth}$ such that the observed visibilities were accurately modelled by \frank down to a S/N of $\approx 2$.   
Two additional hyperparameters, the number of points in radius in a fit $N$ and an outer radius $R_{\rm out}$ beyond which the brightness is treated as 0, were chosen for each source to ensure the full radial extent of disc emission was modelled over a dense grid. Table~\ref{table:hyperparameters} summarises the most appropriate hyperparameters selected to model each disc. 

We note a few caveats related to the \frank fits. Firstly, \frank does not account for the effect of the primary beam, which could bias the profiles of large discs ($>$5--10\,arcsec) to lower surface brightnesses in their outer regions. However, comparison with other modelling methods used in this study shows that this effect is insignificant. 
Secondly, uncertainties on the \frank radial profiles are estimated using the diagonal entries of the covariance matrix of the Gaussian process model evaluated using the best-fit power spectrum \citep{Jennings2020}, which is likely an underestimate of the true uncertainties. 
Finally, the updated \frank code that significantly reduces artefacts by fitting the surface brightness in logarithmic space is only optimised for a scale height of 0 ($h=0$). We find that for the range of scale heights measured in this sample \citep{Zawadzki2025}, the scale height assumption used when fitting the disc does not have a significant effect on the fitted radial profile. Considering that the reduction of oscillatory artefacts and avoidance of negative surface brightness values outweighs the potential inaccuracies of performing the fits under the assumption that the disc is thin, we proceeded with the updated version of \frank assuming $h=0$. The \frank method can nonetheless be applied to constrain the vertical aspect ratio of discs, the details of which are described in \citet{Zawadzki2025}.

\subsection{Non-parametric modelling: \rave} \label{sec:rave_methods}
\rave recovers the radial profile of a disc by fitting models of concentric annuli to its image, solving for the brightness required of each annulus to reproduce the image \citep{Han2022}. To mitigate the dependence of the fitted profile on the placement of annuli boundaries, the fit is repeated 100 times, each time with a different partitioning of annuli boundaries, which are chosen in a Monte Carlo fashion. The degree of scatter between the radial profiles fitted in each iteration was used to estimate the uncertainties on the radial profile \citep{Han2022}. 

Designed to be compatible with observations over a range of wavelengths and imaging modes, \rave works in image space and takes the \clean image as input when fitting to the ARKS sample. The values of the robust parameters used to generate the \clean images that are used by \rave are listed in Table~\ref{table:hyperparameters}. Similar to \frank, stellar emission was subtracted from the \clean images before the radial profiles were fitted. The stellar flux values were assumed to be those fitted in \cite{Marino2025} if they were well-constrained, or the SED-fitted stellar flux otherwise. 

Two versions of \rave, optimised for edge-on \citep{Han2022} and non-edge-on \citep{Han2025} discs, respectively, are available. The former fits directly to the flux profile obtained by summing the image onto the major axis, which more effectively recovers the radial profile when the central cavity of the disc is not resolved, whereas the latter fits to the azimuthally averaged profile of the image, which is more suitable otherwise. Following the approach in \cite{Han2025}, we applied the edge-on version to the only three highly inclined discs in the sample that do not have a resolved central cavity (HD\,39060/$\beta$~Pic, HD\,197481/AU~Mic and HD\,32297). For the other discs, the non-edge-on optimisation was used. To confirm the validity of this choice, we fitted the edge-on discs with the non-edge-on version, and vice versa. Inspection of all discs fitted with both methods confirms that this choice of method produced lower uncertainties and oscillatory artefacts, and only the radials profiles fitted by the more appropriate method is used for the rest of the analysis in this paper. 

\rave requires disc images to be appropriately centred and rotated with a measured (or assumed) inclination for the appropriate model annuli to be simulated. We assumed the same position angle, centring offset and inclination as those fitted in \citet{Marino2025}. Given a radial range over which disc emission is detected, the only hyperparameter of \rave which determines the fitted radial profile was then the number of annuli used to fit the disc, $N$. We experimented with a range of $N$, choosing the maximum value of $N$ that does not generate significant oscillatory artefacts in the fitted profile, beyond which the uncertainties on the radial profile increase, and which generally results in the average annuli being wider than half the beam size. The values of $N$ used in the best-fit profiles are displayed in Table~\ref{table:hyperparameters}. 

We note a few caveats relating to the \rave fits. Firstly, unlike \frank, the \rave fits account for the effect of the primary beam by fitting to the primary beam-corrected \clean images. Secondly, the \rave fit depends on the \clean image. Empirically, the dependence of the fitted profile on the choice of the visibility weighting used to produce the \clean image is found to be low for reasonable robust parameter values. This is confirmed by experimenting with multiple \clean images with different robust parameters, since \rave bins the image into discrete annuli that effectively smooths over resolution elements in the \clean image. Finally, unlike the \frank radial profiles which were fitted assuming $h=0$, the \rave profiles presented here were fitted using the best-fit scale height aspect ratio derived with \rave.
Details of the scale height fitting procedure with \rave are described in \citet{Zawadzki2025}, in which Table 3 lists the best-fit aspect ratios. To briefly summarise, the likelihood of a given aspect ratio is assessed by fitting the radial profile assuming that scale height and computing the squared residuals, and the squared residuals across a range of aspect ratio values were assessed to find the best-fit aspect ratio.

\subsection{Parametric models}
\label{sec:parametric_models}
\subsubsection{Modelling code}
\efm{We parametrically modelled radial structures in the continuum emission of our targets with the debris disc modelling code adapted by \cite{Fehr2023}. This approach complements our non-parametric models by allowing us to directly compare results to theoretical predictions and calculate the significance of substructures. We provide a brief overview of the radial aspects of the parametric modelling code in this section, which is described more fully by the companion paper \cite{Zawadzki2025}.}

\efm{The parametric modelling code creates a temperature profile and mass opacity for each disc, assuming that the dust emission is in blackbody equilibrium with the stellar emission. The code then employs the affine-invariant Markov Chain Monte Carlo (MCMC) implementation in \texttt{emcee} \citep{Foreman-Mackey2013} to fit vertical and radial surface density profiles, obtaining values for the physical parameters. These parameters are used to create a model image of the dust structure from the temperature and surface density profiles. The resulting parametric disc model is inclined and rotated based on free parameters and transformed into a two-dimensional sky-plane projection that, like \frank and \rave models, is axisymmetric. We use Galario \citep{Tazzari2018} to convert the model image into synthetic visibilities sampled at the same antenna separations and orientations as the data. Then, we calculate a $\chi^2$ metric by comparing the synthetic visibilities with observed visibilities at each MCMC step.}

\begin{table*}
    \centering
    \caption{Definitions of functional forms for radial parametric modelling. }
    \label{tab:formdefs}
    \begin{tabular}{l l l}
    \hline \hline
    \multicolumn{1}{c}{Function} & \multicolumn{1}{c}{Surface density profile} & \multicolumn{1}{c}{Parameters} \\
    \hline
    
    
    Double power law & $\Sigma(r) = \left[\left(\dfrac{r}{R_{\rm{c}}}\right)^{-\alpha_{\text{in}} \gamma} + \left(\dfrac{r}{R_{\rm{c}}}\right)^{-\alpha_{\text{out}} \gamma}\right]^{-1/\gamma}$ & $R_{\rm{c}}, \alpha_{\text{in}}, \alpha_{\text{out}}, \gamma=2$ \\
    
    Triple power law & \makecell[tl]{$\Sigma(r) = \left(\dfrac{R_{\text{in}}}{R_{\text{out}}}\right)^{-\alpha_{\text{mid}}} \left[\left(\dfrac{r}{R_{\text{in}}}\right)^{-\alpha_{\text{in}} \gamma_{\text{in}}} + \left(\dfrac{r}{R_{\text{in}}}\right)^{-\alpha_{\text{mid}} \gamma_{\text{in}}}\right]^{-1/\gamma_{\text{in}}}$ \\ $\textcolor{white}{\Sigma(r) =} \times \left[\left(\dfrac{r}{R_{\text{out}}}\right)^{-\alpha_{\text{mid}} \gamma_{\text{out}}} + \left(\dfrac{r}{R_{\text{out}}}\right)^{-\alpha_{\text{out}} \gamma_{\text{out}}}\right]^{-1/\gamma_{\text{out}}}$} & 
    \makecell[tl]{
    $R_{\text{in}}, R_{\text{out}}, \alpha_{\text{in}}, \alpha_{\text{mid}}, \alpha_{\text{out}}, $ \\
    $\gamma_{\text{in}}=2, \gamma_{\text{out}}=2$
    } \\
    
    Power law + error function & $\Sigma(r) = \left[1-\text{erf}\left(\dfrac{R_{\rm{c}}-r}{\sqrt{2} \sigma_{\text{in}} R_{\rm{c}}}\right)\right] \left(\dfrac{r}{R_{\rm{c}}}\right)^{-\alpha_{\text{out}}}$ & $R_{\rm{c}}, \sigma_{\text{in}}, \alpha_{\text{out}}$ \\
    
    
    Gaussian & $\Sigma(r) = \exp\left[-\dfrac{(r-R)^2}{2 \sigma^2}\right]$ & $R, \sigma$ \\
    
    Asymmetric Gaussian &
    \(
    \begin{array}{l} 
        \Sigma(r) = 
        \begin{cases}
            \exp\left[-\dfrac{(r-R_{\rm{c}})^2}{2 \sigma_{\text{in}}^2}\right] & \text{if } r<R_{\rm{c}}\\
    \exp\left[-\dfrac{(r-R_{\rm{c}})^2}{2 \sigma_{\text{out}}^2}\right] & \text{if } r \geq R_{\rm{c}}
        \end{cases}
    \end{array}
    \)
    & $R_{\rm{c}}, \sigma_{\text{in}}, \sigma_{\text{out}}$\\
    
    Double Gaussian & 
    \makecell[tl]{
    $\Sigma(r) = C \times \exp\left[-\dfrac{(r-R_1)^2}{2 \sigma_1^2}\right]$\\ 
    $\textcolor{white}{\Sigma(r) =} + (1-C) \times \exp\left[-\dfrac{(r-R_2)^2}{2 \sigma_2^2}\right]$ 
    } & 
    $R_1, R_2, \sigma_1, \sigma_2, C$\\
    
    Triple Gaussian & 
    \makecell[tl]{
    $\Sigma(r) = C_1 \times \exp\left[-\dfrac{(r-R_1)^2}{2 \sigma_1^2}\right]$ \\
    $\textcolor{white}{\Sigma(r) =} + C_2 \times \exp\left[-\dfrac{(r-R_2)^2}{2 \sigma_2^2}\right]$ \\
    $\textcolor{white}{\Sigma(r) =} + (1-C_1-C_2) \times \exp\left[-\dfrac{(r-R_3)^2}{2 \sigma_3^2}\right]$ 
    } & 
    $R_1, R_2, R_3, \sigma_1, \sigma_2, \sigma_3, C_1, C_2$ \\
   
    Gaussian + double power law & 
    \makecell[tl]{
    $\Sigma(r) = C_1\times\exp\left[-\dfrac{(r-R)^2}{2 \sigma^2}\right]$ \\
    $\textcolor{white}{\Sigma(r) =} + (1-C_1)\times \left[\left(\dfrac{r}{R_{\rm{c}}}\right)^{-\alpha_{\text{in}} \gamma} + \left(\dfrac{r}{R_{\rm{c}}}\right)^{-\alpha_{\text{out}} \gamma}\right]^{-1/\gamma}$ 
    } & 
    $R, \sigma, C_1, R_{\rm{c}}, \alpha_{\text{in}}, \alpha_{\text{out}}, \gamma=2$\\
    
    \hline
    \end{tabular}
    \tablefoot{All surface density profiles are multiplied by $10^{\Sigma_{\rm{c}}}$, the surface density normalisation. We fixed certain parameters at the values noted here to minimise degeneracies; exceptions for specific targets are stated explicitly.}
\end{table*}

\begin{figure*}
    \centering
    \begin{subfigure}[t]{0.24\textwidth}
        \centering
        \includegraphics[width=\textwidth]{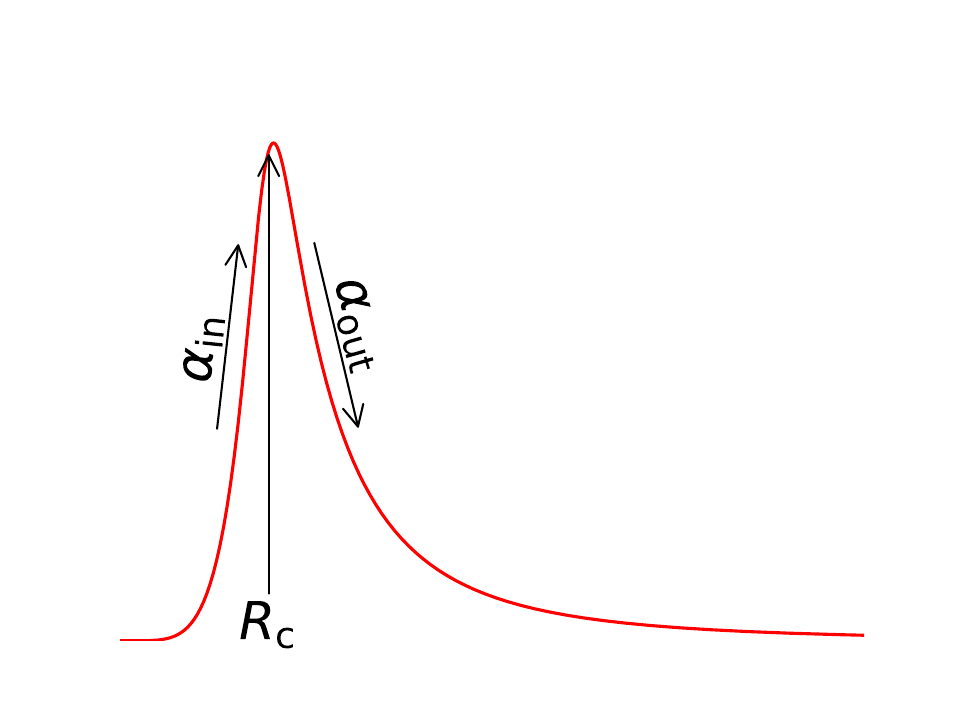}
        \caption{}
    \end{subfigure}
    \begin{subfigure}[t]{0.24\textwidth}
        \centering
        \includegraphics[width=\textwidth]{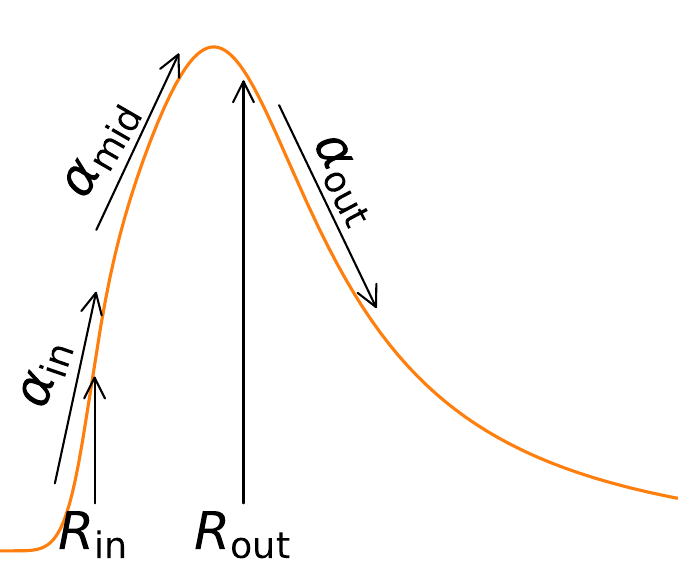}
        \caption{}
    \end{subfigure}
    \begin{subfigure}[t]{0.24\textwidth}
        \centering
        \includegraphics[width=\textwidth]{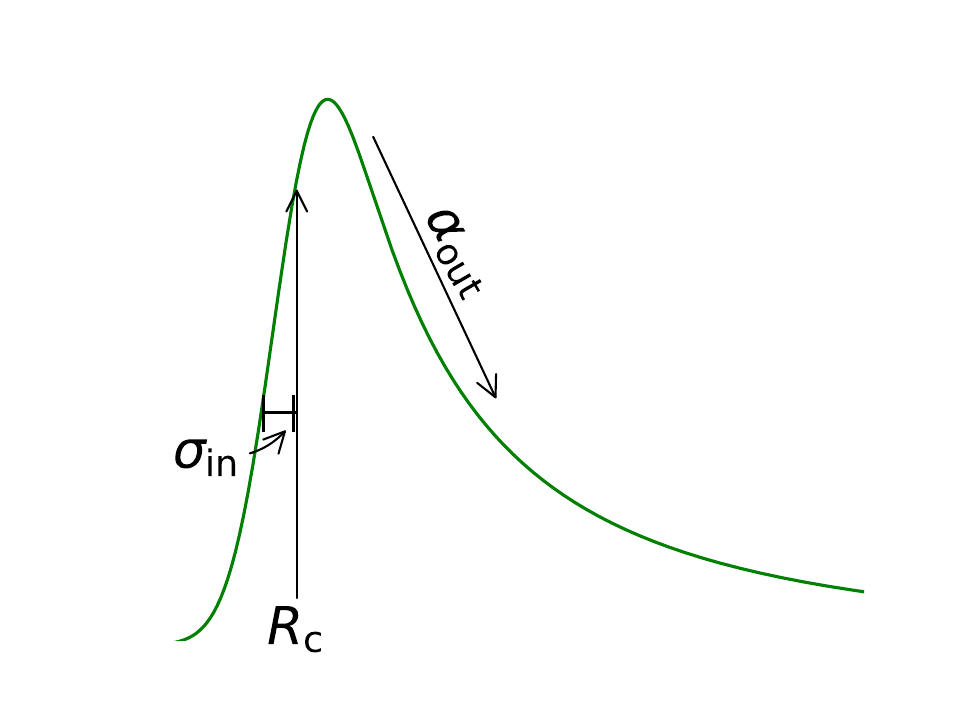}
        \caption{}
    \end{subfigure}
    \begin{subfigure}[t]{0.24\textwidth}
        \centering
        \includegraphics[width=\textwidth]{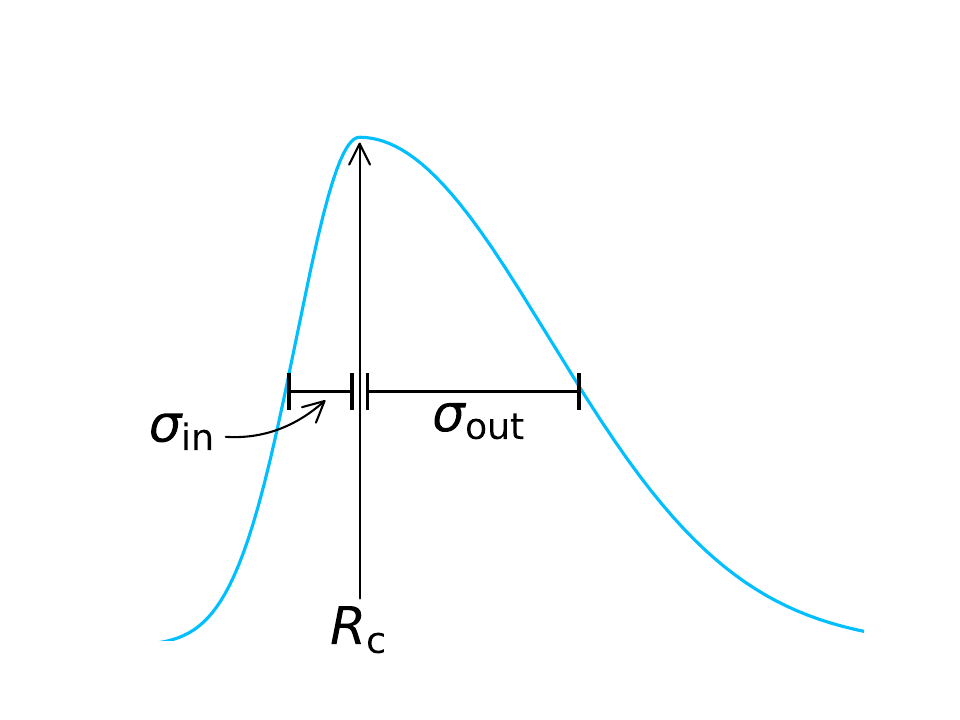}
        \caption{}
    \end{subfigure}
    \begin{subfigure}[t]{0.6\textwidth}
        \centering
        \includegraphics[width=0.45\textwidth]{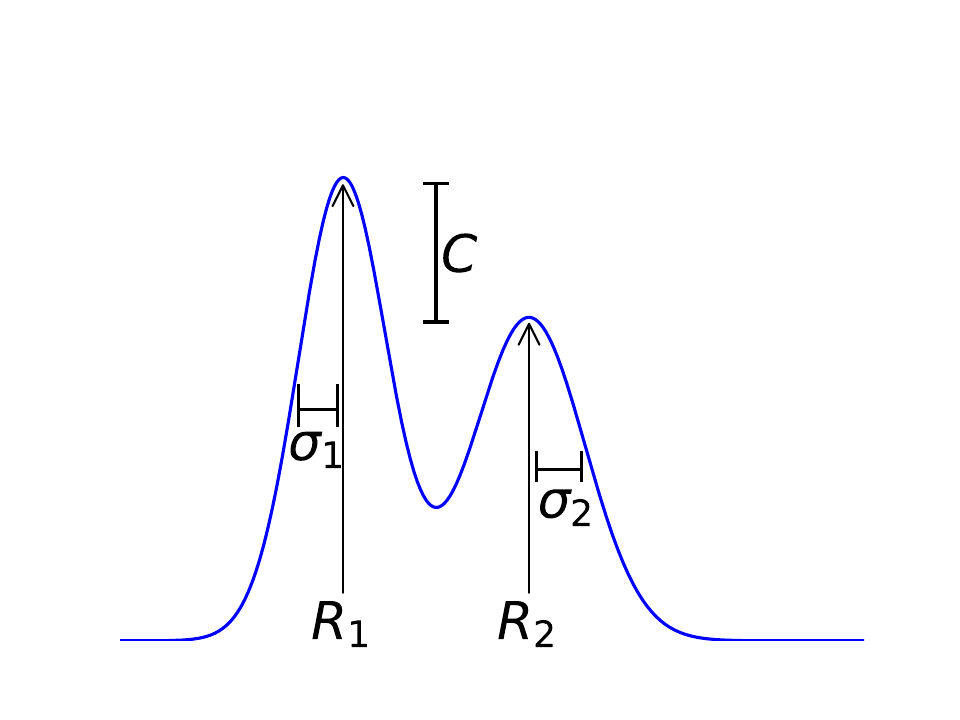}
        \includegraphics[width=0.45\textwidth]{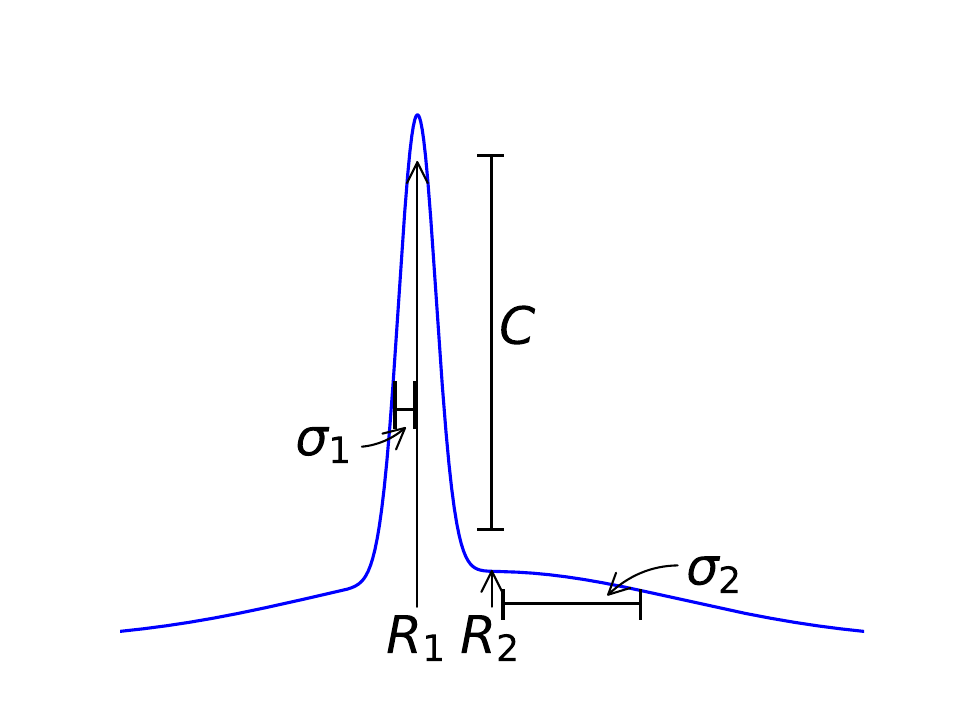}
        \caption{}
    \end{subfigure}
    \begin{subfigure}[t]{0.3\textwidth}
        \centering
        \includegraphics[width=\textwidth]{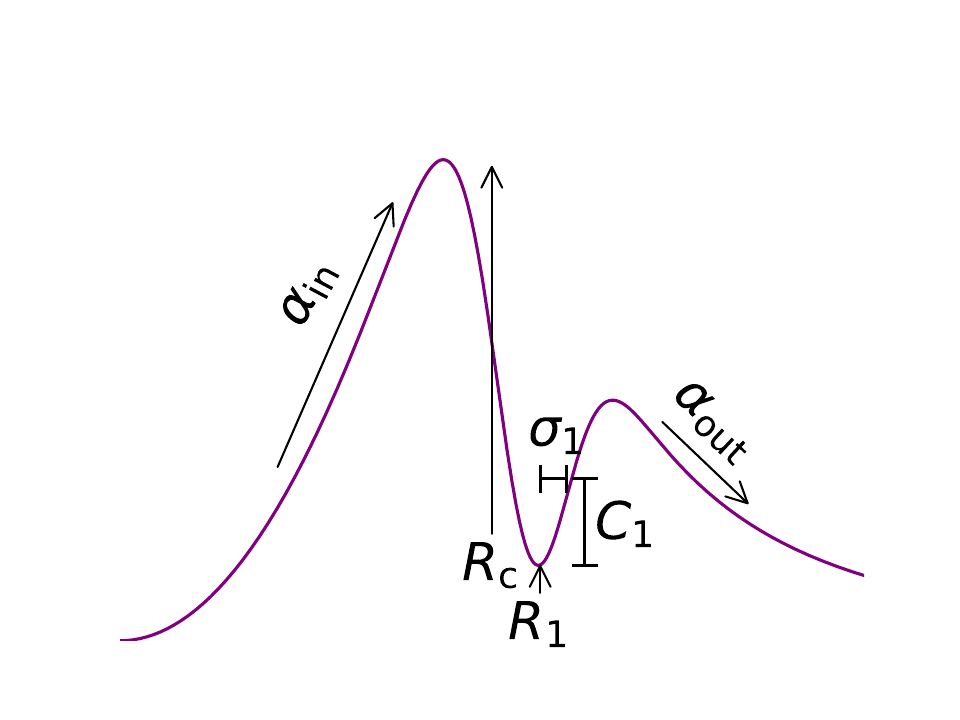}
        \caption{}
    \end{subfigure}
    \caption{Examples of the main radial functional forms from the parametric modelling code and their free parameters: double power law (a), triple power law (b), power law + error function (c), asymmetric Gaussian (d), double Gaussian (e), and double power law with Gaussian gap (f). The double Gaussian functional form is shown twice to demonstrate how it can model two rings or a ring with extended emission.}
    \label{fig:example_profiles}
\end{figure*}

\subsubsection{Radial profiles}
\efm{We fit the radial surface density profile to a variety of functional forms---consisting of combinations of Gaussians, power laws, and error functions---implemented in the parametric modelling code. These functional forms are defined in Table \ref{tab:formdefs} and visualised in Fig. \ref{fig:example_profiles}.}

\efm{We initially fit each disc with a double power law and with two independent Gaussians (hereafter referred to as a ``double Gaussian''), as justified by theoretical models and further supported by the non-parametric results. Power laws allow us to measure the steepness of the inner and outer edges---which are important parameters in models of debris disc structure---while minimising the number of free parameters \citep[e.g.][]{Marino2019,MacGregor2019}. Although Gaussians cannot always reproduce the inner and outer slopes of debris discs, these functions are simple and effective at measuring fractional widths, and are commonly used in the literature \citep[e.g.][]{Marino2016, White2017,Matra2025}.}

\efm{The results of the non-parametric models motivated our choice of additional functional forms. Asymmetric Gaussians and power law + error functions were used to model discs without obvious substructures, while possible gaps were modelled through two approaches. We either modelled the rings with two or three Gaussians, or we modelled any gap by subtracting a Gaussian ($G$, with amplitude $C$) from the best-fitting functional form ($\Sigma_0$) as $\Sigma(r) = \Sigma_0(r) - C \times G(r)$. When we parametrised a gap, we assumed a Gaussian distribution because it modelled the position, width, and depth of the gap with the fewest additional parameters.}

\efm{For every target, we fitted the flux of the host star ($F_\star$), surface density normalisation of the disc ($\Sigma_{\rm{c}}$), and the position of the disc in the sky, including position angle (PA), offset from the phase centre ($\Delta\alpha$ and $\Delta\delta$), and inclination ($i$), assuming that the star was located at the centre of the axisymmetric disc. Additional free parameters for the radial profile depend on the functional form and are outlined in Table~\ref{tab:formdefs}. As an overview of the table, Gaussians were defined by radius ($R$), width ($\sigma$), and amplitude ($C$); power laws were defined by central radius ($R_{\rm{c}}$) or transition radii ($R_{\rm{in}}$ and $R_{\rm{out}}$), slope ($\alpha$), and the sharpness of the transition ($\gamma$); and error functions were defined by central radius ($R_{\rm{c}}$), width ($\sigma$), and slope ($\alpha$).}

\subsubsection{Vertical profiles}
\efm{We minimise degeneracies between radial and vertical structure by conducting separate steps of modelling, one in which we vary only radial parameters and another in which we vary both radial and vertical parameters. This paper is entirely focused on the former round of modelling; the latter is described by \cite{Zawadzki2025}. We therefore assume that the vertical profile is Gaussian for all targets in this paper. We defaulted to fixing the vertical aspect ratio $h_{\rm{HWHM}}$---defined as the half width at half max of the vertical profile and assumed to be constant throughout the disc---at 0.015, adjusting the value for certain discs based on results from \citet{Zawadzki2025}. When $h_{\rm{HWHM}}=0.015$ did not fall between the 16th and 84th percentiles of the posterior distributions of the models that fit the vertical structure, we instead used approximate values obtained from preliminary vertical modelling. These exceptions were HD\,9672 ($h_{\rm{HWHM}}=0.07$), HD\,10647 ($h_{\rm{HWHM}}=0.055$), HD\,14055 ($h_{\rm{HWHM}}=0.0475$), HD\,15115 ($h_{\rm{HWHM}}=0.02$), HD\,32297 ($h_{\rm{HWHM}}=0.012$), HD\,39060 ($h_{\rm{HWHM}}=0.0445$), HD\,61005 ($h_{\rm{HWHM}}=0.02$), and HD\,131488 ($h_{\rm{HWHM}}=0.005$).}

\subsubsection{Parametric model comparison}
\efm{We compared different parametrisations for each disc with the Akaike information criterion (AIC) and the Bayesian information criterion (BIC), which are statistical measures designed to compare the quality of fits for models with differing numbers of parameters and penalise unnecessary complexity. The AIC and the BIC are defined similarly, but the BIC takes the number of data points (in this case, the sum of the real and imaginary components in the visibility domain) into consideration and, therefore, penalises additional parameters more heavily than the AIC. BIC differences are considered to provide strong evidence of a better fit when greater than 10 \citep{Kass1995}, while AIC differences are converted to a Gaussian probability interval to obtain confidence levels.}

\begin{figure*}
    \centering
    \includegraphics[width=1.0\linewidth]{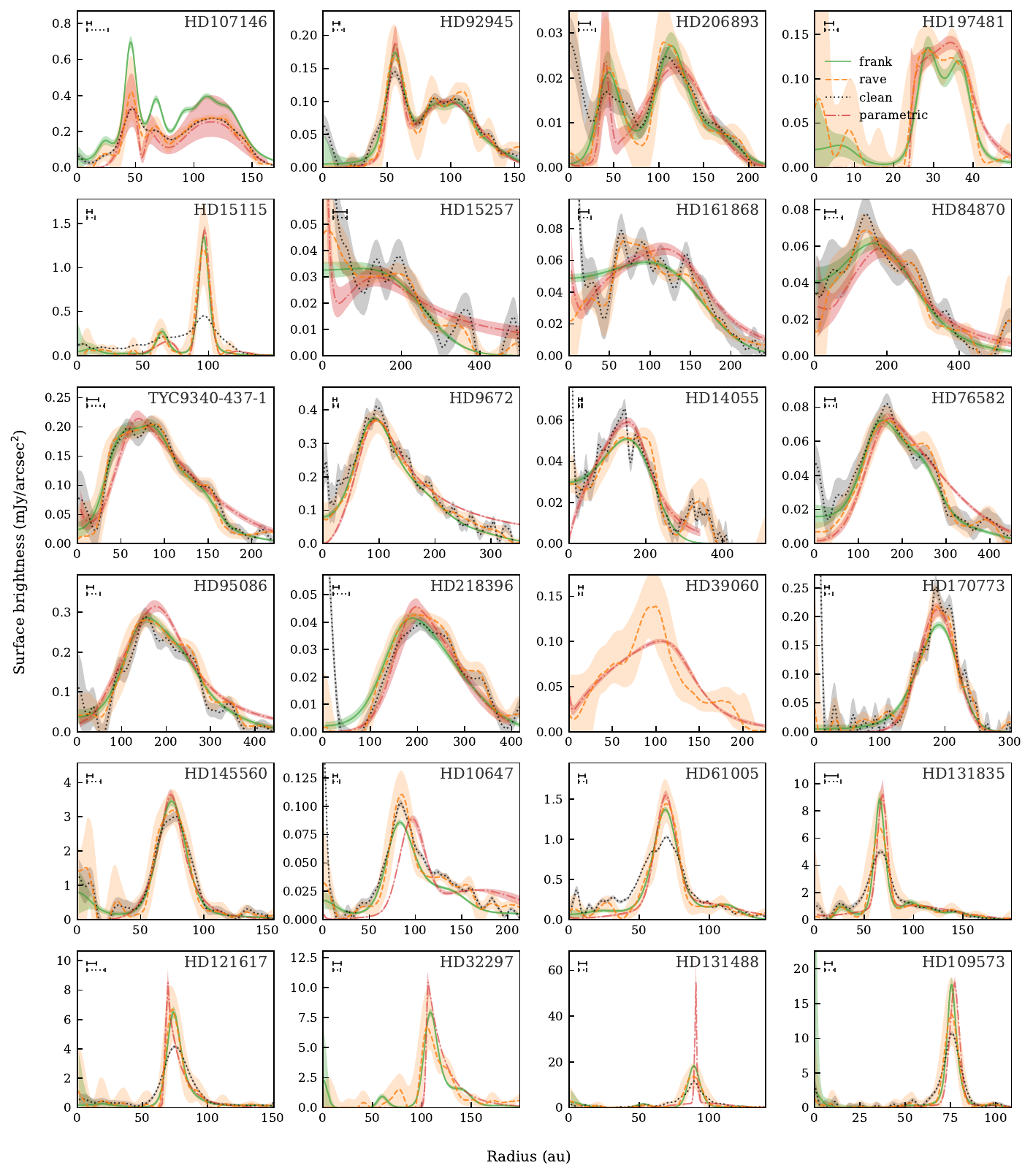}
    \caption{Azimuthally averaged radial profiles derived from the \clean image (dotted black) and the deconvolved and deprojected surface brightness profiles fitted with \frank (solid green), \rave (dashed orange), and parametric models (dash-dotted red). The robust parameters used for \clean imaging are listed in Table~\ref{table:hyperparameters}. The solid bars in the upper left corner of each panel indicate the FWHM resolved by \frank if the radial profile were infinitesimally narrow (i.e. the \frank PSF). The dotted bars indicate the synthesised beam size. In general, oscillations in the \clean profile that are likely due to noise (e.g. as seen in HD\,15257 and HD\,161868) are smoothed out in the \frank, \rave, and parametric models. The deconvolved peaks recovered by the \frank, \rave, and parametric models are often sharper than seen in the (beam-convolved) \clean images;  the most extreme case is the best-fitting parametric model for HD\,131488. We note that not all methods are applicable to every disc (e.g. azimuthal averaging was not applied to edge-on discs to extract the \clean profile, and \frank was not applied to HD\,39060 due to mosaicking being used in the observations). The offset between \rave and \frank for HD\,107146 is due to the two methods fitting at different effective wavelengths.}
    \label{fig:rp_brightness}
\end{figure*}

\efm{Although AIC and BIC values were the primary basis for designating a functional form as the best fit for a given disc, we took into consideration radial profile visualisations (Fig. \ref{fig:rp_brightness}) and residuals (Fig. \ref{fig:parametric_DMR}) when the AIC and the BIC disagreed or when the differences between functional forms were not significant. This required considering parametric results on a case-by-case basis, and we describe the process of designating the best functional form for each disc in Sect. \ref{sec:results}. While we designate only one functional form as the best for each disc, other functional forms often provided a good fit to the data as well. We report multiple results for several discs, and we sometimes use other functional forms to obtain radial parameters of interest. When a functional form with the radial parameter of interest was not available, we extracted it from the radial profile of the functional form designated as the best in Table \ref{tab:AICBIC}. We describe this methodology here.}

\efm{For each disc, we find the inner and outer slope with a double power law or triple power law, sometimes with one or more Gaussian gaps added, and we find the outer transition radius with a triple power law where available or by measuring the radius at 5\% of the peak in brightness from the radial profile of the best functional form. This reliance on power laws limited our ability to measure slopes and outer radii for discs with more complicated structures; for example, we could only measure the inner slope of the innermost ring and the outer slope of the outermost ring for multi-ring discs. For each ring, we find the central radius with a double power law or with multiple Gaussians, and we find $\Delta R$ by multiplying $\sigma$ by 2.355 for discs best fit by multiple Gaussians or by measuring the FWHM of the best functional form from its radial profile. In each case, the choice of a functional form to measure the slope or the central radius was based on the AIC and the BIC differences.} The fitted radial profiles under the different parametrisations attempted for each disc are shown in Fig.~\ref{fig:parametric_profiles}.

\section{Results} \label{sec:results}

\subsection{Radial profiles} \label{sec:results_radialprofiles}
Figure~\ref{fig:rp_brightness} summarises our radial structure modelling results, displaying the surface brightness profiles determined using \clean, \frank, \rave, and the best parametric model for each disc. The gallery provides an overview of the kinds of radial structures suggested by the ARKS observations, based on which parametric models were developed and tested. 
We note that while the \frank, \rave, and parametric profiles are deconvolved, the \clean profiles are convolved with the beam. 
The model images and residuals of the fitted \frank, \rave, and parametric models are displayed in Figs.~\ref{fig:frank_residuals}, \ref{fig:rave_residuals}, and \ref{fig:parametric_DMR}. 

\begin{figure*}[htb!]
    \centering
    \includegraphics[width=1.0\linewidth]{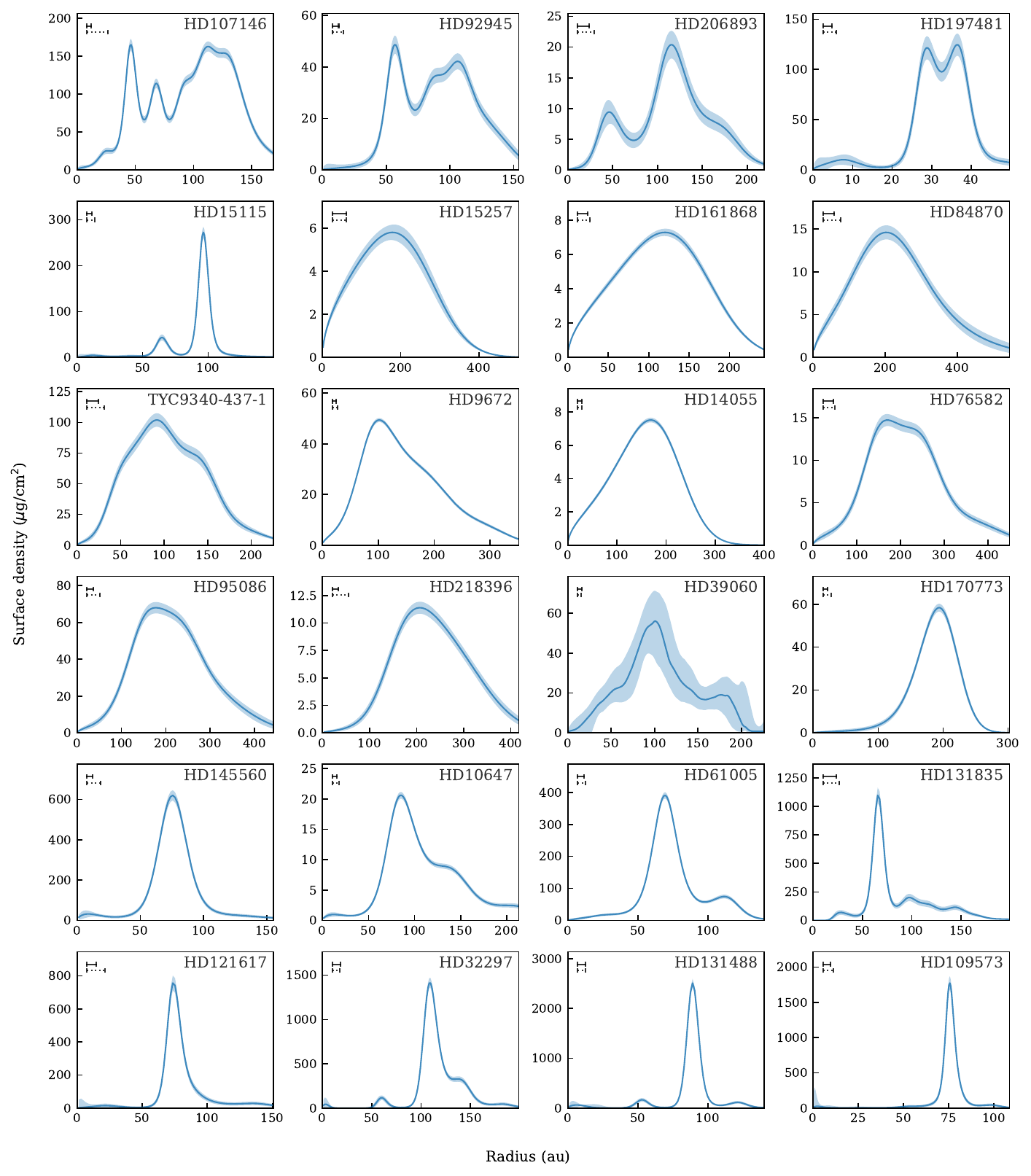}
    \caption{Deconvolved and deprojected surface density profiles fitted with \frank, except for $\beta$~Pic (HD~39060) which was fitted with \rave. These profiles are based on the surface brightness profiles presented in Fig.~\ref{fig:rp_brightness}, but offers a cleaner and simpler alternative to that figure which includes a more comprehensive summary of the different approaches used to model the radial structure in this study. The bars in each panel are the same as those described in Fig.~\ref{fig:rp_brightness}. }
    \label{fig:rp_density}
\end{figure*}

Fig.~\ref{fig:rp_density} displays a gallery of the non-parametric radial surface density profiles, which were converted from the surface brightness profiles in Fig.~\ref{fig:rp_brightness} assuming a blackbody equilibrium temperature profile, stellar luminosities listed in \citet{Marino2025} and the mass opacity described in \citet[1.9\,cm$^2$\,g$^{-1}$ in Band 7, 1.3\,cm$^2$\,g$^{-1}$ in Band 6, and the mean value between the two for the combined Band 6 and 7 data for HD\,39060, or $\beta$~Pic]{Marino2025}. These profiles are based on \frank fitting (except for $\beta$~Pic, for which the \frank model does not appropriately account for multiple pointings and the \rave model was used instead), and are expected to represent the physical distribution of millimetre-sized dust in the disc under these temperature profile and opacity assumptions, and likely also the underlying planetesimal distributions from which the observed dust is produced \citep{Hughes2018}.

The horizontal bars in the lower left corner of each panel in Figs.~\ref{fig:rp_density} and \ref{fig:rp_brightness} provide an indication of the resolution of the radial profiles. Specifically, the solid bar indicates the FWHM of the radial profile recovered by \frank if an infinitesimally narrow ring (i.e. a radial profile described by a Dirac delta function) were to be observed by ALMA (under the same configuration as the data), with the same total flux as the disc, same noise level, and recovered by \frank with the same hyperparameters \citep[as done in][]{Sierra2024}. The longer dotted bar indicates the synthesised beam size when imaged with \clean under Briggs weighting and a robust parameter of 0.5.

The radial profiles obtained from each of the three deprojection and deconvolution approaches (\frank, \rave and parametric modelling) are generally in close agreement, confirming the robustness of the modelling for each system. The fitted non-parametric profiles typically smooth out noise in the \clean profile, and in many cases recover sharper peaks than the beam-convolved image, particularly for the radially narrower discs (e.g. HD\,15115, HD\,121617, HD\,109573, HD\,61005, HD\,107146, HD\,131835, and HD\,131488). For several systems (e.g. HD\,15257, HD\,161868, and HD\,84870), the \frank and \rave profiles differ in their balance between the smoothing of noise and the recovery of sharp features, which are partly controlled by the independent choice of hyperparameter values for each method. 
The \rave uncertainties empirically reflect the range of values within which various possible radial profile shapes could take, and are not conditioned upon a particular radial profile shape. These uncertainties are generally larger than those derived by \frank, as has also been discussed in previous work that compares the two methods \citep{Han2025}. 
The parametric models further explore the significance of each feature suggested by the non-parametric profiles, also yielding close agreement with the \frank and \rave profiles and sometimes finding even sharper features than non-parametric models (e.g. HD\,131488, HD\,121617 and HD\,206893)\efm{, particularly when the width of the parametric model is narrower than the resolution of the data}.

\begin{figure}[htb!]
    \centering
    \includegraphics[width=0.98\linewidth]{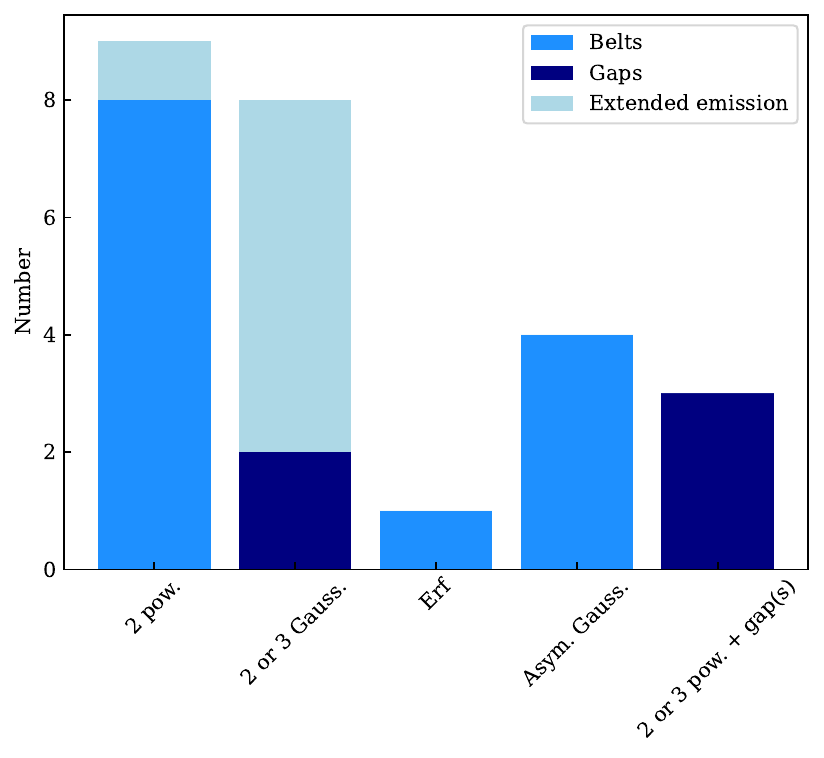}
    \caption{Distribution of best functional forms for all targets, as determined by weighing the AIC confidence intervals, BIC differences, profile plots, and residuals (described in Sect. \ref{sec:parametric_models}).}
    \label{fig:fforms-hist}
\end{figure}

Perhaps one of the strongest conclusions that could be drawn from the galleries of radial profiles is the diversity of radial structures exhibited by debris discs. 
None of the profiles appear to be fully described by a single Gaussian, which is a common assumption of previous works appropriate for lower-resolution observations. 
Figure~\ref{fig:fforms-hist} summarises the best-fit functional forms for the parametric models, demonstrating that double power laws and double Gaussians are the best-fit form for the vast majority of discs in the sample, and that only a minority (7/24) of discs require alternative parametrisation (with 4/24 best fitted by an asymmetric Gaussian and 3/24 requiring a gap to be explicitly parametrised). 

We attempted to broadly classify the range of radial profiles within the sample. We first divided the sample into single-ring discs and multi-ring discs. Within the single-ring discs, we find that a sizeable subset shows evidence of additional low-amplitude features. Given the subtleness of such features, the \frank, \rave, and parametric fits do not always agree or cannot always clearly distinguish between whether these are additional faint rings or halos attached to the main ring. We therefore categorise these single-ring discs with low-amplitude features as a distinct category from those without such features in the radial profile. 

To denote these features, we use an abbreviated notation which appends symbols that represent substructures onto the first-order single-ring (denoted $R$) vs. multi-ring (denoted $M$) classification of a given disc. As a general rule, superscripts denote the number of rings in a given disc (e.g. $M^2$ is a disc with 2 rings separated by a gap), but is omitted for discs with only one ring (e.g. $R$). Subscripts denote any additional structures, including any halos denoted as $h$ either interior (e.g. $_hR$) or exterior (e.g. $R_h$) to a disc, or bumps or ``shoulders'' on the outer edge of a disc (e.g. $R_{s}$). Parentheses may be added to denote that a feature is tentative (e.g. $R_{(s)}$). 
To indicate the presence of gas in any disc for convenience of understanding its basic features, we use the $g_+$ tag to indicate gas-rich discs and $g_-$ to indicate gas-poor discs. No gas tags are added to discs without gas detections. 
We note that this classification is purely phenomenological and is not designed to reflect their underlying physical origins and the fact that the boundaries between categories may be blurry and different features may lie on a continuum of shapes. 

To clarify terminology used in the remainder of this section as we describe the characteristic inner and outer edges and central radius for each classification, we define discs with $\alpha > 15$ in parametric models with power-law edges as a steep slope based on the cut-off for planet sculpting \citep{Pearce2024} and define $R>100$\,au as a disc with a large radius. Emission of a feature is considered low in amplitude if it is lower than 50\% of the main peak, and gas-rich is defined as an estimated gas mass above $10^{-4} \, \mathrm{M}_\oplus$, while gas-poor is defined by a gas detection lower than this threshold. 

The remainder of this section describes the modelling for each individual system. We continue the discussion of the disc properties of  the sample in Sect.~\ref{sec:measurements}. 

\subsubsection{Group [$M^n$]: Multi-ring discs}
This group includes discs with multiple rings separated by radial gaps. 
The non-parametric profiles suggest that eight discs in the ARKS sample have multiple rings based on their surface density profile, \efm{of which five have at least one significant gap when the parametric models are tested for statistical significance with the AIC and the BIC.} These eight discs are listed in this section, with the remainder (parametrically not significant) discs discussed later in the single-ring discs with low-amplitude emission category (Sect.~\ref{sec:Rx}). \efm{The multi-ring discs are typically located at a small radius (mean $\rm{R}=72$\,au), with steeper inner (mean $\alpha_{\rm{in}}=16$) and outer (mean $\alpha_{\rm{out}}=-11$) edges than found among single-ring discs. The gap is often wide and deep in these discs. The inner ring is often narrower than the outer ring, with a mean $\Delta R/R=0.21$ for inner rings and a mean $\Delta R/R = 0.43$ for outer rings. We measure the locations and widths of the peaks from Gaussian parametrisations, and we measure the locations and widths of the gaps from double power law and triple power law functional forms with Gaussian gaps added. The best-fit and median values for the double-Gaussian parametrisation of these discs are described in Table \ref{tab:MCMC-gaussian}.}

\noindent (i) HD\,107146 [$M^3$]

The outer regions of the disc were observed in scattered light by HST and modelled by \citet{Ertel2011}. The disc was then imaged by ALMA and modelled parametrically as a wide belt with one gap in \citet{Marino2018, Marino2021} and with two gaps in \citet{ImazBlanco2023} following the discovery of a bump in the middle of a wide gap using \frank and higher-resolution data. Here, we confirm the double gap structure. The intermediate belt is consistently recovered by both \frank and \rave, which manifests in the beam-convolved \clean profile as a low-amplitude bump and appears to be a real feature. The offset between the \frank and \rave profiles for this system in Fig.~\ref{fig:rp_brightness} is due to the fact that the \rave profile was fitted to the \clean image created by combining band 6 and 7 observations (mean wavelength of 1.0~mm) which improves the S/N but lowers the brightness compared to using band 7 only \citep{Marino2025}, whereas the \frank profile was fitted to the band 7 (mean wavelength of 0.86~mm) visibilities only. 

\efm{We parametrically modelled HD~107146 with five functional forms---a double Gaussian, triple Gaussian, triple power law, triple power law with one gap, and triple power law with two gaps---to determine the significance of these previously observed gaps at the higher resolution of the ARKS observations. We find that the triple power law with two Gaussian gaps is significantly preferred over all other functional forms by both the AIC and the BIC, providing strong evidence that there are three rings in HD~107146. We therefore designate the triple power law with two Gaussian gaps as the best functional form for this disc (MCMC results described in Table \ref{tab:snowflake-params}), although we rely on the triple Gaussian to describe the rings. This is the only disc in the ARKS sample with more than one significant gap}

\efm{As parametrised by the triple Gaussian, the three rings are located at $46.2_{-0.5}^{+0.3}$\,au, $66.3_{-0.7}^{+0.6}$\,au, and $118.07_{-0.11}^{+0.12}$\,au. The inner and centre rings are approximately the same width ($\Delta R=13.2_{-1.2}^{+0.9}$\,au and $\Delta R=8.2_{-1.9}^{+2.4}$\,au, respectively), while the outer ring is much wider ($\Delta R=58.7_{-0.4}^{+0.6}$\,au). The inner ring is the brightest. As parametrised by the triple power law with two Gaussian gaps, the two gaps are located at $56.1_{-0.3}^{+0.5}$\,au and $78.1_{-0.8}^{+0.6}$\,au. The inner gap ($\Delta R = 6.83_{-0.94}^{+1.41}$\,au) is narrower than the outer gap ($\Delta R = 39.1_{-1.9}^{+2.4}$\,au), while the outer gap ($C=0.693_{-0.016}^{+0.025}$) has a slightly greater amplitude than the inner gap ($C=0.57_{-0.9}^{+0.7}$). These results are consistent with the ARKS non-parametric models as well as modelling conducted by \cite{ImazBlanco2023}.}

\efm{As high-resolution multiband data was available for this disc, we additionally fit the emissivity index with the parametric modelling code; results are described in Appendix \ref{app:emissivity}.}

\noindent (ii) HD\,92945 [$M^2$]

The disc has been modelled as a wide belt with a gap \citep{Marino2018, Marino2021}, which is consistent with our non-parametric modelling. On top of that, the \clean profile suggests a possible dip in the relatively wide outer belt, which is deconvolved into a slightly deeper gap by \rave and is also reflected in the \frank profile, though to a lesser extent. Given the slightly lower S/N of the observations compared to HD\,107146, as reflected in the larger uncertainties in the \clean profile, the possibility that there exist three rather than two rings in the system appears to be less certain than in HD\,107146, so we label the disc as $M^2$ for now based on the present dataset. 

\efm{We parametrically modelled HD~92945 with seven functional forms---a double Gaussian, double Gaussian with one gap, triple Gaussian, double power law, double power law with one gap, double power law with two gaps, and triple power law with one gap---to determine the significance of both possible gaps suggested by the non-parametric models. Both the AIC and the BIC indicate that the double power law is significantly worse than the other functional forms, providing strong evidence that there is at least one gap in the disc. The AIC and the BIC consistently prefer the double Gaussian parametrisation, indicating that the additional complexity of parametrising a second gap or a third ring is not justified by a reduction in $\chi^2$. However, we emphasise that higher S/N data are necessary to confirm the presence or absence of a second gap around 100\,au. We therefore designate the double Gaussian as the best functional form for this disc, although we rely on the triple power law with one gap to parametrise the gap itself.}

\efm{As parametrised by the double Gaussian, the two rings are located at $57.0\pm0.8$\,au and $102.7\pm1.5$\,au, overlapping with two rings observed in infrared \citep{Golimowski2011}. The inner ring ($\Delta R = 13_{-3}^{+3}$\,au) is wider than the outer ring ($\Delta R = 59_{-4}^{+5}$\,au). As parametrised by the triple power law with one Gaussian gap, the gap is centred at $R=74.4\pm2.0$\,au, with a width of $\Delta R=21_{-9}^{+7}$\,au and fractional depth of $C=0.52_{-0.06}^{+0.05}$. These results are consistent with previous modelling \citep{Marino2019}.}

\noindent (iii) HD\,206893 [$M^2$]

We find two belts at radial locations consistent with previous non-parametric modelling with \frank, \rave, and parametric modelling \citep{Marino2020, Marino2021, Nederlander2021}. 
\efm{We parametrically modelled HD~206893 with six functional forms---a double power law, double power law with one gap, triple power law, triple power law with one gap, double Gaussian, and triple Gaussian---to explore various parametrisations of the gap. We find that the double Gaussian is strongly preferred over all other functional forms by the BIC, and marginally to moderately preferred by the AIC. We therefore designate the double Gaussian as the best functional form for this disc, although we rely on the double power law with a gap to describe the gap itself.}

\efm{As parametrised by the double Gaussian, the rings are located at $42\pm5$\,au and $127\pm4$\,au. The inner ring ($\Delta R = 14_{-9}^{+21}$\,au) is much narrower than the outer ring ($\Delta R = 85_{-7}^{+9}$\,au), while the outer ring is brighter than the inner ring. As parametrised by the double power law with a Gaussian gap, the gap is centred at $R=83_{-7}^{+6}$\,au, with a width of $\Delta R = 24_{-12}^{+26}$\,au and fractional depth $C=0.50_{-0.20}^{+0.30}$.}

\noindent (iv) HD\,197481 (AU~Mic) [$M^{2}$ ($M^{3}$)]

AU~Mic has been extensively observed by ALMA (e.g. \citealp{MacGregor2013, Daley2019, Vizgan2022}). The main belt at 30~au is resolved into two separate peaks by \frank, and to a lesser extent by \rave, which is more similar to a plateau with relatively sharp edges on either side. Furthermore, our \frank and \rave modelling supports the presence of an inner component \efm{at $<10$\,au}, which has been suggested in previous parametric \citep{MacGregor2013, Daley2019, Marino2021} and non-parametric modelling \citep{Han2022, Terrill2023}. The oscillation of this inner emission in the \rave profile is unlikely to be real, as reflected in the large uncertainty region, and more likely reflects the presence of inner emission that could be relatively sharp. Given the uncertainty of the inner emission, we label this disc as $M^{2}$ but include the $M^{3}$ label in brackets. 

\efm{We parametrically modelled HD~197481 with five functional forms---a double power law, triple power law, double Gaussian, triple power law with one gap, and triple Gaussian---to explore the potential inner component suggested by previous observations and the potential gap at $\sim32$\,au suggested by the ARKS non-parametric models. We find that the addition of the inner component does not reduce the value of $\chi^2$ enough to justify the increased number of parameters, as measured by both the AIC and the BIC. The significance of the gap at $\sim32$\,au is ambiguous, with the BIC significantly preferring the triple power law parametrisation and the AIC moderately preferring the triple power law parametrisation with a Gaussian gap. We therefore report results for both functional forms in Table \ref{tab:snowflake-params} and tentatively classify it as a disc with a gap, defaulting to the triple power law with a Gaussian gap where it is necessary to choose one functional form. We use the double Gaussian parametrisation to describe the rings.}

\efm{As parametrised by the double Gaussian, HD~197481 has two rings located at $29.0_{-0.7}^{+5.3}$\,au and $36.7_{-0.6}^{+18.8}$\,au, with $\Delta R=4.0_{-1.4}^{+7.8}$\,au for the inner ring and $\Delta R=11.1_{-1.4}^{+5.7}$\,au for the outer ring. As parametrised by the triple power law with a Gaussian gap, there is a gap centred at $34_{-2}^{+3}$\,au with $\Delta R=19_{-3}^{+3}$. This gap was not detected by previous observations of HD~197481.}

\efm{Because HD~197481 is a young, magnetically active red dwarf with variable stellar flux with archival observations from three dates \citep{Cranmer2013, Daley2019}, we fit the stellar flux for each observation separately. Our best-fit parametric model---a triple power law with one gap---found that the median stellar fluxes on these dates were $343_{-19}^{+18}$\,$\mu \rm{Jy}$, $140_{-20}^{+30}$\,$\mu \rm{Jy}$, and $205\pm18$\,$\mu \rm{Jy}$, respectively (Table \ref{tab:HD197481-fluxes}). These fluxes are closest to the values obtained with the disc + Ring model run by \cite{Daley2019}, although our median measurement for March, 2014 falls slightly below the uncertainty range calculated by \cite{Daley2019}. The larger fluxes measured by previous studies may be attributable to bias from small flares \cite[e.g.][]{Cranmer2013}.}

\noindent (v) HD\,15115 [$M^2$]

Both \frank and \rave profiles find two narrow rings separated by a deep radial gap, which is found to be clear of emission from parametric modelling. The ring radii found here (at 64 and 96\,au) are slightly exterior to those fitted in prior parametric modelling of observations at lower resolution (with ring outer edges at 51 and 93\,au, \citealp{MacGregor2019}) and are consistent with those found in previous non-parametric fitting \citep{Terrill2023, Han2025}. 

\efm{We parametrically modelled HD~15115 with three functional forms---a double power law, double Gaussian, and double power law with one gap---and find that the double power law with one gap provided a significantly better fit according to both the AIC and the BIC. This disc consists of two narrow rings, which we characterise with the results of the double Gaussian functional form.} 

\efm{As parametrised by the double Gaussian, the two rings are located at $65.8_{-0.3}^{+0.6}$\,au and $97.57_{-0.05}^{+0.07}$\,au. The inner ring ($\Delta R=6_{-2}^{+2}$\,au) is slightly narrower than the outer ring ($\Delta R=11.1_{-0.5}^{+0.4}$\,au), while the outer ring is much brighter than the inner ring. As parametrised by a double power law with a Gaussian gap, the gap is located at $93.9\pm0.3$\,au, with a width of $\Delta R=25.7_{-0.9}^{+0.9}$\,au and fractional depth of $C=0.453_{-0.008}^{+0.007}$ (Table \ref{tab:snowflake-params}}). These results are consistent with \texttt{frank} and \texttt{rave} profiles, although the higher resolution of the ARKS observations indicate that the gap is somewhat wider and further out from the star than previously measured at lower resolution \citep{MacGregor2019}.

\subsubsection{Group [$R$]: Single-ring discs}

This group includes single belts that have either no clear evidence of additional rings or extended halo-like emission. 
A subset of non-parametric profiles show a ``shoulder'', or abrupt steepening in slope, on the outer edge. 
We also find that a number of discs show tentative evidence of a sharp change in slope on the outer edge, giving the appearance of a bump or ``shoulder''. 
Although evidence for these features are tentative, similar features have been identified in protoplanetary discs, for which interactions with massive planets has been suggested to be a possible explanation \citep{Bi2024}. 
We remain cautious, however, since these features were generally not found to be statistically significant when tested with parametric models. Parentheses are therefore added to denote that evidence of a feature is marginal (e.g. $R_{(s)}$). 

\efm{Twelve debris discs in the ARKS sample are characterised by a single, relatively smooth belt and are well described by a double power law, asymmetric Gaussian, or power law + error function. The parametric results for these discs are described in Table \ref{tab:MCMC-simple}. The non-parametric models suggest the presence of possible substructures in several of these discs, including shoulders and plateaus. Although we parametrically fitted a triple power law functional form to 11 targets to describe these changes in slope, the additional parameters were not justified by the improvement in fit according to either the AIC or the BIC. More sensitive observations may be necessary to detect these subtle features at high confidence.}

\efm{The majority of these belts are broad, with a mean fractional width of 0.90, and have shallow inner ($\rm{mean~} \alpha_{\rm{in}}=3.9$) and outer ($\rm{mean~}\alpha_{\rm{out}}=-4.1$) edges. The distributions for both slopes skew toward smaller values, with a median inner power law index of 1.9 and a median outer power law index of $-3.25$. We find that most of these belts without significant substructure are located at large radii, measuring a mean central radius of 160\,au and a median central radius of 181\,au.}

\noindent (i) HD\,15257 [$R$]

A central cavity is not resolved in the \frank or \rave profiles. The peak location in surface density at just under 200~au is slightly interior to the broad belt at 270$^{+60}_{-40}$ fitted based on SMA observations in the REASONS survey \citep{Matra2025}. 

\efm{We parametrically modelled this disc with a double power law and power law + error function. The AIC and the BIC express no significant preference for either model over the other, indicating that both functional forms fit the data well. The power law + error function better recreates the peak, while the double power law adheres more closely to the outer edge (Fig. \ref{fig:parametric_profiles}). We therefore rely on both parametrisations to describe this low S/N disc, although we default to the power law + error function when it is necessary to choose just one functional form.}

\efm{As parametrised by the power law + error function, HD\,15257 is a broad belt ($\Delta R = 330_{-70}^{+40}$\,au), with a large central radius ($R_{\rm{c}}=110\pm30$\,au). As parametrised by the double power law, the disc peaks in emission at a larger radius ($R_c=250_{-40}^{+50}$\,au), which agrees more closely with \frank and REASONS \citep{Matra2025}. The inner edge ($\alpha_{\rm{in}}=0.8_{-0.2}^{+0.4}$) is much shallower than the outer edge ($\alpha_{\rm{out}}=-3.1_{-2.9}^{+1.3}$), which is characteristic of discs in this category.}

\noindent (ii) HD\,161868 ($\gamma$~Oph) [$R_{(s)}$]

A broad disc peaking at $\sim$130~au, the disc radius measured here is significantly smaller than the $\sim$520~au radius based on \textit{Spitzer} 70~$\mu$m observations \citep{Su2008}, but consistent with \textit{Herschel} findings \citep{Moor2015}. The \rave and \clean profiles suggest a possible subtle shoulder at 150~au.

\efm{We parametrically modelled this disc with an asymmetric Gaussian, power law + error function, and double power law to capture the single belt, having previously found that the shoulder suggested by the non-parametric models was not significant when fitted with a triple power law during an earlier iteration of the parametric modelling code. The AIC and the BIC express no significant preference for any of these models over the others, indicating that all three fit the data well. As the radial profiles are similar across the parametric models, we report results for all three, although we default to the double power law when it is necessary to choose just one functional form.}

\efm{As parametrised by the double power law, HD\,161868 is a broad belt ($\Delta R = 160_{-10}^{+30}$\,au, wider than measured by REASONS), with a large central radius ($R_{\rm{c}}=154_{-10}^{+8}$\,au) that is consistent with results from \textit{Herschel} \citep{Moor2015}. The asymmetric Gaussian ($R_{\rm{c}} = 120_{-9}^{+11}$\,au) and power law + error function ($R_{\rm{c}}=99_{-4}^{+3}$\,au) measure a smaller central radius, although the asymmetric Gaussian adheres more closely to the peak found by the non-parametric models. The inner edge ($\alpha_{\rm{in}}=0.99_{-0.11}^{+0.14}$) is shallower than the outer edge ($\alpha_{\rm{out}}=-3.7_{-0.8}^{-0.7}$), which is characteristic of other discs in this category.}

\noindent (iii) HD\,84870 [$R$]

A broad disc peaking at $\sim$200~au with the central cavity just resolved, the peak radius is slightly interior to the 260$\pm$50~au radius derived from REASONS \citep{Matra2025}. 

\efm{We parametrically modelled this disc with a double power law and a power law + error function to capture the single belt, having previously found that the shoulder suggested by \rave was not significant when fitted with a triple power law during an earlier iteration of the parametric modelling code. The AIC and the BIC moderately prefer the double power law, indicating that this model provides a better fit to the data with the same number of parameters as the power law + error function. We therefore designate the double power law as the best functional form for this disc.} 

\efm{As parametrised by the double power law, HD~84870 is a broad belt ($\Delta R=250_{-30}^{+70}$\,au, in agreement with REASONS), with a large central radius ($R_{\rm{c}}=220\pm30$\,au) that is consistent with both the non-parametric results and REASONS. The inner ($\alpha_{\rm{in}} = 1.5\pm0.3$) and outer edges ($\alpha_{\rm{out}} = -2.0_{-0.7}^{+0.5}$) are smooth and fairly symmetric.}

\noindent (iv) TYC\,9340-437-1 [$R_{(s)}$]

The disc has previously been resolved and modelled as a large belt centred at 96~au with \textit{Herschel} \citep{Tanner2020} and at 130$\pm$20~au by ALMA in the REASONS survey \citep{Matra2025}. Here, we find the radial profile to plateau between 50 and 100~au in surface brightness in both the \frank and \rave profiles, with the surface density peaking at 90~au and exhibiting a shoulder, or secondary bump, on the outer edge.

\efm{We parametrically modelled this disc with four functional forms: a double power law and an asymmetric Gaussian to capture the single belt, a triple power law to recreate the plateau suggested by \frank and \rave, and a double Gaussian to reflect any additional complexity. The AIC expresses moderate preference for the triple power law, but the BIC expresses significant preference for the asymmetric Gaussian and the double power law, indicating that all three models fit the data well. The triple power law recreates the plateau, while the asymmetric Gaussian adheres more closely to inner edge and the double power law adheres more closely to the peak. We therefore rely on all three functional forms to describe the disc, although we default to the asymmetric Gaussian when it is necessary to choose just one functional form.}

\efm{As parametrised by the asymmetric Gaussian, TYC\,9340 is broad ($\Delta R=98\pm6$\,au), with a central radius ($R_{\rm{c}}=99_{-7}^{+6}$\,au) that is somewhat smaller than measured by REASONS. The disc has an inner edge that is likely steeper than average for discs in this category ($\alpha_{\rm{in}}=14_{-8}^{+11}$) and a characteristically shallow outer edge ($\alpha_{\rm{out}}=-3.3_{-0.7}^{+0.5}$), as measured by the triple power law parametrisation. The plateau has a slightly increasing slope ($\alpha_{\rm{mid}}=0.41\pm0.14$) and is located between $R_{\rm{in}}=36_{-4}^{+5}$\,au and $R_{\rm{out}}=128\pm12$\,au.}

\noindent (v) HD\,9672 (49~Cet) [$R, g_+$]

Both the \frank and \rave profiles suggest a very smooth inner and outer edge, with the outer edge being significantly shallower in absolute terms than the inner edge, though they are similar in steepness when divided by the respective location of the edges (defined as emission at 50\% of the peak). Our modelling is consistent with the ring radius in previous ALMA observations \citep{Hughes2017, Nhung2017}. A radial profile was not derived based on \textit{Herschel} observations \citep{Moor2015}, but the position angle and inclination derived from the PACS\,100\,$\mu$m image are consistent with those derived from ARKS.

\efm{We parametrically modelled HD~9672 with a double power law and a double Gaussian. The AIC moderately prefers and the BIC significantly prefers the double power law, indicating that the additional parameters fit by the double Gaussian do not result in a significantly better fit to the data. We therefore designate the double power law as the best functional form for this disc.}

\efm{As parametrised by the double power law, HD~9672 is broad ($\Delta R=121\pm3$\,au, narrower than measured by REASONS), with a central radius ($R_{\rm{c}} = 96\pm4$\,au) that is consistent with the peak measured by \cite{Hughes2017}. The disc has a potentially shallow, but uncertain, inner edge ($\alpha_{\rm{in}} > 1.79$) and a shallow outer edge ($\alpha_{\rm{out}} = -1.11_{-0.09}^{+0.08}$) that are typical for discs in this category and consistent with the power law indices previously fitted by \cite{Hughes2017}.}

\noindent (iv) HD\,14055 ($\gamma$~Tri) [$R$]

The observations resolve a central cavity that is not completely clear of emission. The peak location is consistent with the belt radius of 180$\pm$10~au from REASONS \citep{Matra2025} and the radius of $>$170~au from DEBRIS \citep{Booth2013}. The \rave profile suggests a possible plateau in surface brightness between 120 and 210~au, but this is marginal and not detected in \frank. The \clean and \rave profiles suggest the presence of an outer ring, but this is not detected in \frank or favoured by parametric fits, so we label this disc as  consisting of one rather than two rings for now. 

\efm{We parametrically modelled this disc with a double power law and a double Gaussian. There is strong preference for the double power law over the double Gaussian as measured by the BIC but only marginal preference as measured with the AIC,  indicating that the double power law fits the data well with fewer parameters. We note that the double Gaussian fits a single belt without additional features in this case, providing a profile that is very similar to the double power law. We therefore designate the double power law as the best functional form for this disc.}

\efm{HD~14055 is broad ($\Delta R = 176_{-9}^{+10}$\,au, in agreement with REASONS), with a central radius of $R_{\rm{c}}=181_{-13}^{+9}$\,au that is consistent with measurements by REASONS and DEBRIS \citep{Booth2013}. The disc has shallow inner ($\alpha_{\rm{in}} = 1.20_{-0.11}^{+9}$) and outer ($\alpha_{\rm{out}}=-3.1_{-1.0}^{+0.7}$) slopes, which are characteristic of discs in this category. The parametric models were not able to constrain the significance of the plateau suggested by \rave, while the failure of the double Gaussian to fit for a central cavity suggests that this non-parametric feature is not significant given the S/N of the observations.}

\noindent (vii) HD\,76582 [$R_s$]

Previous modelling based on far-infrared imaging by \textit{Herschel}, submillimetre imaging by JCMT and the SED suggested two resolved components at 80 and 270~au, respectively, in addition to an unresolved inner component \citep{Marshall2016}. Our modelling based on the ALMA images supports the finding of resolved substructures, with the \frank surface brightness profile suggesting a shoulder, which translates to almost a plateau in surface density between approximately 150 and 250~au. The \rave surface brightness profile suggests a bump on the outer edge that is reminiscent of an additional ring peaking at approximately 250~au, in addition to the ring peaking at 150~au, although the uncertainties are sufficiently large such that a broad plateau rather than two rings is also possible.

\efm{We parametrically modelled this disc with a triple power law to capture the shoulder suggested by \frank and with an asymmetric Gaussian, double power law, and power law + error function to capture the single belt. The BIC significantly prefers the asymmetric Gaussian, which the AIC moderately prefers over the double power law and significantly prefers over the power law + error function. Although the AIC only marginally prefers the asymmetric Gaussian over the triple power law, we cannot rule out this functional form and emphasise that higher-resolution observations are necessary to constrain the radial structure of this disc. We therefore report results for the triple power law here, although we default to the asymmetric Gaussian when it is necessary to choose just one functional form.}

\efm{As parametrised by the asymmetric Gaussian, HD\,76582 is broad ($\Delta R=192_{-7}^{+8}$\,au), with a large central radius ($R_{\rm{c}}=181_{-8}^{+9}$\,au). The disc is symmetric, with characteristically shallow inner ($\alpha_{\rm{in}}=3.3_{-0.7}^{+3.0}$) and outer ($\alpha_{\rm{out}}=-3.4\pm0.4$) edges, as measured by the triple power law. We note that the triple power law opted to fit a plateau rather than a shoulder, suggesting that the shoulder is not significant at the S/N of the ARKS observations. If present, the plateau is extremely shallow ($\alpha_{\rm{mid}}=0.1\pm0.2$), extending from $R_{\rm{in}}=137_{-29}^{+19}$\,au to $R_{\rm{out}}=269_{-16}^{+19}$\,au.}

\noindent (viii) HD\,95086 [$R_{(s)}$]

HD\,95086 hosts a broad disc with a well-resolved central cavity. The belt radius is consistent with prior \textit{Herschel} \citep{Moor2015} and ALMA \citep{Su2017} observations. Both the \rave and \frank profiles suggest a subtle shoulder on the outer edge; however, this profile could be biased by a background source which could be undersubtracted \citep{Marino2025}. 

\efm{We parametrically modelled this disc with a double power law to fit the single belt, a triple power law to recreate the subtle shoulder, and a double Gaussian to capture additional complexity. The AIC marginally prefers the triple power law, while the BIC significantly prefers the double power law. We note that both the double Gaussian and the triple power law do not reproduce any substructures in this case but rather closely follow the profile of the double power law, indicating that a simpler structure is present at the S/N of the observations. We therefore designate the double power law as the best functional form for this disc.}

\efm{As parametrised by the double power law, HD\,95086 is broad ($\Delta R = 180_{-11}^{+11}$\,au) with a large central radius ($R_{\rm{c}}=198_{-7}^{+11}$\,au), in agreement with REASONS \citep{Matra2025}, \textit{Herschel} \citep{Moor2015}, and \cite{Su2017}. The inner $\alpha_{\rm{in}}=1.93_{-0.18}^{+0.20}$ and outer $\alpha_{\rm{out}}=-2.5_{-0.3}^{+0.2}$ edges are shallow, as characteristic of discs in this category. We note that we measure a much larger outer power law index than previous lower-resolution observations \citep{Su2017}.}

\noindent (ix) HD\,218396 (HR\,8799) [$R$]

We find a broad disc with a large central cavity with the \frank and \rave profiles, which was previously modelled with a range of parametric profiles \citep{Faramaz2021}. Any potential bumps on the outer edge suggested by the \clean profile appear to be low in amplitude and do not exhibit a clear and sudden change in slope at the sensitivity of these observations. 

\efm{We parametrically modelled this disc with a double power law, triple power law, asymmetric Gaussian, and double Gaussian. We find that the BIC strongly prefers the asymmetric Gaussian to the three other functional forms tested, while the AIC moderately prefers the asymmetric Gaussian and the double Gaussian to the other functional forms. We note that the double Gaussian recreates a shoulder on the outer edge, but this feature is not significant at the S/N of the observations. We therefore designate the asymmetric Gaussian as the best functional form for this disc, although we note that a more complex structure is possible.}

\efm{As parametrised by the asymmetric Gaussian, HD~218396 is broad ($\Delta R=150_{-20}^{+10}$\,au), with a large central radius ($R_{\rm{c}}=208_{-10}^{+11}$\,au). These values are smaller than those measured by REASONS at lower resolution. The inner ($\alpha_{\rm{in}}=9_{-3}^{+4}$) and outer ($\alpha_{\rm{out}}=-4.77_{-0.17}^{+0.30}$) edges are steeper than average for discs in this category, as measured by the triple power law.}

\noindent (x) HD\,39060 ($\beta$~Pic) [$R_{(s)}, g_-$]

$\beta$~Pic has been extensively observed by ALMA (e.g. \citealp{Dent2014, Matra2017, Hull2022}), and its dust continuum radial profile has been modelled as a Gaussian peaking at 105~au \citep{Matra2019}, consistent with the \rave profile's peak (the \frank profile is not displayed as the method is not optimised for mosaic observations at present), although the \rave profile suggests a slightly more triangular shape than Gaussian, possibly with an additional bump on the outer edge. 

\efm{We parametrically modelled this disc with a double power law and double Gaussian functional forms. While the AIC marginally prefers the double Gaussian, the BIC significantly prefers the double power law. We therefore report results for both functional forms, although we default to the double power law when it is necessary to choose just one functional form.}

\efm{As parametrised by the double power law, $\beta$~Pic is broad ($\Delta R = 122_{-5}^{+5}$\,au, slightly wider than previously measured by REASONS), with a central radius ($R_{\rm{c}} = 130\pm5$\,au) that is larger than the peak found by \rave and \cite{Matra2019}. The double Gaussian captures the triangular shape indicated by \rave, with a narrow Gaussian ($\sigma_1=16_{-6}^{+14}$\,au) centred at $R_1=115_{-6}^{+4}$\,au and a broad Gaussian ($\sigma_2=54_{-3}^{+18}$\,au) centred at $R_2=106.4_{-2.6}^{+1.9}$\,au. The inner edge ($\alpha_{\rm{in}}=1.09_{-0.07}^{+0.10}$) is shallower than the unusually steep outer edge ($\alpha_{\rm{out}}=-4.8\pm0.8$), as measured by the double power law.}

\noindent (xi) HD\,170773 [$R$]

We find a smooth single belt with a shallower inner edge than outer edge. The peak location at 200\,au is consistent with \textit{Herschel} \citep{Moor2015} and previous ALMA \citep{Sepulveda2019} observations. 
\efm{We parametrically modelled this disc with an asymmetric Gaussian and a double power law. The AIC and the BIC prefer neither functional form significantly. We therefore report results for both, although we default to the asymmetric Gaussian when it is necessary to choose just one functional form.}

\efm{As parametrised by the asymmetric Gaussian, HD~170773 is narrow ($\Delta R=66_{-5}^{+6}$\,au, in agreement with REASONS), with a large central radius ($R_{\rm{c}}=194_{-4}^{+7}$\,au). The double power law finds a similar central radius ($R_c=200_{-7}^{+13}$\,au). HD~170773 has an inner edge ($\alpha_{\rm{in}}=5.6_{-1.2}^{+1.0}$) and outer edge ($\alpha_{\rm{out}}=-11_{-5}^{+2}$) that are steeper than average for discs in this category, as measured by the double power law. The asymmetric Gaussian is fairly symmetric, with an inner width of $\sigma_{\rm{in}}=29_{-3}^{+6}$\,au and an outer width of $\sigma_{\rm{out}}=26_{-4}^{+3}$\,au.}

\noindent (xii) HD\,145560 [$R$]

We find a smooth and symmetric single belt at a radius consistent with the value derived from previous observations \citep{Matra2025}. Both the \frank and \rave profiles suggest possible low-amplitude inner emission. 
\efm{We parametrically modelled this disc with a double power law and a single Gaussian to capture a single belt and a double Gaussian to capture the possible low-amplitude emission. Both criteria agree that the single Gaussian does not provide a good fit to the data, despite having the fewest number of parameters. The AIC strongly prefers the double Gaussian over the double power law, while the BIC strongly prefers the double power law over the double Gaussian. We therefore report results for both functional forms and tentatively classify the disc as a single belt, although we emphasise that higher-resolution observations are necessary to determine whether the disc has a halo, as suggested by the AIC's preference for the double Gaussian functional form. We default to the double power law when necessary to choose just one functional form.}

\efm{As parametrised by the double power law, HD~145560 is narrow ($\Delta R=23.6_{-1.0}^{+1.2}$\,au, consistent with measurements by REASONS at a lower resolution). The belt is centred at $\sim 75$\,au according to both the double power law and double Gaussian parametrisations, which is consistent with the peak radius found by REASONS \citep{Matra2025}. The inner ($\alpha_{\rm{in}}=6_{-0.5}^{+0.6}$) and outer ($\alpha_{\rm{out}}=-7.1_{-0.5}^{+0.4}$) edges are steeper than average for discs in this category. If present, the halo is centred at $88_{-12}^{+13}$\,au with a width of $180_{-50}^{+30}$\,au, as measured by the double Gaussian parametrisation.}

\subsubsection{Group [$R_x$]: Single-ring discs with additional low-amplitude emission (halo or possible ring)}
\label{sec:Rx}

This group includes single-ring discs with evidence of more extended low-amplitude emission, which could be in the form of a halo attached to the main ring or an additional low-amplitude ring separated from the main ring by a radial gap. The distinction between the two is often unclear given the low surface brightness of these features relative to the sensitivity of the observations, and the different modelling approaches may provide different suggestions on whether a halo or ring is more likely. 
Unlike the multi-ring discs which have shallow gaps, any additional rings appear to be separated by deeper gaps that appear to be largely clear of emission.  

\efm{We define halos as a smooth, radially extended distribution of millimetre emission, typically reaching a larger radius than the main planetesimal ring. The parametric models indicate that halos are present in six of the seven discs in this category and find that the extended emission described by the non-parametric models is not significant in the remaining disc (HD\,32297). These halos are best reproduced parametrically by the sum of two overlapping Gaussians: a narrow Gaussian reproduces the main ring, while a broad Gaussian describes the extended emission of the halo (see Fig. \ref{fig:example_profiles}). The narrow component (mean $\Delta R/R=0.18$) often has an extremely steep inner edge (mean $\alpha_{\rm{in}}=56$) and shallower outer edge (mean $\alpha_{\rm{out}}=-8.6$), while the broad component (mean $\Delta R/R=0.75$) has gradual slopes that extend to larger radii. Many of the discs described in this section have fractional widths that are much smaller than previously measured, likely because the low surface brightness of the broad halos skewed measurements at lower resolution. Although we report the FWHM of both the main belt and the halo throughout this section, as measured by the parametric models, we emphasise that the narrow/broad structure of these discs limits the efficacy of FWHM as a quantification of width.}

\efm{We ran a single Gaussian for a selection of discs in this category with the lowest levels of extended emission to better understand the significance of this feature. Despite the single Gaussian parametrisation having the fewest number of parameters of any functional form in this paper, we find that the double Gaussian is strongly preferred over it in all cases that we tested. This indicates that the extended emission, while ambiguous in nature, is a significant feature in these discs and should be studied further.}

\efm{We devote the rest of this section to discussing millimetre halos, although we reiterate that there could be unresolved low-amplitude rings in some of these systems. Theoretical models have not typically predicted the narrow rings surrounded by extended millimetre emission that we find in this subset of the ARKS sample, although halos have been detected with optical and scattered-light observations in a large fraction of debris discs. Scattered-light halos are expected to consist primarily of small dust grains that are excited to high eccentricities by radiation pressure, which has only a negligible effect on the large dust grains observed in thermal emission. The detection of halos at longer wavelengths therefore suggests that a different dynamical process is responsible for these mm halos \citep{Thebault2023}.}

\efm{Halos were first observed in millimetre emission by \cite{Marino2016} and \cite{MacGregor2018}, which prompted \cite{Thebault2023} to re-analyse observations of halos in scattered light. They found that approximately 25\% of scattered-light radial profiles with halos could not be explained by their models, which indicates that dynamical interactions may be responsible for the substructure, and predicted that higher-resolution ALMA observations would find more halos in thermal emission as well.}

\efm{With the higher resolution of the ARKS sample, we now have evidence of narrow rings surrounded by millimetre halos in several targets, although more sensitive observations are necessary to confirm the nature of these features. All discs with low-level extended emission (7/7) described in this section are detected in scattered light as well, whereas a minority of single-ring discs (4/12) and multi-ring discs (2/5) are detected in scattered light \citep{Milli2025}. These detection rates may indicate that there are more small grains in systems with low-level extended emission. Approximately half (4/7) of discs with millimetre halos have scattered-light halos as well \citep[HD~32297, HD~61005, HD~121617, and HD~131835,][]{Thebault2023}. Additionally, we find that this two-component radial structure is reminiscent of the vertical structure of the Kuiper Belt, in which two dynamical populations are present \citep{Morbidelli2008}. This may indicate that dynamical processes have shaped the discs, although further research is necessary to understand the mechanisms responsible for this substructure, which appears in approximately one-third of the ARKS sample.}

\noindent (i) HD\,10647 (q$^1$~Eri) [$R_h$]

The main features of our non-parametric profiles are broadly consistent with modelling based on prior observations \citep{Lovell2021}; however, we find that the outer edge slope abruptly becomes shallower at approximately 110~au, resulting a halo-like structure. This is notably different from the single-ring discs with potential shoulders, which show a steepening in slope from a plateau-like region to the outer edge. 
We note that the system hosts a giant planet at 2\,au \citep{Butler2006}.

\efm{We parametrically modelled HD~10647 with a double power law, double Gaussian, and double power law with one gap. The AIC moderately prefers the double Gaussian, while the BIC moderately prefers the double power law. The double power law with a gap is not preferred by either criterion. Although we designate the double Gaussian as the best functional form and classify HD~10647 as a belt with a halo-like substructure, we emphasise that this previously unconstrained feature is ambiguous and that the eccentricity of the disc likely impacts our modelling efforts \citep{Lovell2025}. The main belt is centred at $89.2_{-0.8}^{+0.7}$\,au with a width of $\Delta R=28_{-2}^{+2}$\,au, while the halo is centred at $164_{-7}^{+6}$\,au with a width of $\Delta R=108\pm7$\,au. These results are consistent with observations by \cite{Lovell2021}.}

\noindent (ii) HD\,61005 [$_hR_h$]

The disc appears to show emission both interior and exterior to the relatively narrow main belt, which is more similar to a ``halo'' than an additional ring according to our non-parametric models, although the \rave profile and uncertainties suggest an additional faint outer and inner ring are also possible. Evidence of an outer halo \citep{MacGregor2018, Han2025} or ring \citep{Terrill2023} has also been suggested based on prior ALMA observations. 

\efm{We parametrically modelled HD~61005 with a double power law, double Gaussian, and triple Gaussian. Both the AIC and the BIC find that the double power law is a significantly poorer fit. We find that the BIC significantly prefers the double Gaussian, while the AIC has no preference between the double Gaussian and triple Gaussian. The triple Gaussian recreates the same radial profile as the double Gaussian, which suggests that the faint outer and inner ring suggested by \rave are not significant. We therefore designate the double Gaussian as the best functional form for this disc, which indicates the presence of a halo. Although \cite{MacGregor2018} previously observed the presence of a halo in HD~61005 with ALMA using shorter-baseline data, the higher resolution of the ARKS observations allows us to further constrain the locations and widths of the planetesimal belt and halo. Our parametric models find that the belt is narrower ($\Delta R=17.4_{-1.2}^{+1.2}$\,au) and centred at larger radius ($R=69.4_{-0.4}^{+0.5}$\,au) than \cite{MacGregor2018}, while the halo is centred at $95_{-3}^{+2}$\,au with a width of $\Delta R=75\pm4$\,au, extending across the planetesimal belt from $\sim20$\,au to $\sim140$\,au. This suggests that the halo detected by ARKS observations is different than the halo measured by \cite{MacGregor2018}, perhaps because of the difference in baselines.}

\noindent (iii) HD\,131835 [$R^3, g_+$ ($_hR_h, g_+$)]

Both the \frank and \rave fits suggest a narrow main ring with a low-amplitude inner ring and extended outer emission. The 83.7$^{+0.9}_{-1.1}$~au radius fitted in REASONS \citep{Matra2025} is further out from the main belt than found from the non-parametric ARKS profiles, which peaks at 66\,au, and is likely due to the presence of the extended outer emission. 

It is not clear whether the outer emission is a separate ring or a halo attached to the main ring \efm{based on the non-parametric models}. The \frank profile suggests a shallow radial gap that separates the extended outer emission from the main ring; however, this gap is subtler in the \rave profile and is contained within a region of large uncertainties. 
The outer emission at 100\,au has been resolved to be the most prominent ring in scattered light using SPHERE \citep{Feldt2017}, whereas the main ring in ALMA at 65\,au is significantly fainter in scattered light. The possible origin of this difference is discussed and modelled in detail in \citet{Jankovic2025}. We tentatively label the disc as $R^3$, but note a likely alternative of $_hR_h$, given that parametric models (described below) favour describing the disc as a main belt with an extended halo.

\efm{We parametrically modelled HD~131835 with a double Gaussian, triple Gaussian, and double power law to explore three possible structures: an inner ring, an outer ring, and a halo, with the double power law acting as a benchmark for significance. The AIC and the BIC significantly prefer the other models to the double power law, indicating that the extended emission is significant. We find that the BIC significantly prefers and the AIC marginally prefers the double Gaussian over the triple Gaussian, indicating that the inner ring is not significant. Additionally, the double and triple Gaussian parametrisations both fit the inner emission at $\sim 25$\,au as a part of the halo rather than a separate component, resulting in very similar radial profiles. This provides further evidence that the inner and outer components are not distinct from the main ring but rather part of a broad halo. We therefore designate the double Gaussian as the best functional form for this disc. The main ring is centred at $R=68.6\pm0.4$\,au with a width of $\Delta R=11.5_{-1.4}^{+1.4}$\,au, while the halo is centred at $R=105.8_{-1.7}^{+2.0}$\,au with a width of $\Delta R=102_{-4}^{+3}$\,au. The main ring is much narrower than measured by REASONS \citep[$\Delta R=87\pm4$\,au,][]{Matra2025}, likely due to the previously unresolved broad component.}

\noindent (iv) HD\,121617 [$R_h, g_+$]

Our non-parametric profiles fitted with both \frank and \rave suggest a single narrow belt. Low-level emission possibly extends out to at least twice the belt radius on the outer edge. The belt radius is consistent with modelling based on previous observations \citep{Moor2017}. 

\efm{We parametrically modelled HD~121617 with a double Gaussian, single Gaussian, power law + error function, double power law, and asymmetric Gaussian. Although we find that the asymmetric Gaussian and single Gaussian provide significantly poorer fits to the data according to both the AIC and the BIC, the statistical criteria differ on the significance of the other three functional forms: the AIC strongly prefers the double Gaussian (which models a halo), but the BIC strongly prefers the power law + error function. The double power law is not ruled out either. We report results for all three functional forms, although we designate the double Gaussian as the best parametric model for the disc based on a visual evaluation of the radial profiles (Fig. \ref{fig:parametric_profiles}). This indicates that the low-level emission found by the non-parametric profiles and by the double Gaussian functional form is of ambiguous significance at the S/N of the data.}

\efm{As parametrised by the double power law, this disc has steep inner ($\alpha_{\rm{in}}>20$) and outer ($\alpha_{\rm{out}}=-6.9_{-0.4}^{+0.2}$) edges, similar to those found in scattered light by \cite{Perrot2023}. We find comparable central radii with the double power law ($R_{\rm{c}} = 68.5_{-0.5}^{+0.7}$\,au), power law + error function ($R_{\rm{c}}=67.8_{-0.4}^{+0.6}$\,au), and double Gaussian ($R=75.3_{-0.3}^{+0.4}$\,au). As parametrised by the double Gaussian, the main ring is centred at $R=75.3_{-0.3}^{+0.4}$\,au with a width of $\Delta R=14.1_{-1.2}^{+1.2}$\,au, while the halo is centred at $R=103_{-5}^{+8}$\,au with a width of $\Delta R=100_{-20}^{+20}$\,au. The main ring is much narrower than measured by REASONS \citep[][]{Matra2025}, likely due to the previously unresolved broad component.}

\noindent (v) HD\,32297 [$R^2_h, g_+$ ($R_h, g_+$)]

The \frank and \rave profiles show a narrow ring with a faint outer halo, consistent with suggestions based on prior ALMA observations \citep{MacGregor2018}. The non-parametric models also suggest the presence of inner emission, likely as a separate faint ring component interior to the main belt, \efm{although it is not significantly detected in parametric models. }

\efm{We parametrically modelled this disc with a double power law as a default model, a double Gaussian to explore the presence of a halo, and a sum of a Gaussian and a double power law to explore the presence of an inner component suggested by non-parametric modelling. We find strong evidence that the double Gaussian does not fit the data well, according to both the AIC and the BIC. Although the AIC marginally prefers a Gaussian with a double power law, the BIC significantly prefers the simpler structure provided by a double power law. We therefore designate the double power law as the best parametric model, but we note that higher-resolution observations may be necessary to further constrain the radial substructures present in this disc, including the outer halo and inner emission suggested by the non-parametric models. As parametrised by the double power law, HD~32297 is centred at $R_{\rm{c}}=105.2_{-0.5}^{+0.7}$\,au with a width of $\Delta R = 14.1_{-0.6}^{+0.6}$\,au, significantly narrower than measured by REASONS \citep[][]{Matra2025}. The discrepancy in width is more characteristic of discs with halos, and the low-amplitude emission found by the non-parametric models---although not significantly detected by the parametric models---may have skewed measurements at lower resolution toward a broader width. The inner ($\alpha_{\rm{in}}>27$) and outer edges ($\alpha_{\rm{out}}=-6.3\pm0.12$) are among the steepest found in the ARKS sample.}

\noindent (vi) HD\,131488 [$R^2, g_+$ ($_hR_h, g_+$)]

Previous continuum maps observed at lower resolution were modelled by a ring at 84$\pm$3~au \citep{Moor2017}, consistent with the radius of the very narrow ring resolved here. Both \rave and \frank resolve a small bump at approximately 50~au, interior to the inner edge of the main belt. A slight outer bump in the \frank profile is not reproduced in the \rave fit. 

\efm{We parametrically modelled HD~131488 with a double power law to recreate a simple belt, a double Gaussian to recreate a halo, and a triple Gaussian to recreate a halo as well as the inner bump resolved by \rave and \frank. Both the AIC and the BIC strongly prefer the double Gaussian parametrisation to the double power law and to the triple Gaussian, indicating that the halo is significant. We therefore designate the double Gaussian as the best functional form for this disc. The profile produced by the double Gaussian (Fig. \ref{fig:parametric_profiles}) suggests that the inner bump resolved by the non-parametric models and the outer bump found by \frank is part of a broad halo. The main ring is centred at $R_c=90.84_{-0.08}^{+0.18}$\,au with a width of $\Delta R=2.3_{-0.4}^{+0.7}$\,au, while the halo is centred at $R=96.1_{-0.7}^{+0.6}$\,au with a width of $\Delta R=45\pm4$\,au. Although the location of the main ring is consistent with previous modelling in both thermal emission and scattered light \citep{Moor2017,Pawellek2024,Matra2025}, the ring is much narrower with the ARKS modelling, likely due to the previously unresolved broad component. However, we note that the width of the main ring measured by ARKS is smaller than the angular resolution of the data \citep[with a beam size of 6\,au,][]{Marino2025}. We therefore consider the width to be marginally resolved, as it appears to be smaller than the resolution of the data.}

\noindent (vii) HD\,109573 (HR\,4796) [$R_h$]

We find a narrow belt consistent with previous modelling based on lower-resolution observations \citep{Kennedy2018}, with possible very low-level emission on the outer edge. 
\efm{We parametrically modelled HD~109573 with a double Gaussian, single Gaussian, and double power law. Both the AIC and the BIC strongly prefer the double Gaussian parametrisation over the double power law and the single Gaussian, indicating the presence of a halo. The main ring is centred at $R=77.61_{-0.16}^{+0.04}$\,au with a width of $\Delta R=6.8_{-0.4}^{+0.4}$\,au, while the halo is centred at $R=84_{-1.8}^{+2.1}$\,au with a width of $\Delta R=52\pm9$\,au. The main ring is much narrower than measured by REASONS \citep[][]{Matra2025}, likely due to the previously unresolved broad component.}

\subsection{Quantifying disc features} \label{sec:measurements}
The results of Sect.~\ref{sec:results_radialprofiles} show that there is a great diversity of radial profiles observed across the sample. One way to systematically characterise the radial properties of the sample is to quantify features of interest, including the width of each ring and its inner and outer edges, in a way that is analogous to what has been performed for protoplanetary discs \citep{Huang2018}. We performed these measurements with both the non-parametric profiles, as represented by the \frank surface density profiles, and the best fitting parametric model selected with the AIC and BIC. \efm{We describe the approach used for the non-parametric models here; for the use of parameters from the parametric models; see Sect. \ref{sec:parametric_models}.}

For discs with a single ring, we measured the peak radius and the radii where the inner and outer edges reach 25\%, 50\%, and 75\% of the peak, respectively. For discs with multiple rings, it may not be appropriate to assume that the base of the belt is at a surface density of 0, given that neighbouring belts could superimpose. For each ring within a multi-belt disc, we measured its peak radius and treated the inner and outer edges separately when measuring the radius of 25\%, 50\%, and 75\% surface density for either edge. For either edge of the ring, if the edge immediately borders a radial gap, the baseline surface density value was assumed to be the local minimum value of the gap; otherwise, the edge of the ring is assumed to be the edge of the entire disc and the baseline surface density value was assumed to be 0. The radii of 25\%, 50\% and 75\% surface density were measured as percentages of the difference between the peak surface density of the gap and the baseline. 

We used the 50\% radii at the inner and outer edges to calculate the FWHM of the disc, and the difference between the 25\% and 75\% radii of either edge to define the width of the edge (the inverse of it being the slope of the edge). The ring location was defined using the centroid radius of the disc if the disc only has a single ring, or the peak radius of the ring if the ring is in a multi-belt system. The inner and outer edge locations were defined using the 50\% radii of the edge. These measurements also enable the characterisation of the radial gaps found between neighbouring rings. We used the local minimum within the gap to define its location, and the 50\% radius of the outer/inner edge of the inner/outer neighbouring rings of each gap to define the width of the gap. The measurements of these disc features are summarised in Table~\ref{table:rings} and Table~\ref{table:gaps}, which will be studied at the population level in the following section. Fig.~\ref{fig:rp_labelled} provides a visualisation of the slope and width measurements on the radial profile.

\subsection{The prevalence of substructures and distribution of fractional widths} \label{sec:substructures}

The radial profiles described in Sect.~\ref{sec:results_radialprofiles} demonstrate that substructures are prevalent in the sample. We find that 5/24 discs show strong evidence of multi-ring structures and 7/24 are single rings with evidence of additional low-amplitude rings or extended halos. The two categories together account for 50\% of the sample, meaning that the radial structure of half the sample cannot be simply described by a single ring. Even for the single-ring discs, potential features such as shoulders and inner and outer edge asymmetries suggest that the incidence rate of substructures could be even higher. 

These findings provide an updated understanding of debris disc structures, both in terms of the basic properties such as the fractional width of rings in debris discs and the prevalence of substructures.
The REASONS sample contains the most complete set of debris discs resolved by ALMA (though biased towards bright discs), which have been modelled by assuming that they follow Gaussian radial profiles, which is appropriate given its more limited sensitivity and resolution \citep{Matra2025}. 

\begin{figure}
    \centering
    \includegraphics[width=1.0\linewidth]{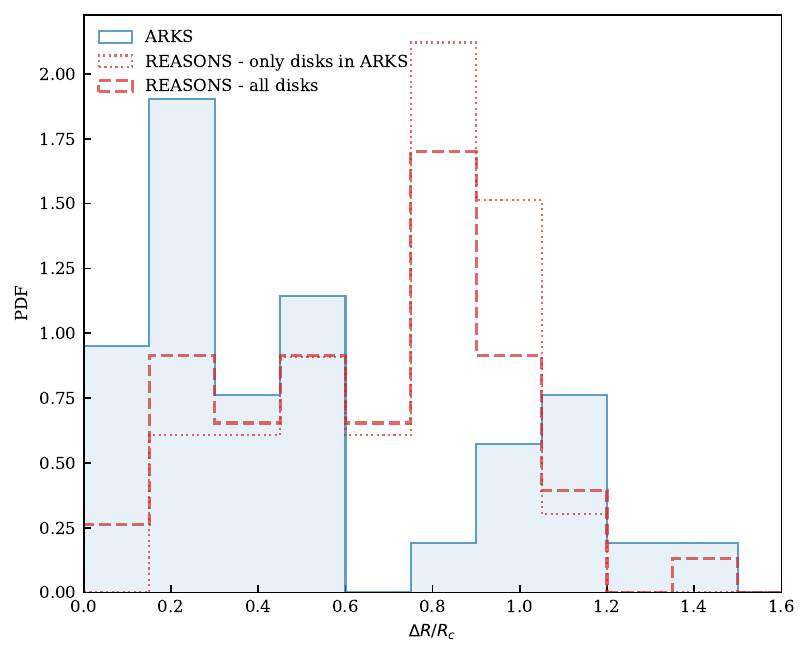}
    \caption{Comparison between the distribution of fractional widths found from the REASONS \citep{Matra2025} and ARKS datasets. The vertical axis shows the probability density function (PDF). Discs found to have multiple rings in ARKS are separated into their constituent rings. Two distributions are shown for REASONS, which correspond to the distribution of all REASONS discs, and the REASONS values for discs that overlap with those in ARKS respectively. REASONS discs with fractional uncertainties greater than 0.5 have been excluded from the REASONS distributions. }
    \label{fig:REASONS_vs_ARKS_dr_r_hist}
\end{figure}

Fig.~\ref{fig:REASONS_vs_ARKS_dr_r_hist} shows the distribution of the fractional width of debris discs based on the REASONS and ARKS samples. Discs in the REASONS sample with relative uncertainties on the fractional width that are larger than 50\% (i.e. HD\,15257 and TYC\,9340-437-1) have been excluded. HD\,36546 was also excluded as the disc was centrally peaked and a Gaussian fit did not adequately measure its fractional width \citep{Matra2025}. 
The subset of REASONS values for discs that are also covered by ARKS is distributed similarly to the full REASONS sample. The updated distribution based on the ARKS dataset, however, is shifted significantly towards the narrower end while also retaining the wide end, almost exhibiting a bimodal distribution overall.
The different between the REASONS and ARKS fractional widths is largely because previously unresolved (sub)structure such as multiple rings could result in REASONS measuring wider rings, and the presence of low surface brightness, extended emission could cause the same effect when observing at lower resolution. We discuss the implications of these findings in more detail in Sect.~\ref{sec:discussion}.

\section{Discussion} \label{sec:discussion}
In this section we discuss several implications of our findings in the context of both debris disc formation from protoplanetary discs and the dynamical interaction between planets and the debris disc in mature planetary systems. Section~\ref{sec:ppd} compares the rings seen in debris discs with those in protoplanetary discs. Section~\ref{sec:correlations} explores any potential correlations between structural features among debris discs. Sections~\ref{sec:gaps} and \ref{sec:inner-slopes} place constraints on potential perturbing planets based on any radial gaps and the slope of the inner edge of debris discs in the sample. Section~\ref{sec:outer-slopes} constrains the eccentricity dispersion of planetesimals based on the outer edge slope. We note that we expect follow-up studies to perform more detailed dynamical modelling to interpret the disc structures more wholistically than is considered here. 

\subsection{Comparison with protoplanetary discs} \label{sec:ppd}

Prior to the analysis based on the ARKS sample, the distribution of the fractional widths of rings in debris disc appeared to peak at approximately 0.8 to 1.0, with a median of 0.71 and 70\% of the discs with fractional widths of above 0.5 (Fig.~\ref{fig:REASONS_vs_ARKS_dr_r_hist}, \citealp{Matra2025}). This is significantly larger than the fractional widths of rings in protoplanetary discs, which exhibit a median fractional width of 0.29,
with 24\% of rings characterised by a fractional width larger than 0.5 based on the sample of resolved protoplanetary discs compiled by \citet{Bae2023}.
It has been unclear whether this trend is due to the low resolution of these debris disc observations or reflects an evolutionary process occurring between the protoplanetary and debris stages. 

\begin{figure}
    \centering
    \includegraphics[width=1.0\linewidth]{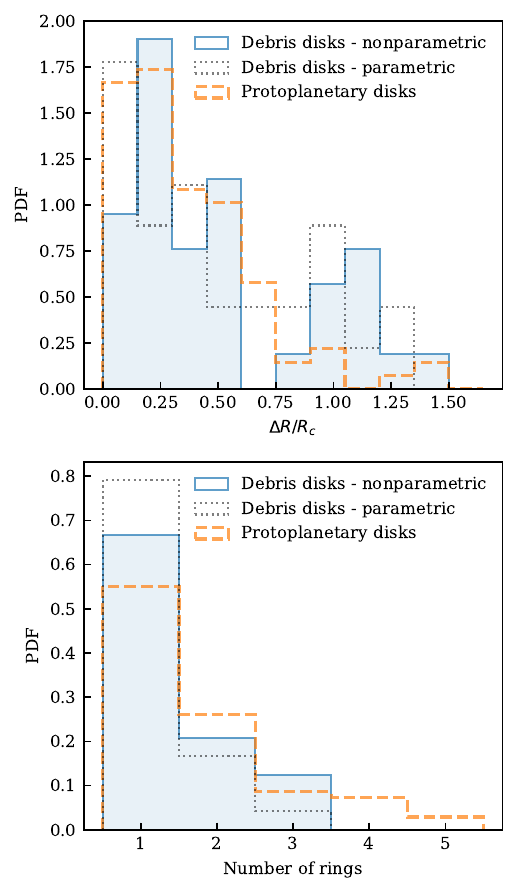}
    \caption{Comparison between rings in debris discs from ARKS and the sample of resolved protoplanetary discs compiled by \citet{Bae2023}. Discs with multiple rings are separated into their constituent rings. The top panel shows the distribution of fractional widths, where both the distributions derived from \frank and parametric modelling are shown for debris discs. The bottom panel shows the distribution of the number of rings in a disc, where the ring counts in debris discs are based on the \frank profiles, as labelled in Fig.~\ref{fig:rp_labelled}.}
    \label{fig:PPD_DD_summary}
\end{figure}

The higher-resolution observations in the ARKS dataset have significantly shifted this distribution towards smaller fractional widths (Fig.~\ref{fig:REASONS_vs_ARKS_dr_r_hist}). A comparison of the fractional widths between the protoplanetary and debris ring populations is shown in Fig.~\ref{fig:PPD_DD_summary}. Based on our non-parametric (\frank) modelling, we find a median fractional width of 0.40 for debris disc rings, with 37\% being wide (13/35 rings across the 24 discs with $\Delta R / R_c > 0.5$). 
The parametric modelling offers a slightly more conservative classification of whether a disc hosts multiple belts via rigorous model comparison with statistical criteria such as the AIC and BIC, \efm{finding a median fractional width of 0.33, with 40\% of the rings (12/30 rings) being wide. The distribution of fractional widths derived from the non-parametric and parametric approaches are in broad agreement, as shown in Fig.~\ref{fig:PPD_DD_summary}.}
These results suggest that, compared to the REASONS distribution, which is abundant in fractionally wide discs, the updated distribution based on ARKS is significantly shifted towards fractionally narrower discs, and is more similar to the distribution of protoplanetary discs than previously thought, particularly within the range of fractional widths between 0 and 0.8.

The large overlap between debris disc and protoplanetary disc fractional width suggests that in some cases, the debris rings may inherit the general distribution seen in the protoplanetary phase. 
The prevalent concentric rings in protoplanetary discs \citep{Andrews2018} have been interpreted as either gaps carved by newly formed planets \citep{Dong2015}, or local pressure bumps formed through instabilities (e.g. \citealp{Flock2015}) but could nonetheless be conducive to planetesimal formation \citep{Stammler2019, Miller2021} by trapping dust and undergoing streaming instabilities \citep{Youdin2005}. These rings could therefore be the predecessors of the planetesimals producing mm dust that we observe in debris discs, resulting in debris discs that inherit protoplanetary disc structures \citep{Marino2018, Marino2019}.

However, given the presence of a population of debris disc rings which are significantly fractionally wider than their protoplanetary counterparts, this is unlikely to explain all discs. 
Planets likely play a significant role in setting the location and width of the planetesimal belts that form from protoplanetary rings \citep{Wyatt2015}. Simulations have suggested that dust traps could produce planetesimals while moving, which could occur due to planet migration, leaving a trail of planetesimals in their path that span a wide range of radii and form a planetesimal belt with a large fractional width \citep{Miller2021}. Even in the absence of planet migration in protoplanetary discs, other simulations have found that planets could eventually scatter the ring and leave a fractionally wider planetesimal belt, whereas inward migration of the newly formed planet away from the ring could sustain the ring for longer and potentially leave a fractionally narrower planetesimal belt \citep{Jiang2023}. 
The exact effect of planet migration therefore appears to be nuanced as suggested by both the differing setups in these theoretical studies and the large range of debris ring widths observed in reality, and one that would require further theoretical work to more clearly explain.

We discuss a few caveats related to the comparison above. Firstly, the debris disc and protoplanetary disc samples are not studied with uniform resolution. However, typical resolutions in protoplanetary disc surveys (e.g. 5\,au for high-resolution observations such as DSHARP, \citealp{Andrews2018}) can be comparable to ARKS resolutions (which lie between 4\,au and 36\,au). The fractional widths derived for debris discs in this study are deconvolved, whereas the protoplanetary disc values are generally not \citep{Huang2018, Bae2023}. \citet{Jennings2022a} found that the ring widths recovered by \frank in the DSHARP sample are on average 26\% narrower than by \clean. This could mean that the protoplanetary disc fractional widths used here are biased to be slightly larger than they are. However, this should not fundamentally shift the conclusions of the comparison discussed here, since the fractional ring width distribution of protoplanetary discs is relatively wide, and reducing their values by approximately one quarter would still result in a distribution comparable to that of the fractionally narrow population of debris discs. 

Secondly, while there exists the possibility that the discs in our sample could still consist of more unresolved rings, our \frank resolution tests (benchmarking the fractional width recovered when observing an infinitesimally thin ring, as described in Sect.~\ref{sec:results_radialprofiles}) suggest that if the broad discs were to consist of 2--3 very narrow rings instead, our observations should have resolved these radial substructures (although if they were to consist of a large number of rings, e.g. $\gtrsim$10 rings, these very narrow rings would not have been individually resolved). The presence of the fractionally broad population of debris disc rings is thus likely to be real. 

Thirdly, the protoplanetary disc sample of \citet{Bae2023} contains a higher fraction of lower-mass stars (with a median stellar mass of 0.9\,$M_\odot$) than the ARKS sample (with a median stellar mass of 1.6\,$M_\odot$), which could bias the protoplanetary discs in this comparison to be at smaller radii. However, we find that when comparing only the A stars (for which the ARKS sample is most abundant in) to control for the effect of stellar mass (including only protoplanetary disc host stars with masses equivalent to those of main-sequence A stars), the findings based on the comparison of fractional width distribution still hold, although in a less robust way given the more limited sample size. 

Relatedly, this comparison loosely assumes that the protoplanetary and debris discs observed are from the same population and offset only in time, i.e. the protoplanetary discs in the \citet{Bae2023} sample will evolve into debris discs analogous to those in the ARKS sample. In practice, the evolutionary pathway could be diverse. For example, a number of possible formation and evolutionary pathways of debris discs could exist \citep{Michel2021, Najita2022}, and the observed debris discs could represent only the population of systems emerging from bright and extended protoplanetary discs. A broader range of evolutionary pathways could, therefore, also exist, but may not be captured by the sample studied here. 

Finally, it is important to note that this comparison relies on defining the width of a belt based on its FWHM. Low-amplitude extended emission, which we have found to be common in the sample as discussed in Sect.~\ref{sec:results_radialprofiles}, is not accounted for by the FWHM metric. If we are to more wholistically connect protoplanetary rings to debris rings, the formation of the broad faint distributions in debris discs will also need to be accounted for. 

\subsection{Ring width correlations}  \label{sec:correlations}

\begin{figure*}
    \centering
    \includegraphics[width=1.0\linewidth]{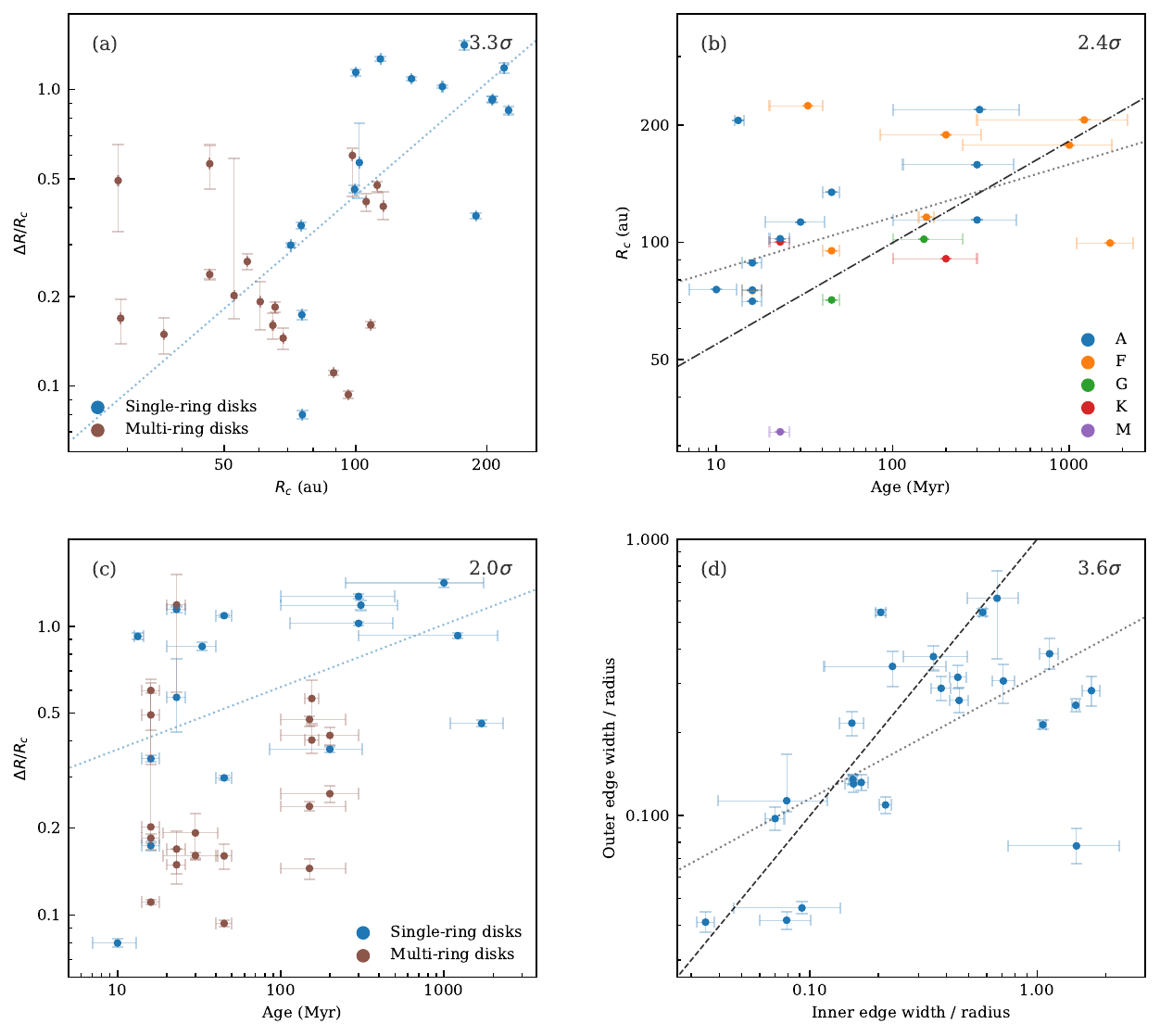}
    \caption{Radial properties across the ARKS sample. Linear models were fitted in linear or log space (as appropriate for the way the data in each panel are plotted) and displayed as dotted lines. The significance is evaluated as the ratio of the slope to its uncertainties, as displayed in the upper right corner of each panel. In panels (a) and (c), the linear model shown is that fitted to single-ring discs only, whereas in panels (b) and (d), the line is fitted to all points. In panel (b), the dash-dotted line has a slope of $6/23$, as predicted for self-stirred discs \citep{Kennedy2010}. In panel (d), the dashed line indicates where the fractional inner and outer edge widths are equal. The width of an edge is defined as the number of au for the radial profile to rise (fall) from 25\% (75\%) to 75\% (25\%) of the peak, and the radius of an edge is defined as where the radial profiles crosses 50\% of the peak. }
    \label{fig:radial_analysis}
\end{figure*}

In addition to the comparison between debris discs and protoplanetary discs, the relatively large span in age of the ARKS sample allows us to investigate the evolution of debris discs subsequent to their formation from protoplanetary discs. 
Fig.~\ref{fig:radial_analysis}(a) displays the fractional widths of all discs in the ARKS sample measured from the \frank profiles as a function of the centroid radius. A positive correlation is seen among the single-ring systems (at 3.3$\sigma$, which is important to isolate from the multi-ring systems given that some of the low-amplitude rings from non-parametric modelling are of lower significance than the main belts), and at lower significance when all rings in multi-ring systems suggested by non-parametric (\frank) fits are also considered (at 1.9$\sigma$), where the significance was estimated using the ratio between the least-squares slope and its standard error when fitting a linear model in logarithmic space. This suggests that wider discs are not only wider in absolute terms, but they are also fractionally wider, i.e. with the larger radius adjusted for. 

Age appears to be a basis for this correlation. 
Figs.~\ref{fig:radial_analysis}(b) and (c) display the radius and fractional widths as a function of age, and both tentatively increase with age (at 2.4$\sigma$ and 2.0$\sigma$ respectively, where the significance is effectively unchanged even if the only M star, AU~Mic, is removed in panel b). The increase in age of the fractional width in panel c also holds when the rings from multi-ring systems are also included (at 2.6$\sigma$). 
In Fig.~\ref{fig:radial_analysis}(b), the black dashed line indicates the evolution of the radius of peak optical depth expected in self-stirred discs \citep{Kennedy2010}. A positive correlation between fractional width and age has previously been observed among relatively well-resolved ALMA images \citep{Han2025}, though not among the full REASONS sample \citep{Matra2025}. Analysing the REASONS fractional width measurements based on only its subsample that constitute the ARKS sample (excluding the 2 discs with relative uncertainties above 50\%), there is no correlation between the fractional width and age for the ARKS single rings (at 0.1$\sigma$) and only a weak correlation for all rings (modelled as single Gaussians, at 1.5$\sigma$). 
The lower significance among the REASONS sample could either reflect the fact that the REASONS modelling included a sizeable fraction of marginally resolved discs, or that the tentative correlation found in ARKS is affected by a biased sample. 

Also seen across this sample (but not plotted) is a weak negative correlation between fractional width and dust mass, which could be explained by the decreasing dust mass with age due to collisional depletion \citep{Dominik2003, Wyatt2007b, Loehne2008, Holland2017}. The anti-correlation may also be consistent with theoretical expectations based on planet-induced secular stirring of self-gravitating discs \citep{Sefilian2024}. Tentative positive correlations are also seen between the fractional inner and outer widths and the fractional width of the belt (and therefore with age), resulting in an apparent coordinated evolution of the radial profile. However, as noted in earlier discussion, the majority of radial profiles appear to be asymmetric. Although the fractional inner and outer edge widths appear to be correlated with each other across the sample, we find that the fractional inner edge width appears to be more typically broader than the fractional outer edge width, as shown in Fig.~\ref{fig:radial_analysis}(d). We note that these fractional edge widths correspond to their absolute width divided by the edge location, rather than the centroid location of the whole belt, thereby accounting for the difference in location of the inner and outer edges. This finding could arise from a higher dynamical excitation close to the inner than outer edges which would make inner edges appear smoother than the outer edges \citep{Marino2021}. 

The overall picture in this sample therefore appears to be that the disc radially broadens while its centroid moves further out over timescales of the order of $\sim$100\,Myr. Collisional evolution could cause the peak emission to propagate to larger radii over time as the inner regions of the disc are collisionally depleted over faster timescales \citep{Kenyon2008, Kennedy2010, ImazBlanco2023}. Similarly, collisional evolution could also cause the smaller discs to collisionally deplete below the detection threshold faster, resulting in the average radius of a population of narrow belt debris discs to increase with age \citep{Wyatt2007}. However, it is not clear that this could also explain any potential increase in fractional widths observed. It is possible that the fractional width increase could relate to the scattering between planetesimals or by planets. The fractionally shallower inner edge relative to outer edge could reflect a larger eccentricity dispersion closer to the inner system, or that the effect of collisional evolution, which tends to make the surface density converge to a power law index of 2 \citep{Marino2017, ImazBlanco2023}. However, it is important to note that the more extended halos, which could correspond to higher-eccentricity components on the outer edge, may not be fully captured by the edge width measurements, which are not affected by low-level emission under 25\% of the peak surface density of the belt. 

Of the 5 gas-rich debris discs (HD\,121617, HD\,131835, HD\,131488, HD\,32297, and HD\,9672), 4 are classified as narrow discs. Conversely, among the 5 fractionally narrowest single-ring discs ($R$ and $R_x$), 4 are gas-rich. 
These gas-bearing discs are young ($<$50~Myr). On top of the potential fractional width and age correlation, this tentative correlation between narrowness and the presence of gas could potentially also relate to the concentration of dust due to gas drag \citep{Olofsson2022}. 
Gas has also been detected in $\beta$~Pic's relatively broad disc, although the gas mass is relatively low. The gas-rich debris discs are discussed in more detail in an accompanying paper \citep{MacManamon2025}.

\subsection{Planet constraints from gaps} \label{sec:gaps}

\begin{table*}
    \centering
    \caption{Gap location and widths, and inferred single planet masses. }
    \label{table:gaps}
\begin{tabular}{llllll}
\hline \hline
Target & Gap & $r_\text{gap}$ (au) & $\Delta r_\text{gap}$ (au) & $M_\text{pl}$ ($M_\text{J}$) & Consistent with direct imaging \\
\hline
HD\,15115 & 1 & 79.3$^{+4.0}_{-4.3}$ & 22.2$^{+0.7}_{-0.6}$ & 0.37$^{+0.09}_{-0.07}$ & Y\\

HD\,32297 & 1 & 79.9$^{+6.4}_{-8.5}$ & 35.4$^{+1.3}_{-1.2}$ & 2.0$^{+0.9}_{-0.6}$ & Y \\

HD\,92945 & 1 & 72.6$^{+4.7}_{-4.1}$ & 19.6$^{+2.3}_{-1.7}$ & 0.20$^{+0.10}_{-0.07}$ & Y \\

HD\,107146 & 1 & 57.3$^{+2.8}_{-1.8}$ & 12.3$^{+0.6}_{-0.6}$ & 0.11$^{+0.03}_{-0.02}$ & Y \\
 & 2 & 79.2$^{+3.1}_{-2.6}$ & 18.9$^{+2.7}_{-1.4}$ & 0.15$^{+0.07}_{-0.05}$ & Y \\

HD\,131488 & 1 & 69.4$^{+3.6}_{-6.2}$ & 26.1$^{+1.0}_{-0.8}$ & 1.3$^{+0.4}_{-0.3}$ & Y \\

HD\,131835 & 1 & 42.2$^{+6.8}_{-9.1}$ & 25.2$^{+3.3}_{-3.9}$ & 6.3$^{+8.6}_{-3.5}$ & Y \\
 & 2 & 84.8$^{+4.9}_{-4.2}$ & 18.9$^{+2.3}_{-1.8}$ & 0.20$^{+0.10}_{-0.07}$ & N \\

HD\,197481 & 1 & 17.0$^{+3.6}_{-6.0}$ & 14.1$^{+2.1}_{-1.4}$ & 7.1$^{+16.7}_{-4.5}$ & N \\
 & 2 & 32.5$^{+1.8}_{-1.8}$ & 4.1$^{+1.3}_{-1.2}$ & 0.010$^{+0.015}_{-0.007}$ & Y \\

HD\,206893 & 1 & 73.5$^{+10.8}_{-12.0}$ & 40.4$^{+4.6}_{-4.3}$ & 3.7$^{+3.6}_{-1.8}$ & Y \\
\hline
\end{tabular}
\tablefoot{Gap location and widths are measured from the \frank surface density profiles. The planet masses correspond to a single planet required to clear the measured widths (see Eq. \ref{eq:MMR_Delta_a_gap}). These values are not tested for consistency with the gap depth. The rightmost column lists whether such a planet is consistent with current constraints from the available direct imaging observations based on the detection probability maps in \citet{Milli2025}, i.e. `Y' means not expected to be detected by available observations at 50\%. }
\end{table*}

The radial profiles derived in this study enable detailed dynamical modelling in each individual system to infer their planetary architecture. We expect dedicated work in future studies to model potential interactions between the disc and unseen planets in more detail; however, it is possible to make some simple predictions based on several basic assumptions. 

\begin{figure}
    \centering
    \includegraphics[width=1.0\linewidth]{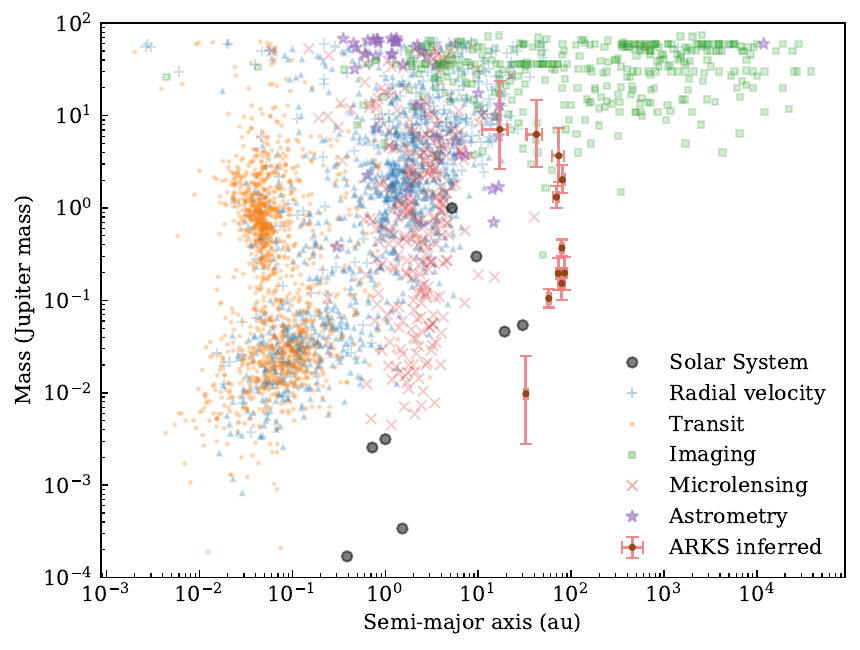}
    \caption{Planet population inferred from the ARKS sample (red points) assuming all gaps suggested by non-parametric (\frank) profiles are cleared by a single planet due to overlap of first-order mean motion resonances, overplotted on the sample of confirmed exoplanets observed with common detection methods as of March 2025 (retrieved from exoplanet.eu). }
    \label{fig:planets}
\end{figure}

In part motivated by the recent example of a potential planet in a debris disc gap \citep{Lagrange2025, Crotts2025}, if we were to assume that all gaps identified in the sample are created by a single planet which formed in situ on a circular orbit, and that the mechanism of gap clearing is scattering driven by the chaotic evolution due to the overlap of first-order mean-motion resonances, then it is possible to use the width of each gap to estimate the mass of the corresponding planet clearing the gap. The total width of the chaotic zone (i.e. summed over the chaotic regions both interior and exterior to the planet's orbit) is described by
\begin{equation}
    \Delta a \approx 3 a_\mathrm{pl}\left( \frac{M_\mathrm{pl}}{M_*} \right) ^{2/7},
    \label{eq:MMR_Delta_a_gap}
\end{equation}
where $a_\mathrm{pl}$ is the semi-major axis of the planet, $M_\mathrm{pl}$ is the mass of the planet and $M_*$ is the mass of the star \citep{Wisdom1980, Morrison2015}. Assuming this relation, we mapped the disc widths listed in Table~\ref{table:gaps} to the planet masses required to produce the gaps under these assumptions, which are overplotted on the population of known exoplanets\footnote{Retrieved from \href{https://exoplanet.eu}{exoplanet.eu}.} in Fig.~\ref{fig:planets}. The orbit of these hypothetical planets lie between 20 and 80\,au and their masses range widely from a few Earth masses to a few Jupiter masses. They are typically near the edge of or below current direct imaging limits. 

It is important to emphasise a few caveats on the nuances that this basic analysis ignores. Firstly, these mass estimates are not tested for compatibility of the depth of the gap and are based solely on its width. It is possible that some of the masses estimated from the gap width are high enough to clear a deeper gap than that observed over the age of the system, thus making it unlikely for such a planet to exist in the gap. For these shallower gaps, a chain of lower-mass planets may be more likely than a single higher-mass planet \citep{Shannon2016}; however, planetesimal-driven migration of planets could also complicate such a picture \citep{Morrison2018, Friebe2022}.

Secondly, it is also important to recognise that the multi-ring structures that we find in the ARKS sample could also be inherited from the protoplanetary disc phase in which the dynamical evolution is shaped by gas, rather than purely by planet clearing as has been assumed here \citep[see discussion in][]{Marino2019}. Mean motion resonances beyond the chaotic zone \citep{Tabeshian2016}, secular interactions \citep{Pearce2015, Zheng2017, Yelverton2018, Sefilian2021, Sefilian2023} and low-mass planets migrating through the disc \citep{Morrison2018, Friebe2022} could also play a major role in clearing material both close to the planet and at more distant locations, thereby further complicating the picture. 
It is, therefore, important for future work to systematically model the radial profile, rather than only the gap location and its width as performed in this study, to reach a more comprehensive conclusion on potential planetary configurations consistent with the disc structures observed. 

Finally, available direct imaging observations already provide constraints on the range of possible planets within these systems. Based on the detection probability maps computed in \citet{Milli2025}, one hypothetical planet in Table~\ref{table:gaps} (in the inner gap of HD\,197481, or AU~Mic) should have been detected by SPHERE with 95\% probability, although considering the uncertainties on the planet mass and orbit, the detection probability could be at 50\% or lower. Another possible planet (in the inner gap of HD\,131835) should have been detected by SPHERE with a probability between 50\% and 95\%, and the detection probability could be lower than 50\% considering uncertainties on the constraints on such a planet. The detection probabilities of all other such gap-inferred planets in Fig.~\ref{fig:planets} are lower than 50\%.

\subsection{Trends in the steepness of the inner edge}
\label{sec:inner-slopes}

\efm{The sharpness of the inner edges of debris discs can indicate sculpting by planets \citep[e.g.][]{Quillen2006,Chiang2009,Pearce2024}. We compare the slopes of the ARKS targets based on the results of power law parametrisations, with gaps added as necessary. Although double power laws and triple power laws did not provide the best fit to the data for several discs, we find that the models can still place constraints on the slopes. Because the implementation of the double power law described in Table \ref{tab:formdefs} is smooth across radii, we additionally tested whether introducing a sharp cut-off on the inner and outer edges would affect the power law indices. We found that this did not impact the values obtained for the indices. We therefore take $\alpha_{\rm{in}}$ and $\alpha_{\rm{out}}$ values from double power law fits in most cases, except for HD~92945, HD~107146, HD~197481, TYC~9340-437-1, and HD~218396, where a triple power law provided a better fit to the data according to the AIC and the BIC.}

\begin{table}[h]
    \centering
    \caption{Planet sculpting constraints for the seven discs with the steepest inner edges in the ARKS sample. }
    \label{tab:slopes}
    \begin{tabular}{l|ccccc}
        \hline \hline
        Disc & \multicolumn{2}{c}{$\alpha_{\rm{in}}$} & $M_{\rm{pl}}$ & $a_{\rm{pl}}$ \\
        & \texttt{frank} & Parametric & $(M_{\rm{J}})$ & $(\rm{au})$ \\
        \hline
        HD\,10647 & 5.4 & $>18$ & $<2.5$ & $<48$\\
        HD\,32297 & 22 & $>27$ & $<1.0$ & $<74$ \\
        HD\,107146 & 14 & $>13.91$ & $<5.0$ & $<32$\\
        HD\,109573 & 28 & $34_{-3}^{+5}$ & $0.68_{-0.18}^{+0.30}$ & $<50$\\
        HD\,121617 & 14 & $>20$ & $<3.1$ & $<49$\\
        HD\,131488 & 22 & $28_{-4}^{+5}$ & $1.2_{-0.5}^{+0.6}$ & $<63$\\
        HD\,197481 & 14 & $>14$ & $<3.2$ & $<19$\\\hline
    \end{tabular}
    \tablefoot{Non-parametric slopes are derived from the inner edge definitions in Table \ref{table:rings}.}
\end{table}

\begin{figure*}[h]
    \centering
    \begin{subfigure}{0.48\textwidth}
        \includegraphics[width=\textwidth]{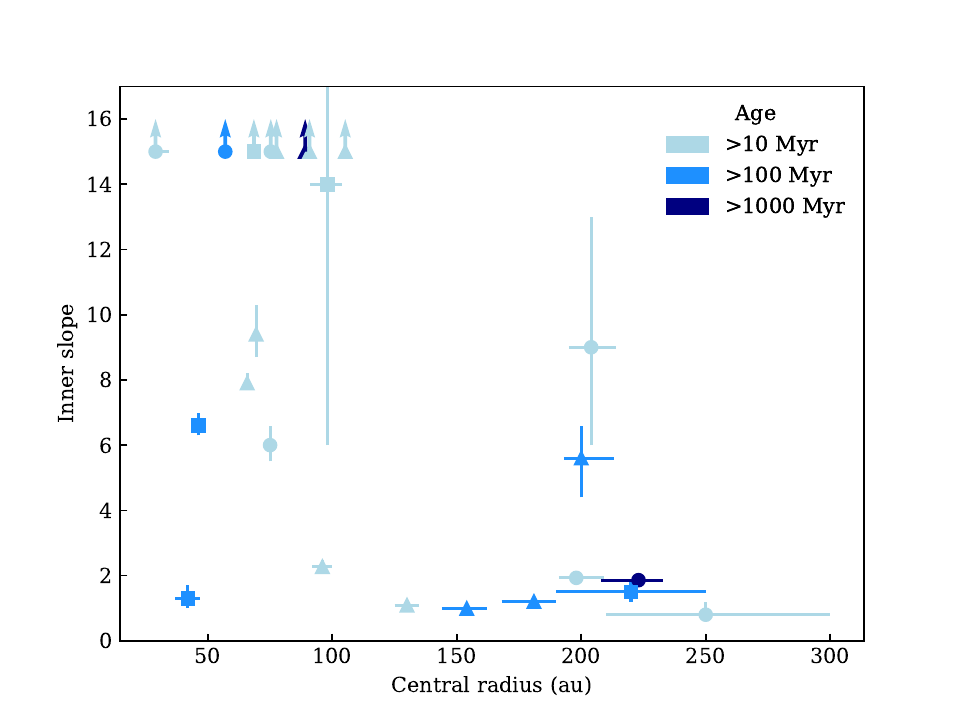}
    \end{subfigure}
    \begin{subfigure}{0.48\textwidth}
        \includegraphics[width=\textwidth]{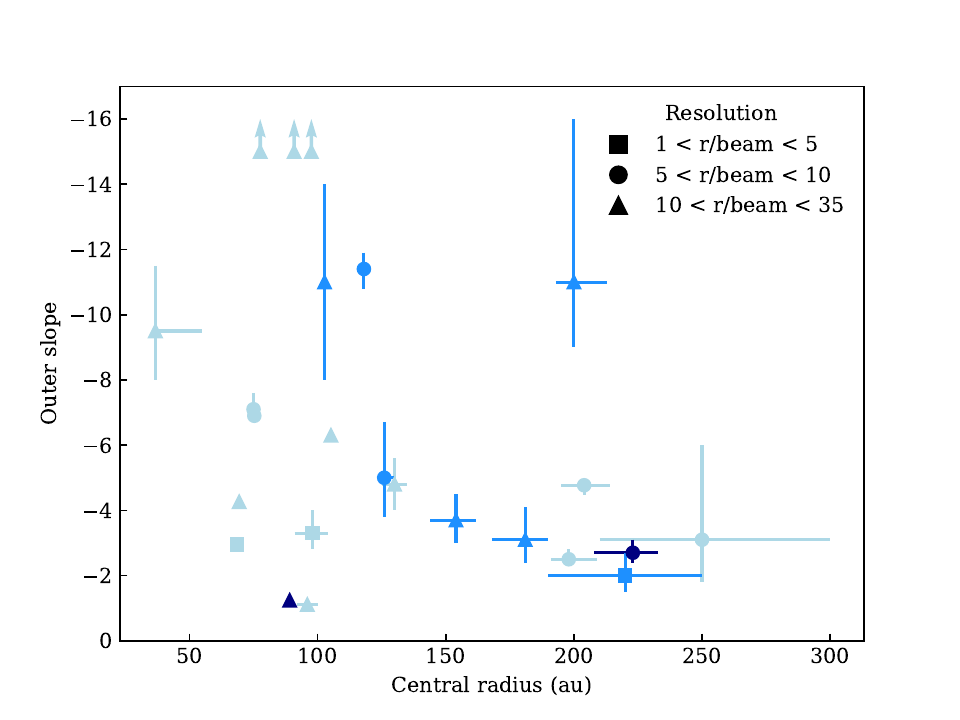}
    \end{subfigure}
    \caption{Sharpness of inner and outer slopes as a function of distance from the host star. The slopes are measured with a double power law or triple power law ($\alpha_{\rm{in}}$ and $\alpha_{\rm{out}}$), while the central radii are measured with a double power law ($R_{\rm{c}}$) or multiple Gaussians ($R$). For discs with multiple rings, only the innermost ring is plotted on the left and only the outermost ring is plotted on the right; for discs with halos, only the main planetesimal belts are plotted. This is due to the limitations of power-law parametrisations. The colours indicate the age of the system \citep[Table A.2 in][]{Matra2025}, while the marker shapes indicate the resolution of the ARKS observations \citep[Table 2 in][]{Marino2025}. Discs with median power-law indices greater than 15---the cut-off for pure planet sculpting of the inner edge---are indicated by arrows for both plots. We note that the x-error bars on the steepest discs are too small to be seen.}
    \label{fig:inner-outer-slopes}
\end{figure*}

\efm{We find a bimodal distribution of inner slopes: discs that are closer to their host stars ($R_{\rm{c}} < 100$\,au) have sharper inner edges ($\alpha_{\rm{in}}>15$), while discs that are further out have shallower inner edges (Fig. \ref{fig:inner-outer-slopes}). 
We propose four explanations for these results, although we emphasise that self-stirring and secular interactions may also play an important role in shaping inner edges.}

\efm{First, planets may be more likely to have orbits closer to their host stars, so the sharp inner edges observed in discs at smaller radii might simply reflect the underlying distribution of planet semi-major axes. Direct imaging has shown that planets at large distances from their host stars are rare \citep{Vigan2021,Currie2023}, so discs at large radii may not experience planet truncation to the same extent as discs at small radii, or the timescale for sculpting inner edges at large radii may simply be too large: For example, a Jupiter-mass planet located just interior to a disc will truncate an inner edge at 50\,au in 3.9\,Myr and an inner edge at 200\,au in 31\,Myr \citep[from eq. 3 in][]{Pearce2024}.}

\efm{Second, secular resonances can allow planets to sculpt the inner edges of discs at large radii \citep{Zheng2017, Sefilian2021,Sefilian2023,Pearce2024}, but these discs may experience excitation that flattens the inner edge or may not have visibly sharp inner edges due to delays related to the distance from the host star.}

\efm{Third, collisional evolution may be responsible for the shallow inner edges we observe in debris discs at large radii. Models by \cite{Marino2017} and \cite{ImazBlanco2023} have shown that the inner regions of debris discs lose material faster, resulting in a power law index of $\alpha_{\rm{in}}\approx2$. As seen in Fig.~\ref{fig:inner-outer-slopes}, most discs centred at large radii in the ARKS sample have inner slopes consistent with those expected from collisional evolution.}

\efm{Fourth, these discs with large radii and shallow slopes are the only systems in the ARKS sample with directly imaged planets. It is possible that the presence of a massive planet shaped the radial structure of these discs, creating shallow slopes.}

\efm{The higher resolution of the ARKS observations finds much steeper inner edges than previous studies of debris discs. Although \cite{Pearce2024} found that many debris discs observed at lower resolutions have inner edges that are too flat to be explained by planet sculpting alone, we calculate that seven discs in the ARKS sample have inner edges that are steep enough to constrain planet masses. The maximum flatness of $\sigma_i \leq 0.05$ derived by \cite{Pearce2024} for pure planet sculpting corresponds to a power-law index of $\alpha_{\rm{in}} \geq 15$, due to the inverse relationship between the parameters:
    \begin{equation}
        \sigma_i = \sqrt{\frac{2}{\pi}} \frac{1}{\alpha_{\rm in}}  .
    \end{equation}
We additionally calculate a power-law slope from the steepest part of the inner edge described by \texttt{frank} profiles for each target to compare the results of the parametric and non-parametric models. While parametric models suggest steeper slopes than those derived from \texttt{frank}, the seven discs described in Table \ref{tab:slopes} generally meet the criterion of $\alpha_{\rm{in}} \geq 15$ with both methods.}

\efm{We note that the seven discs that are likely candidates for planet sculpting include five of the seven discs with extended, low-amplitude emission and two of the five discs with multiple statistically significant rings described in Sect. \ref{sec:results_radialprofiles}. This result suggests that the presence of gaps or extended emission in debris discs may indicate the influence of planets, assuming that sharp inner edges are caused by planet sculpting. Additionally, we find that the seven candidates for planet sculpting have some of the smallest widths in the ARKS sample. Because these discs are centred at small radii, this suggests that there may be a correlation between planet sculpting and disc width. Perhaps truncation by planets cleared the inner regions of discs centred at small radii, removing enough dust to decrease the widths of these objects. However, we emphasise that this explanation is extremely tentative and that the correlation observed here requires theoretical exploration beyond the scope of this paper.}

\efm{We estimate upper limits on the masses and semi-major axes of the predicted planets sculpting these seven discs with Eq. (17) in \cite{Pearce2024} and with the publicly available planet-truncation model from \cite{Pearce2022},\footnote{https://github.com/TimDPearce/SculptingPlanet/tree/main} respectively. These constraints rely on the assumption that a planet is located just interior to the inner edge of the disc and has cleared debris out to the inner edge. We report these calculations in Table \ref{tab:slopes}, and we compare our results with constraints from other studies in Appendix \ref{appendix:planet-sculpting}. We find that planets around these discs need to be $<5\,M_{\rm{J}}$ and at $<74$\,au to sculpt the inner edge.}

\subsection{The outer edge and eccentricity}
\label{sec:outer-slopes}

\begin{table*}[h]
    \centering
    \caption{Outer slopes, mean particle eccentricities, and fractional widths for the ARKS sample.}
    \label{tab:excitation}
    \begin{tabular}{l|llll}
        \hline \hline
        Disc & $\alpha_{\rm{out}}$ & \multicolumn{2}{c}{$e_{\rm{rms}}$} & $\Delta R/R$\\
        & & Median & Upper limit & \\
        \hline
        HD\,9672 & $-1.11_{-0.09}^{+0.08}$ & $0.69_{-0.06}^{+0.05}$ & $0.86$ & $1.27_{-0.06}^{+0.06}$ \\
        \hline
        HD\,10647 & $-1.24_{-0.08}^{+0.04}$ & 
        & 
        & $0.34_{-0.02}^{+0.02}$\\
        \hline
        HD\,14055 & $-3.1_{-1.0}^{+0.7}$ & $0.24_{-0.08}^{+0.06}$ & 0.56 & $0.99_{-0.09}^{+0.08}$ \\
        \hline
        \multirow{2}{*}{HD\,15115} & \multirow{2}{*}{$-16.3_{-0.8}^{+0.6}$} & \multirow{2}{*}{$0.047_{-0.002}^{+0.002}$} & \multirow{2}{*}{0.052} & $0.114_{-0.005}^{+0.004}$ \\
                                  & & & & $0.086_{-0.029}^{+0.036}$ \\
        \hline
        HD\,15257 & $-3.1_{-2.9}^{+1.3}$ & $0.25_{-0.23}^{+0.10}$ & 1.2 & $1.3_{-0.4}^{+0.3}$ \\
        \hline
        HD\,32297 & $-6.30_{-0.12}^{+0.12}$ & 
        & 
        & $0.134_{-0.006}^{+0.006}$ \\
        \hline
        HD\,39060 & $-4.8_{-0.8}^{+0.8}$ & $0.16_{-0.03}^{+0.03}$ & 0.24 & $0.94_{-0.06}^{+0.05}$\\
        \hline
        HD\,61005 & $-4.26_{-0.09}^{+0.08}$ & $0.181_{-0.004}^{+0.003}$ & 0.19 & $0.269_{-0.015}^{+0.017}$ \\
        \hline
        HD\,76582 & $-2.7_{-0.4}^{+0.3}$ & $0.29_{-0.04}^{+0.03}$ & 0.40 & $0.86_{-0.07}^{+0.05}$ \\
        \hline
        HD\,84870 & $-2.0_{-0.7}^{+0.5}$ & $0.39_{-0.13}^{+0.10}$ & 1.1 & $1.1_{-0.2}^{+0.4}$ \\
        \hline
        \multirow{2}{*}{HD\,92945} & \multirow{2}{*}{$-11_{-3}^{+3}$} & \multirow{2}{*}{$0.07_{-0.02}^{+0.02}$} & \multirow{2}{*}{0.14} & $0.23_{-0.05}^{+0.05}$ \\
                                  & & & & $0.58_{-0.04}^{+0.05}$ \\
        \hline
        HD\,95086 & $-2.5_{-0.3}^{+0.2}$ & $0.31_{-0.04}^{+0.02}$ & 0.41 & $0.89_{-0.06}^{+0.07}$ \\
        \hline
        \multirow{3}{*}{HD\,107146} & \multirow{3}{*}{$-11.4_{-0.5}^{+0.6}$} & \multirow{3}{*}{$0.068_{-0.003}^{+0.003}$} & \multirow{3}{*}{3.0} & $0.29_{-0.03}^{+0.02}$ \\
                                   & & & & $0.497_{-0.003}^{+0.005}$\\
                                   & & & & $0.12_{-0.03}^{+0.04}$ \\
        \hline
        HD\,109573 & $-21.7_{-1.5}^{+1.2}$ & $0.035_{-0.002}^{+0.002}$ & 0.041 & $0.081_{-0.003}^{+0.002}$ \\
        \hline
        HD\,121617 & $-6.9_{-0.4}^{+0.2}$ & $0.112_{-0.006}^{+0.003}$ & 0.12 & $0.117_{-0.018}^{+0.017}$ \\
        \hline
        HD\,131488 & $-16.6_{-0.7}^{+0.8}$ & 
        & 
        & $0.072_{-0.003}^{+0.003}$ \\
        \hline
        HD\,131835 & $-2.96_{-0.05}^{+0.05}$ & 
        & 
        & $0.182_{-0.015}^{+0.016}$ \\
        \hline
        HD\,145560 & $-7.1_{-0.5}^{+0.4}$ & $0.108_{-0.008}^{+0.006}$ & 0.13 & $0.315_{-0.015}^{+0.017}$ \\
        \hline
        HD\,161868 & $-3.7_{-0.8}^{+0.7}$ & $0.21_{-0.04}^{+0.04}$ & 0.38 & $1.02_{-0.10}^{+0.19}$ \\
        \hline
        HD\,170773 & $-11_{-5}^{+2}$ & $0.07_{-0.03}^{+0.01}$ & 0.12 & $0.33_{-0.03}^{+0.04}$ \\
        \hline
        \multirow{2}{*}{HD\,197481} & \multirow{2}{*}{$-9.5_{-2.0}^{+1.5}$} & \multirow{2}{*}{$0.08_{-0.02}^{+0.01}$} & \multirow{2}{*}{0.13} & $0.14_{-0.05}^{+0.27}$ \\
                                   & & & & $0.30_{-0.04}^{+0.21}$ \\
        \hline
        \multirow{2}{*}{HD\,206893} & \multirow{2}{*}{$-5.0_{-1.7}^{+1.2}$} & \multirow{2}{*}{$0.15_{-0.05}^{+0.04}$} & \multirow{2}{*}{0.30} & $0.3_{-0.2}^{+0.5}$ \\
                                   & & & & $0.67_{-0.06}^{+0.08}$ \\
        \hline
        TYC\,9340-437-1 & $-3.3_{-0.7}^{+0.5}$ & $0.23_{-0.05}^{+0.04}$ & 0.35 & $1.00_{-0.09}^{+0.09}$ \\
        \hline
        HD\,218396 & $-4.77_{-0.17}^{+0.30}$ & $0.161_{-0.006}^{+0.010}$ & 0.20 & $0.72_{-0.10}^{+0.08}$ \\
        \hline
    \end{tabular}
    \tablefoot{Multiple fractional widths ($\Delta R/R$) are reported for discs with multiple rings, but the mean particle eccentricity ($e_{\rm{rms}}$) and the outer power-law index ($\alpha_{\rm{out}}$) are reported only for the outermost ring. Median eccentricities are calculated from the 50th percentile of $\alpha_{\rm{out}}$ with 1\,$\sigma$ uncertainties, and upper limits on eccentricities are calculated from the 99.7th percentile of $\alpha_{\rm{out}}$. Eccentricities are not reported for discs with $e_{\rm{rms}}>\Delta R/R$ as the relation between outer slope and eccentricity no longer holds in this case.}
\end{table*}

\efm{The steepness of the outer edges of debris discs can provide insights into the levels of stirring, which can reveal the dynamical history of planetary system \citep{Marino2021}. We find a variety of outer power-law indices among the discs in the ARKS sample, ranging from $-1$ to $-22$, which we use to estimate the mean particle eccentricities ($e_{\rm{rms}}$) near the outer edges. The limitations of the power law parametrisations mean that eccentricity can be calculated only for the outermost ring of multi-ring discs and only for the main planetesimal belt of discs with halos. We use the approximation derived by \cite{Marino2021} for a model of a stirred disc that assumes a Rayleigh distribution of particle eccentricities and inclinations \citep[see also][]{Rafikov2023},
\begin{equation}
    e_{\rm{rms}}=0.77\frac{l_{\rm{out}}}{R_{\rm{out}}}  ,
\end{equation}
where $R_{\rm{out}}$ is either the parameter fit by the triple power law or the radius of the disc at 0.5\% of the peak surface density and where $l_{\rm{out}}=R_{\rm{out}}/
\,\alpha_{\rm{out}}$. Because this relation breaks down as $e_{\rm{rms}}$ approaches $\Delta R/R$, we only report $e_{\rm{rms}}$ for discs with $\Delta R/R < e_{\rm{rms}}$ in Table \ref{tab:excitation}. A higher $e_{\rm{rms}}$ corresponds to a shallower outer edge, with particles distributed across a wider range of radii.}

\begin{figure}
    \centering
    \includegraphics[width=0.95\linewidth]{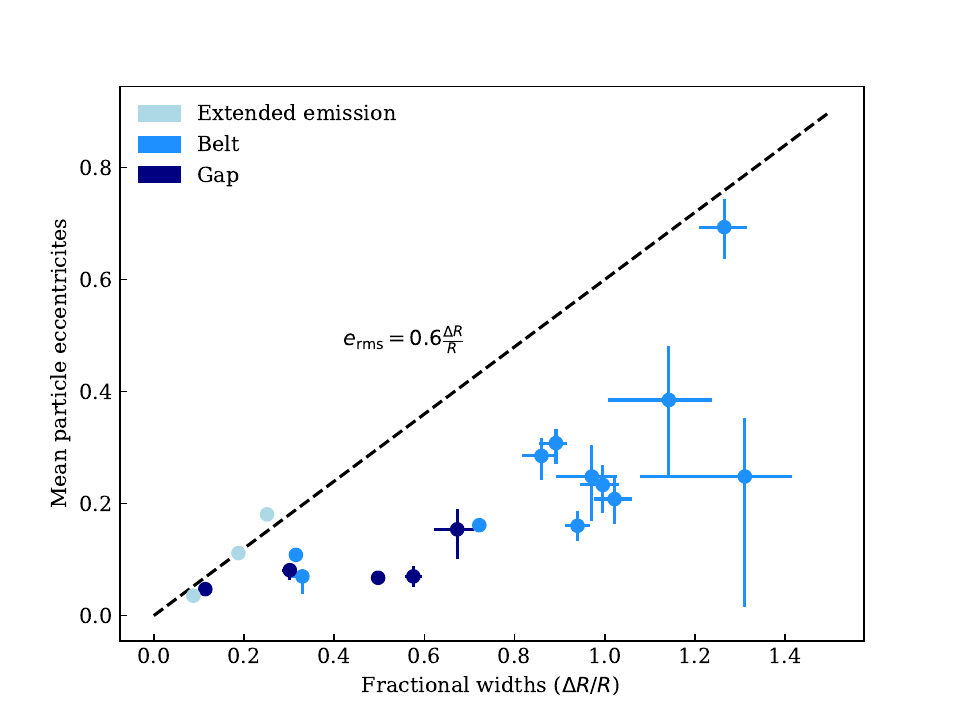}
    \caption{Eccentricities and fractional widths of the discs with $e_{\rm{rms}}<\Delta R/R$. The maximum eccentricity \citep{Marino2021} is represented by the black dashed line, and the error bars correspond to 1\,$\sigma$. For discs with gaps, we only plot $\Delta R/R$ for the outer ring as this is the only outer edge modelled by power-law parametrisations. For discs with low-amplitude emission, the eccentricities are derived from the outer slope of the planetesimal belt, not the faint emission.}
    \label{fig:eccentricities}
\end{figure}

\efm{Our calculated particle eccentricities for HD~107146, HD~206893, HD~218396, and HD~197481 are systematically smaller than those calculated by \cite{Marino2021} by a factor ${\sim}2$. These differences are likely due to the different parametric model chosen here (edges modelled as a power law instead of hyperbolic tangents or error functions) and how the values of $R_{\rm{out}}$ and $\alpha_{\rm{out}}$ are approximated to an equivalent $l_{\rm out}$.}

\efm{We find that discs with multiple statistically significant rings or extended emission are often narrower and have steeper outer edges that correspond to low particle eccentricities, as shown in Fig. \ref{fig:inner-outer-slopes}. This may indicate that the possible halos observed in the ARKS sample are not due to scattering, which can produce eccentricities greater than 1 \citep{Gladman2001}. However, we note that the observed eccentricities for these discs with extended emission or multiple significant rings are extremely close to the maximum constraint placed by their FWHM (represented by the dashed line in Fig. \ref{fig:eccentricities}) and that we do not have outer slopes for the extended emission itself. We additionally find that discs with $\Delta R/R > 0.65$ shown in Fig. \ref{fig:eccentricities} are the same as the discs with $\alpha_{\rm{in}}\leq4$ shown in Fig. \ref{fig:inner-outer-slopes}, with the exception of TYC~9340-437-1 and HD~218396. This suggests two distinct populations of discs: narrow belts with significant radial substructures---either gaps or halos---located at small radii with steep inner edges and low mean particle eccentricities, and wide belts without significant substructures located at large radii with shallow inner edges and high mean particle eccentricities. The correlation between wide belts and smooth outer edges may suggest an evolutionary explanation. If debris discs widen over time, then their outer edges will become smoother. However, further research is necessary to confirm the existence of this bimodal population, determine the origins of these substructures, and understand the underlying dynamical processes.}

\section{Conclusions} \label{sec:conclusions}
We summarise the main findings of this study in the following points. 

\begin{itemize}
    \item We modelled the 24 debris discs in the ARKS sample and obtained deprojected and deconvolved radial profiles with the visibility-space non-parametric method, \frank, the image-space non-parametric method, \rave, and a range of parametric models. The results from the different modelling approaches generally yielded close agreement, confirming the robustness of the modelling of the main radial features. 

    \item Among the sample of 24 discs, we found that 8 discs  exhibit multiple rings from non-parametric models, 5 of which were found to be statistically significant when tested with parametric modelling. Additionally, 7 discs show low-amplitude extended emission on the inner or outer edge, which could correspond to either a halo or additional low-amplitude rings (which are often difficult to distinguish between). Substructures that deviate from a single symmetric ring are therefore abundant in this sample. 

    \item The majority of discs appear to be well-fitted by either a double power law or a double-Gaussian radial profile, the latter of which can account for either two rings separated by a gap or a single ring with a radial profile that deviates from a Gaussian. 
    Both the radius and the fractional width of the debris discs show a tentative positive correlation with age. 

    \item The distribution of fractional widths of individual rings in the sample is significantly more skewed towards the narrower end than  was previously understood to be the case in debris discs. The distribution that we find appears to be similar to known protoplanetary discs for fractional widths below 0.8, but we also find a small population of broad belts that have no counterpart among the existing population of protoplanetary discs. These widths are based on the FWHM of rings and do not account for the low-amplitude extended emission found in 7 of the 24 debris discs in the ARKS sample.  

    \item Assuming that the gaps found in the debris discs are carved by planets via scattering (driven by the chaotic overlap of mean-motion resonances near the planet), we inferred a population of planets at tens of au with masses that mostly lie between a few Neptune masses and a few Jupiter masses. We used the inner edge to provide additional constraints based on planet sculpting models and the outer edge to constrain the eccentricities within the disc. Future work may wish to model the dynamical scenario in each disc more holistically by taking into account the full radial profile and age of each system. 
\end{itemize}

\begin{acknowledgements}
This study made use of 
\texttt{NumPy} \citep{doi:10.1109/MCSE.2011.37}, 
\texttt{SciPy} \citep{virtanen2019scipy},
\texttt{Matplotlib} \citep{doi:10.1109/MCSE.2007.55},
\texttt{Astropy} \citep{astropy:2013, astropy:2018, astropy:2022},
\casa \citep{2007ASPC..376..127M}, 
\frank \citep{2020MNRAS.495.3209J}, 
\rave \citep{Han2022, Han2025}, 
\mpol \citep{mpol, 2023PASP..135f4503Z},
\arksia \citep{arksia}, NASA's Astrophysics Data System Bibliographic Services and Karl Stapelfeldt's Catalog of Circumstellar discs. 
This paper makes use of the following ALMA data: ADS/JAO.ALMA\# 2022.1.00338.L, 2012.1.00142.S, 2012.1.00198.S, 2015.1.01260.S, 2016.1.00104.S, 2016.1.00195.S, 2016.1.00907.S, 2017.1.00167.S, 2017.1.00825.S, 2018.1.01222.S and 2019.1.00189.S. 
ALMA is a partnership of ESO (representing its member states), NSF (USA) and NINS (Japan), together with NRC (Canada), MOST and ASIAA (Taiwan), and KASI (Republic of Korea), in cooperation with the Republic of Chile. The Joint ALMA Observatory is operated by ESO, AUI/NRAO and NAOJ. The National Radio Astronomy Observatory is a facility of the National Science Foundation operated under cooperative agreement by Associated Universities, Inc. 
The project leading to this publication has received support from ORP, that is funded by the European Union’s Horizon 2020 research and innovation programme under grant agreement No 101004719 [ORP]. 
We are grateful for the help of the UK node of the European ARC in answering our questions and producing calibrated measurement sets. 
This research used the Canadian Advanced Network For Astronomy Research (CANFAR) operated in partnership by the Canadian Astronomy Data Centre and The Digital Research Alliance of Canada with support from the National Research Council of Canada the Canadian Space Agency, CANARIE and the Canadian Foundation for Innovation.
This research made use of the Wesleyan University high-performance computing cluster. 
YH is supported by a Caltech Barr Fellowship and this work was made possible by support from a Gates Cambridge Scholarship enabled by the Bill \& Melinda Gates Foundation (OPP1144). 
EM acknowledges support from the NASA CT Space Grant. 
SM acknowledges funding by the Royal Society through a Royal Society University Research Fellowship (URF-R1-221669) and the European Union through the FEED ERC project (grant number 101162711). 
AMH, AJF and EM gratefully acknowledge support from the National Science Foundation through Grant No. AST-2307920.  
Support for BZ was provided by The Brinson Foundation. 
This material is based upon work supported by the National Science Foundation Graduate Research Fellowship under Grant No. DGE 2140743. 
MB acknowledges funding from the Agence Nationale de la Recherche through the Ddisc project (grant No. ANR-21-CE31-0015). 
AB acknowledges research support by the Irish Research Council under grant GOIPG/2022/1895. 
CdB acknowledges support from the Agencia Estatal de Investigación del Ministerio de Ciencia, Innovación y Universidades (MCIU/AEI) under grant WEAVE: EXPLORING THE COSMIC ORIGINAL SYMPHONY, FROM STARS TO GALAXY CLUSTERS and the European Regional Development Fund (ERDF) with reference PID2023-153342NB-I00/10.13039/501100011033, as well as from a Beatriz Galindo Senior Fellowship (BG22/00166) from the MICIU. The University of La Laguna (ULL) and the Department of Economy, Knowledge, and Employment of the Government of the Canary Islands are also gratefully acknowledged for the support provided to CdB (2024/347). 
EC acknowledges support from NASA STScI grant HST-AR-16608.001-A and the Simons Foundation. 
S.E. is supported by the National Aeronautics and Space Administration through the Exoplanet Research Program (Grant No. 80NSSC23K0288, PI: Faramaz). 
This material is based upon work supported by the National Science Foundation Graduate Research Fellowship under Grant No. DGE 2140743. 
MRJ acknowledges support from the European Union's Horizon Europe Programme under the Marie Sklodowska-Curie grant agreement no. 101064124 and funding provided by the Institute of Physics Belgrade, through the grant by the Ministry of Science, Technological Development, and Innovations of the Republic of Serbia. 
This work was also supported by the NKFIH NKKP grant ADVANCED 149943 and the NKFIH excellence grant TKP2021-NKTA-64. Project no.149943 has been implemented with the support provided by the Ministry of Culture and Innovation of Hungary from the National Research, Development and Innovation Fund, financed under the NKKP ADVANCED funding scheme. 
JBL acknowledges the Smithsonian Institute for funding via a Submillimeter Array (SMA) Fellowship, and the North American ALMA Science Center (NAASC) for funding via an ALMA Ambassadorship. 
SMM and LM acknowledge funding by the European Union through the E-BEANS ERC project (grant number 100117693), and by the Irish research Council (IRC) under grant number IRCLA- 2022-3788. Views and opinions expressed are however those of the author(s) only and do not necessarily reflect those of the European Union or the European Research Council Executive Agency. Neither the European Union nor the granting authority can be held responsible for them. 
JPM acknowledges research support by the National Science and Technology Council of Taiwan under grant NSTC 112-2112-M-001-032-MY3. 
JM acknowledges funding from the Agence Nationale de la Recherche through the DDISK project (grant No. ANR-21-CE31-0015) and from the PNP (French National Planetology Program) through the EPOPEE project. 
TDP is supported by a UKRI Stephen Hawking Fellowship and a Warwick Prize Fellowship, the latter made possible by a generous philanthropic donation. 
SP acknowledges support from FONDECYT Regular 1231663 and ANID -- Millennium Science Initiative Program -- Center Code NCN2024\_001. 
AAS is supported by the Heising-Simons Foundation through a 51 Pegasi b Fellowship. 
PW acknowledges support from FONDECYT grant 3220399 and ANID -- Millennium Science Initiative Program -- Center Code NCN2024\_001. 
\end{acknowledgements}

\bibliographystyle{aa}
\bibliography{references}

\begin{appendix}

\section{Emissivity index of HD 107146} \label{app:emissivity}
\efm{HD~107146 was observed in Band 6 in April 2017 and in Band 7 in October through December 2016 and in May 2021 \citep{Marino2018,ImazBlanco2023}. Because these data are comparable in resolution and sensitivity, we can include both in our parametric models to improve our interpretation of the structure of the disc. The availability of multi-frequency data for HD~107146 additionally provides an opportunity to measure the emissivity index, $\beta$. Because the spectral index and emissivity index depend on the density of the material and the distribution of dust grain sizes \citep{Lohne2020}, we can relate these indices to the tensile strength of the dust and better constrain the masses of the bodies producing dust in debris discs \citep{Pan2012}.}

\efm{While the emissivity index is incorporated into the parametric modelling code, it is a fixed variable used to calculate the dust mass opacity $\kappa$ \citep{Zawadzki2025}. We modified the parametric modelling code to accept $\beta$ as an additional free parameter for HD~107146. Although spectral index has been measured with ALMA for several discs \citep[e.g.][]{Ricci2015,MacGregor2016,Vizgan2022}, the emissivity index has not often been fit directly. Fitting $\beta$ allows us to then calculate the grain-size distribution parameter $q$ \citep{Draine2006}, defined as
    \begin{equation}
        q = \frac{\beta}{\beta_s} + 3   ,
    \end{equation}
where $\beta_s$ is the dust opacity spectral index of small particles, assumed to be 1.8 \citep{MacGregor2016}. We obtain a best-fit and median value of $\beta=0.3\pm0.2$ for the emissivity index, which corresponds to a grain-size distribution slope of $q=3.17\pm0.11$.}

\efm{We compare this result to previous studies of spectral indices and emissivity indices in other debris discs here. \cite{MacGregor2016} measured the spectral indices of 15 debris discs and calculated the grain-size distribution slope for each. They found a weighted mean of $\langle q \rangle = 3.36\pm0.02$, with slopes in the sample ranging from 2.84\textendash3.64. Our measured grain-size distribution slope for HD~107146 is somewhat smaller than this weighted mean but within the range observed by \cite{MacGregor2016}. We note that HD~107146 has a grain-size distribution slope that is among the smallest measured in debris discs: HD~141569 has a slope of $2.84\pm0.05$ \citep{MacGregor2016} and HD~197481 has a slope of $q=3.03\pm0.02$ \citep{Vizgan2022}.}

\efm{Our measurement of the grain-size distribution slope for HD~107146 places this disc within the gravity regime of $3<q<3.21$ described by \citep{Pan2012}, in which small bodies have greater velocities than large bodies and catastrophic collisions significantly damp the velocities of all bodies in the disc. The relatively small $\beta$ value we measure for HD~107146 could also result if material strength does not dominate colliding bodies or from inhomogeneities in densities and fragmentation energies \citep{MacGregor2016}. However, additional research is necessary to understand the implications for the collisional cascade in HD~107146.}

\section{Planet constraints from the inner edge}
\label{appendix:planet-sculpting}
\efm{We compare the planet masses and semi-major axes calculated from the properties of the inner edges in Sect. \ref{sec:inner-slopes} to previous detections or constraints. All masses are below the detection limits at both the 99.7\% and 50\% confidence levels calculated by the companion paper \cite{Milli2025}, unless these limits are unavailable due to observation sensitivities.}

\efm{HD~10647 has an inner radius of 79\,au and an inner power law index of $>18$, corresponding to an upper limit of $2.5\,M_{\rm{J}}$ on a planet sculpting its inner edge and an upper limit of 48\,au on the semi-major axis of its orbit. The 1\,$\rm{M_{Jup}}$ planet previously observed in this system \citep{Butler2006} has an orbit of $\sim2$\,au that is too small to affect the inner edge of the disc. \citep{Pearce2024}.}

\efm{HD~32297 has an inner radius of 102\,au and an inner power law index of $>27$, corresponding to an upper limit of 1\,$M_{\rm{J}}$ on a planet sculpting its inner edge and an upper limit of 74\,au on the semi-major axis of its orbit. Although no planets have been previously observed in this system, the planet mass derived here is consistent with both the detection limit of 1\,$M_{\rm{J}}$ based on previous scattered light observations of the discs \citep{Bhowmik2019} and the minimum planet mass of $0.04_{-0.10}^{+0.03}\,M_{\rm{J}}$ for secular stirring \citep{Pearce2022}.} The latter estimate assumes a massless disc; adopting a maximum disc mass of $100 M_{\oplus}$ increases the minimum required planet mass to $0.63\,M_{\rm{J}}$ \citep{Sefilian2024}.

\efm{HD~107146 has an inner radius of 44\,au and an inner power law index of $>13.91$, corresponding to an upper limit of 5\,$M_{\rm{J}}$ on a planet sculpting its inner edge and an upper limit of 32\,au on the semi-major axis of its orbit. No planets have been previously observed in this system.}

\efm{HD~109573 has an inner radius of 69\,au and a median inner power law index of $34_{-3}^{+5}$, corresponding to an upper limit of $0.68_{-0.18}^{+0.30}\,M_{\rm{J}}$ on a planet sculpting its inner edge and an upper limit of 50\,au on the semi-major axis of its orbit. Although no planets have been previously observed in this system, the planet mass derived here further constrains the maximum planet mass of $0.4_{-0.4}^{+8.4}\,M_{\rm{J}}$ calculated by \cite{Milli2017} based on truncation of the inner slope observed in scattered light.}

\efm{HD~121617 has an inner radius of 65\,au and an inner power law index of $>20$, corresponding to an upper limit of 3.1\,$M_{\rm{J}}$ on a planet sculpting its inner edge and an upper limit of 49\,au on the semi-major axis of its orbit. Although no planet has been previously observed in this system, the planet mass derived here is consistent with the minimum planet mass of $0.3\pm0.2\,M_{\rm{J}}$ necessary for secular stirring \citep{Pearce2022}.}  The latter estimate assumes a massless disc; adopting a maximum disc mass of $100 M_{\oplus}$ increases the minimum required planet mass to $1 \,M_{\rm{J}}$ \citep{Sefilian2024}.

\efm{HD~131488 has an inner radius of 79\,au and a median inner power law index of $28_{-4}^{+5}$, corresponding to an upper limit of $1.2_{-0.5}^{+0.6}\,M_{\rm{J}}$ on a planet sculpting its inner edge and an upper limit of 63\,au on the semi-major axis of its orbit. No planet has been previously observed in this system.}

\efm{HD~197481 has an inner radius of 25\,au and an inner power law index of $>14$, corresponding to an upper limit of 3.2\,$M_{\rm{J}}$ on a planet sculpting its inner edge and an upper limit of 19\,au on the semi-major axis of its orbit. Although two Neptune-sized planets have been detected in this system \citep{Plavchan2020,Gilbert2022}, these objects have orbits that are too small to affect the inner edge of the disc according to the model of \citet{Pearce2024}.}

\section{Best-fit parameters from parametric modelling}
This appendix lists the fitted model parameters from parametric modelling. Details of the fitting methods are described in Sect.~\ref{sec:parametric_models}.

\begin{sidewaystable*}[ph!]
    \centering
    \caption{MCMC results for discs with a clear single peak. }
    \label{tab:MCMC-simple}
    \setlength{\tabcolsep}{2.4pt} 
    \begin{tabular}{l|lr|lr|lr|lr|lr|lr|lr|lr|lr}
        \multicolumn{19}{c}{Double power law} \\
        \hline \hline
        Disc & \multicolumn{2}{c}{$R_{\rm{c}}$ (au)} & \multicolumn{2}{c}{$\rm{\alpha_{in}}$} & \multicolumn{2}{c}{$\rm{\alpha_{out}}$} & \multicolumn{2}{c}{$\Sigma_{\rm{c}}$} & \multicolumn{2}{c}{PA $(^\circ)$} & \multicolumn{2}{c}{$\Delta \alpha$ (mas)} & \multicolumn{2}{c}{$\Delta \delta$ (mas)} & \multicolumn{2}{c}{$i$ $(^\circ)$} & \multicolumn{2}{c}{$F_*$ $(\rm{\mu Jy})$}\\
        \hline
         HD\,9672 & $95$ & $96_{-4}^{+4}$ & $2.4$ & $2.27_{-0.19}^{+0.21}$ & $-1.1$ & $-1.11_{-0.09}^{+0.08}$ & $-4.188$ & $-4.191_{-0.009}^{+0.009}$ & $108.02$ & $107.95_{-0.15}^{+0.12}$ & $46$ & $51_{-15}^{+16}$ & $-38$ & $-39_{-6}^{+7}$ & $79.1$ & $79.13_{-0.11}^{+0.12}$ & $16$ & $14_{-9}^{+10}$ \\
         HD\,10647 & $74.5$ & $73.7_{-0.7}^{+0.8}$ & $120$ & $140_{-90}^{+100}$ & $-1.22$ & $-1.24_{-0.08}^{+0.04}$ & $-4.584$ & $-4.578_{-0.010}^{+0.015}$ & $57.48$ & $57.04_{-0.07}^{+0.46}$ & $118$ & $123_{-10}^{+10}$ & $-23$ & $-22_{-8}^{+8}$ & $79.69$ & $79.37_{-0.13}^{+0.37}$ & $142$ & $138_{-9}^{+10}$\\
         HD\,14055 & $189$ & $181_{-13}^{+9}$ & $1.19$ & $1.20_{-0.11}^{+0.14}$ & $-3.7$ & $-3.1_{-1.0}^{+0.7}$ & $-5.074$ & $-5.075_{-0.014}^{+0.014}$ & $161.93$ & $161.60_{-0.14}^{+0.27}$ & $-56$ & $-62_{-10}^{+9}$ & $-67$ & $-66_{-12}^{+12}$ & $80.2$ & $80.0_{-0.3}^{+0.2}$ & $100$ & $101_{-10}^{+9}$\\
         HD\,15257 & $240$ & $250_{-40}^{+50}$ & $0.9$ & $0.8_{-0.2}^{+0.4}$ & $-2.3$ & $-3.1_{-2.9}^{+1.3}$ & $-5.11$ & $-5.12_{-0.06}^{+0.05}$ & $51$ & $49_{-4}^{+3}$ & $100$ & $100_{-70}^{+80}$ & $130$ & $100_{-90}^{+90}$ & $65$ & $63_{-3}^{+3}$ & $100$ & $100_{-30}^{+30}$ \\
         HD\,32297 & $105.1$ & $105.2_{-0.5}^{+0.7}$ & $110$ & $100_{-50}^{+60}$ & $-6.33$ & $-6.30_{-0.12}^{+0.12}$ & $-2.719$ & $-2.722_{-0.014}^{+0.013}$ & $47.45$ & $47.47_{-0.03}^{+0.05}$ & $20.2$ & $20.5_{-1.4}^{+1.6}$ & $15.3$ & $15.3_{-1.3}^{+1.3}$ & $88.46$ & $88.39_{-0.05}^{+0.07}$ & $23$ & $23_{-11}^{+13}$\\
         HD\,39060 & $134$ & $130_{-5}^{+5}$ & $1.07$ & $1.09_{-0.07}^{+0.10}$ & $-5.4$ & $-4.8_{-0.8}^{+0.8}$ & $-4.202$ & $-4.201_{-0.011}^{+0.012}$ & $29.66$ & $29.73_{-0.06}^{+0.07}$ & $29$ & $26_{-12}^{+13}$ & $3$ & $-4_{-18}^{+19}$ & $86.06$ & $86.06_{-0.07}^{+0.04}$ & $73$ & $66_{-13}^{+12}$\\
         HD\,84870 & $220$ & $220_{-30}^{+30}$ & $1.5$ & $1.5_{-0.3}^{+0.3}$ & $-1.9$ & $-2.0_{-0.7}^{+0.5}$ & $-4.70$ & $-4.70_{-0.04}^{+0.03}$ & $134$ & $137_{-3}^{+4}$ & $390$ & $360_{-80}^{+70}$ & $-360$ & $-280_{-100}^{+100}$ & $55$ & $53_{-2}^{+3}$ & $0$ & $8_{-6}^{+10}$\\
         HD\,95086 & $199$ & $198_{-7}^{+11}$ & $1.84$ & $1.93_{-0.18}^{+0.20}$ & $-2.6$ & $-2.5_{-0.3}^{+0.2}$ & $-3.966$ & $-3.972_{-0.017}^{+0.015}$ & $99$ & $97_{-2}^{+3}$ & $-120$ & $-110_{-40}^{+40}$ & $90$ & $90_{-30}^{+30}$ & $30$ & $31_{-2}^{+3}$ & $22$ & $20_{-13}^{+16}$\\
         HD\,121617 & $68.5$ & $68.5_{-0.5}^{+0.7}$ & $70$ & $60_{-20}^{+30}$ & $-6.9$ & $-6.9_{-0.4}^{+0.2}$ & $-2.993$ & $-2.995_{-0.021}^{+0.019}$ & $58.7$ & $58.4_{-0.6}^{+0.7}$ & $-1$ & $0_{-3}^{+2}$ & $-4$ & $-3_{-3}^{+2}$ & $44.5$ & $44.6_{-0.7}^{+0.4}$ & $22$ & $27_{-10}^{+10}$\\
         HD\,145560 & $74.5$ & $75.0_{-1.2}^{+1.2}$ & $6.1$ & $6_{-0.5}^{+0.6}$ & $-6.8$ & $-7.1_{-0.5}^{+0.4}$ & $-3.058$ & $-3.056_{-0.016}^{+0.015}$ & $40$ & $39.2_{-0.8}^{+0.8}$ & $16$ & $17_{-3}^{+3}$ & $-19$ & $-18_{-3}^{+4}$ & $47.5$ & $47.5_{-0.5}^{+0.5}$ & $24$ & $26_{-8}^{+9}$\\
         HD\,161868 & $154$ & $154_{-10}^{+8}$ & $0.97$ & $0.99_{-0.11}^{+0.14}$ & $-3.9$ & $-3.7_{-0.8}^{+0.7}$ & $-4.97$ & $-4.98_{-0.03}^{+0.02}$ & $55.1$ & $56.1_{-1.5}^{+1.9}$ & $10$ & $0_{-20}^{+30}$ & $-90$ & $-100_{-20}^{+30}$ & $65.1$ & $65.7_{-1.1}^{+1.2}$ & $151$ & $152_{-20}^{+20}$\\
         HD\,170773 & 197 & $200_{-7}^{+13}$ & 6.3 & $5.6_{-1.2}^{+1.0}$ & $-10$ & $-11_{-5}^{+2}$ & $-3.99$ & $-4.00_{-0.03}^{+0.02}$ & 113.4 & $113.4_{-9.8}^{+1.8}$ & $-130$ & $-130_{-40}^{+30}$ & 20 & $20_{-30}^{+30}$ & 32.9 & $31.7_{-1.6}^{+1.9}$ & 90 & $80_{-20}^{+20}$\\
         TYC\,9340 & $99$ & $98_{-7}^{+6}$ & $1.73$ & $1.77_{-0.14}^{+0.17}$ & $-1.9$ & $-1.8_{-0.3}^{+0.3}$ & $-3.834$ & $-3.840_{-0.020}^{+0.020}$ & $152$ & $147_{-16}^{+8}$ & $-10$ & $-20_{-60}^{+50}$ & $-30$ & $-50_{-50}^{+40}$ & $29$ & $32_{-4}^{+3}$ & $1$ & $7_{-5}^{+11}$\\
         \hline
         \multicolumn{19}{c}{}\\
         \multicolumn{19}{c}{Asymmetric Gaussian} \\
         \hline \hline
         disc & \multicolumn{2}{c}{$R_{\rm{c}}$ (au)} & \multicolumn{2}{c}{$\sigma_{\rm{in}}$ (au)} & \multicolumn{2}{c}{$\sigma_{\rm{out}}$ (au)} & \multicolumn{2}{c}{$\Sigma_{\rm{c}}$} & \multicolumn{2}{c}{PA $(^\circ)$} & \multicolumn{2}{c}{$\Delta \alpha$ (mas)} & \multicolumn{2}{c}{$\Delta \delta$ (mas)} & \multicolumn{2}{c}{$i$ $(^\circ)$} & \multicolumn{2}{c}{$F_*$ $(\rm{\mu Jy})$}\\
         \hline
         HD\,76582 & $184$ & $181_{-8}^{+9}$ & $55$ & $53_{-6}^{+6}$ & $143$ & $145_{-10}^{+10}$ & $-4.825$ & $-4.822_{-0.014}^{+0.013}$ & $104.8$ & $104.7_{-0.5}^{+0.3}$ & $0$ & $-20_{-50}^{+50}$ & $-40$ & $-50_{-20}^{+30}$ & $73.8$ & $73.7_{-0.7}^{+0.4}$ & $53$ & $57_{-17}^{+18}$\\
         HD\,161868 & 115 & $120_{-9}^{+11}$ & 50 & $54_{-8}^{+9}$ & 74 & $72_{-11}^{+10}$ & $-5.12$ & $-5.12_{-0.03}^{+0.02}$ & 55.1 & $57.0_{-1.9}^{+2.3}$ & 0 & $10_{-30}^{+30}$ & $-100$ & $-110_{-20}^{+30}$ & 65.4 & $66.4_{-1.4}^{+1.2}$ & 150 & $140_{-30}^{+20}$\\
         HD\,170773 & $191$ & $194_{-4}^{+7}$ & $27$ & $29_{-3}^{+6}$ & $27$ & $26_{-4}^{+3}$ & $-4.185$ & $-4.191_{-0.018}^{+0.018}$ & $115.1$ & $113.3_{-2.9}^{+1.7}$ & $-120$ & $-130_{-30}^{+40}$ & $10$ & $20_{-20}^{+30}$ & $32.1$ & $32.2_{-1.6}^{+1.6}$ & $90$ & $80_{-20}^{+20}$\\
         TYC\,9340 & 78 & $81_{-5}^{+4}$ & 25 & $26_{-3}^{+2}$ & 91 & $88_{-9}^{+9}$ & $-4.012$ & $-4.017_{-0.020}^{+0.020}$ & 151 & $137_{-13}^{+16}$ & $-50$ & $-60_{-50}^{+60}$ & 20 & $40_{-60}^{+60}$ & 28 & $30_{-5}^{+4}$ & 1 & $9_{-7}^{+12}$\\
         HD\,218396 & 205 & $208_{-10}^{+11}$ & 45 & $48_{-6}^{+5}$ & 89 & $96_{-9}^{+14}$ & $-4.91$ & $-4.93_{-0.02}^{+.03}$ & 95 & $92_{-13}^{+3}$ & $-30$ & $-60_{-50}^{+50}$ & $-90$ & $-100_{-70}^{+60}$ & 12 & $12_{-6}^{+16}$ & 58 & $55_{-10}^{+10}$ \\
         \hline
         \multicolumn{19}{c}{}\\
         \multicolumn{19}{c}{Power law + error function} \\
         \hline \hline
         disc & \multicolumn{2}{c}{$R_{\rm{c}}$ (au)} & \multicolumn{2}{c}{$\sigma_{\rm{in}}$ (au)} & \multicolumn{2}{c}{$\alpha_{\rm{out}}$} & \multicolumn{2}{c}{$\Sigma_{\rm{c}}$} & \multicolumn{2}{c}{PA $(^\circ)$} & \multicolumn{2}{c}{$\Delta \alpha$ (mas)} & \multicolumn{2}{c}{$\Delta \delta$ (mas)} & \multicolumn{2}{c}{$i$ $(^\circ)$} & \multicolumn{2}{c}{$F_*$ $(\rm{\mu Jy})$}\\
         \hline
         HD\,15257 & 130 & $110_{-30}^{+30}$ & 0.46 & $0.51_{-0.12}^{+0.35}$ & 0.9 & $0.6_{-0.4}^{+0.5}$ & $-5.33$ & $-5.40_{-0.12}^{+0.10}$ & 47 & $47
         _{-3}^{+3}$ & 130 & $80_{-70}^{+70}$ & 80 & $80_{-90}^{+80}$ & 61.4 & $62.4_{-2.1}^{+1.9}$ & 50 & $50_{-30}^{+40}$\\
         HD\,121617 & 67.5 & $67.8_{-0.4}^{+0.6}$ & 0.007 & $0.018_{-0.011}^{+0.014}$ & 6.7 & $6.9_{-0.3}^{+0.4}$ & $-3.268$ & $-3.264_{-0.016}^{+0.017}$ & 58.6 & $58.4_{-0.6}^{+0.7}$ & 0 & $0_{-3}^{+2}$ & $-4$ & $-3_{-3}^{+2}$ & 44.4 & $44.6_{-0.6}^{+0.4}$ & 22 & $27_{-10}^{+10}$\\
         HD\,161868 & 100 & $99_{-4}^{+3}$ & 0.299 & $0.297_{-0.014}^{+0.015}$ & 1.89 & $1.90_{-0.17}^{+0.16}$ & $-5.13$ & $-5.12_{-0.03}^{+0.02}$ & 58.9 & $58.5_{-1.2}^{+1.1}$ & 60 & $70_{-30}^{+30}$ & $-160$ & $-150_{-30}^{+20}$ & 67.7 & $66.8_{-1.1}^{+1.1}$ & 20 & $27_{-19}^{+29}$\\
         \hline
    \end{tabular}
    \tablefoot{The left column under each parameter is the best fit---defined as having the lowest $\chi^2$ value---while the right column is the median. Uncertainties are calculated from the 16th and 84th percentiles of the posterior distribution. We report multiple parametrisations for discs where the significance between these models, as measured by the AIC and the BIC, was minimal or ambiguous. The surface density normalisation $\Sigma_{\rm{c}}$ is parametrised as an exponent, with $10^{\Sigma_{\rm{c}}}$ in units of $\rm{g\;cm^{-2}}$.}
\end{sidewaystable*}

\begin{sidewaystable*}[ph!]
    \centering
    \setlength{\tabcolsep}{6pt} 
    \caption{MCMC results for discs with gaps or extended emission fit by a double Gaussian. }
    \label{tab:MCMC-gaussian}
    \begin{tabular}{l|llllllllllllll}
        \multicolumn{12}{c}{Best Fits} \\
        \hline \hline
        Disc & $R_1$ (au) & $R_2$ (au) & $\sigma_1$ (au) & $\sigma_2$ (au) & $C$ & $\Sigma_{\rm{c}}$ & PA $(^\circ)$ & $\Delta \alpha$ (mas) & $\Delta \delta$ (mas) & $i$ $(^\circ)$ & $F_*$ $(\rm{\mu Jy})$ \\
        \hline
         HD\,10647 & 89.2 & 168 & 12 & 48 & 0.659 & $-4.433$ & 56.93 & 121 & $-17$ & 79.08 & 128 \\
         HD\,15115 & 65.8 & 97.54 & 1.6 & 4.75 & 0.26 & $-3.40$ & 98.466 & 37 & $-14$ & 86.7 & 23 \\
         HD\,61005 & $69.7$ & $94$ & $7.3$ & $32.3$ & $0.836$ & $-3.358$ & $70.29$ & $2$ & $-16$ & $86.2$ & $1$\\
         HD\,92945 & 57.1 & 102.7 & 5.6 & 24.1 & 0.52 & $-4.09$ & 100.9 & $-70$ & $-87$ & 66.3 & 39\\
         HD\,109573 & $77.65$ & $85$ & $2.98$ & $24$ & $0.979$ & $-2.77$ & $26.52$ & $13.8$ & $-39.9$ & $76.54$ & $52$\\
         HD\,121617 & 75.2 & 100 & 6 & 42 & 0.948 & $-3.17$ & 58.0 & 0 & $-2$ & 43.9 & 5 \\
         HD\,131488 & $90.8$ & $95.1$ & $0.69$ & $17.3$ & $0.97$ & $-1.96$ & $97.305$ & $1.1$ & $-7.3$ & $84.84$ & $40$\\
         HD\,131835 & $68.7$ & $104.2$ & $4.9$ & $44.7$ & $0.866$ & $-2.9$ & $58.94$ & $1.4$ & $-5.9$ & $74.47$ & $4$\\
         HD\,206893 & 42 & 125 & 6 & 36 & 0.43 & $-4.53$ & 64 & $-50$ & 20 & 44 & 17\\
         \hline
         \multicolumn{12}{c}{Medians}\\
         \hline\hline
         Disc & $R_1$ (au) & $R_2$ (au) & $\sigma_1$ (au) & $\sigma_2$ (au) & $C$ & $\Sigma_{\rm{c}}$ & PA $(^\circ)$ & $\Delta \alpha$ (mas) & $\Delta \delta$ (mas) & $i$ $(^\circ)$ & $F_*$ $(\rm{\mu Jy})$ \\
         \hline
         HD\,10647 & $89.1_{-0.8}^{+0.7}$ & $164_{-8}^{+6}$ & $11.9_{-0.9}^{+1.0}$ & $46_{-3}^{+3}$ & $0.664_{-0.010}^{+0.009}$ & $-4.438_{-0.012}^{+0.012}$ & $56.93_{-0.05}^{+0.06}$ & $121_{-10}^{+10}$ & $-23_{-7}^{+9}$ & $79.00_{-0.18}^{+0.12}$ & $130_{-10}^{+10}$\\
         HD\,15115 & $65.8_{-0.3}^{+0.6}$ & $97.57_{-0.05}^{+0.07}$ & $2.4_{-0.8}^{+1.0}$ & $4.73_{-0.20}^{+0.18}$ & $0.19_{-0.04}^{+0.06}$ & $-3.44_{-0.02}^{+0.04}$ & $98.460_{-0.013}^{+0.011}$ & $36_{-3}^{+2}$ & $-14_{-0.9}^{+0.9}$ & $86.71_{-0.02}^{+0.03}$ & $28_{-6}^{+5}$\\
         HD\,61005 & $69.4_{-0.4}^{+0.5}$ & $95_{-3}^{+2}$ & $7.4_{-0.5}^{+0.5}$ & $31.7_{-1.5}^{+1.6}$ & $0.837_{-0.012}^{+0.012}$ & $-3.359_{-0.015}^{+0.016}$ & $70.29_{-0.03}^{+0.02}$ & $0_{-5}^{+5}$ & $-17_{-2}^{+2}$ & $86.17_{-0.07}^{+0.05}$ & $5_{-4}^{+7}$\\
         HD\,92945 & $57.0_{-0.8}^{+0.8}$ & $102.7_{-1.5}^{+1.5}$ & $5.5_{-1.1}^{+1.2}$ & $25.1_{-1.9}^{+2.0}$ & $0.52_{-0.03}^{+0.04}$ & $-4.10_{-0.03}^{+0.04}$ & $100.5_{-0.5}^{+0.5}$ & $-80_{-30}^{+30}$ & $-78_{-16}^{+17}$ & $66.0_{-0.6}^{+0.5}$ & $36_{-15}^{+15}$\\
         HD\,109573 & $77.61_{-0.16}^{+0.04}$ & $84_{-1.8}^{+2.1}$ & $2.89_{-0.19}^{+0.18}$ & $22_{-4}^{+4}$ & $0.977_{-0.009}^{+0.006}$ & $-2.77_{-0.02}^{+0.02}$ & $26.51_{-0.07}^{+0.05}$ & $13.9_{-0.8}^{+0.8}$ & $-39.6_{-1.2}^{+1.2}$ & $76.65_{-0.09}^{+0.09}$ & $38_{-17}^{+17}$\\
         HD\,121617 & $75.3_{-0.3}^{+0.4}$ & $103_{-5}^{+8}$ & $6_{-0.5}^{+0.5}$ & $42_{-7}^{+8}$ & $0.952_{-0.010}^{+0.008}$ & $-3.17_{-0.02}^{+0.03}$ & $58.3_{-0.6}^{+0.6}$ & $-1_{-2}^{+2}$ & $-2_{-2}^{+3}$ & $44.1_{-0.5}^{+0.6}$ & $12_{-8}^{+11}$ \\
         HD\,131488 & $90.84_{-0.08}^{+0.18}$ & $96.1_{-0.7}^{+0.6}$ & $0.96_{-0.19}^{+0.28}$ & $19.3_{-1.3}^{+1.9}$ & $0.966_{-0.006}^{+0.006}$ & $-2.08_{-0.10}^{+0.08}$ & $97.301_{-0.029}^{+0.012}$ & $1.2_{-0.6}^{+0.6}$ & $-7.1_{-0.3}^{+0.3}$ & $84.87_{-0.03}^{+0.06}$ & $38_{-8}^{+8}$\\
         HD\,131835 & $68.5_{-0.4}^{+0.4}$ & $105.8_{-1.7}^{+2.0}$ & $4.9_{-0.6}^{+0.6}$ & $43.5_{-1.5}^{+1.3}$ & $0.869_{-0.010}^{+0.011}$ & $-2.939_{-0.04}^{+0.04}$ & $58.95_{-0.18}^{+0.18}$ & $1.6_{-1.8}^{+1.8}$ & $-5.6_{-1.4}^{+1.3}$ & $74.41_{-0.20}^{+0.20}$ & $10_{-7}^{+11}$\\
         HD\,206893 & $42_{-5}^{+5}$ & $126_{-3}^{+4}$ & $6_{-4}^{+9}$ & $36_{-3}^{+4}$ & $0.38_{-0.17}^{+0.22}$ & $-4.57_{-0.12}^{+0.20}$ & $62_{-3}^{+3}$ & $-10_{-60}^{+60}$ & $30_{-50}^{+50}$ & $46_{-2}^{+3}$ & $14_{-6}^{+6}$\\
         \hline
    \end{tabular}
    \tablefoot{The top half of the table describes the best fits, while the bottom half describes the median values and their uncertainties, as calculated from the 16th and 84th percentiles of the posterior distribution. The surface density normalisation $\Sigma_{\rm{c}}$ is parametrised as an exponent, with $10^{\Sigma_{\rm{c}}}$ in units of $\rm{g\;cm^{-2}}$. Although HD\,15115, HD\,107146, and HD\,197481 are included in this table, these discs are better fit by variations of the triple power law or double power law functional forms. These results can be found in Table \ref{tab:snowflake-params}.}
\end{sidewaystable*}

\begin{table*}
    \centering
    \caption{MCMC results from the best parametric models for HD\,15115, HD\,107146, and HD\,197481. }
    \label{tab:snowflake-params}
    \begin{tabular}{l|cc|cc|cc|cc}
        \hline \hline
        & \multicolumn{2}{c}{HD\,15115} & \multicolumn{2}{c}{HD\,107146} & \multicolumn{4}{c}{HD\,197481} \\
        & \multicolumn{2}{c}{2 pow. + 1 gap} & \multicolumn{2}{c}{3 pow. + 2 gaps} & \multicolumn{2}{c}{3 pow. + 1 gap} & \multicolumn{2}{c}{3 pow.} \\
        Parameter & Best fit & Median & Best fit & Median & Best fit & Median & Best fit & Median \\
        \hline
        $R_{\rm{c}}$ (au) & 98.11 & $98.09_{-0.19}^{+0.16}$ & & & & & & \\
        $R_{\rm{in}}$ (au) & & & 44.4 & $43.8_{-0.9}^{+0.8}$ & 25.4 & $25.1_{-1.0}^{+0.8}$ & 26.7 & $26.9_{-0.7}^{+0.5}$ \\
        $R_{\rm{out}}$ (au) & & & 137.8 & $137.5_{-1.0}^{+1.0}$ & 38.7 & $38.1_{-1.4}^{+1.2}$ & 38.3 & $39.0_{-1.3}^{+1.1}$ \\
        $\alpha_{\rm{in}}$ & 7.9 & $7.9_{-0.2}^{+0.3}$ & 6.7 & $6.6_{-0.3}^{+0.4}$ & 25 & $28_{-11}^{+14}$ & 28 & $25_{-5}^{+4}$ \\
        $\alpha_{\rm{mid}}$ & & & 0 & $0.02_{-0.05}^{+0.04}$ & 0.9 & $0.8_{-0.4}^{+0.3}$ & 0.8 & $0.5_{-0.3}^{+0.4}$ \\
        $\alpha_{\rm{out}}$ & $-16.2$ & $-16.3_{-0.8}^{+0.6}$ & $-11.3$ & $-11.4_{-0.5}^{+0.6}$ & $-9.3$ & $-9.5_{-2.0}^{+1.5}$ & $-11.1$ & $-11.4_{-2.1}^{+1.4}$ \\
        $R_1$ (au) & 94 & $93.9_{-0.3}^{+0.3}$ & 55.9 & $56.1_{-0.3}^{0.5}$ & 36 & $34_{-2}^{+3}$ \\
        $\sigma_1$ (au) & 11 & $10.9_{-0.4}^{+0.4}$ & 2.7 & $2.9_{-0.4}^{+0.6}$ & 7.5 & $8_{-1.1}^{+1.2}$ \\
        $C_1$ & 0.448 & $0.452_{-0.008}^{+0.007}$ & 0.64 & $0.57_{-0.9}^{+0.7}$ & 0.8 & $0.6_{-0.3}^{+0.3}$ \\
        $R_2$ (au) & & & 77.9 & $78.1_{-0.8}^{+0.6}$ & & & & \\
        $\sigma_2$ (au) & & & 16.2 & $16.6_{-0.8}^{+1.0}$ & & & & \\
        $C_2$ & & & 0.692 & $0.693_{-0.016}^{+0.025}$ & & & & \\
        $\Sigma_c$ & $-3.04$ & $-3.02_{-0.02}^{+0.02}$ & $-4.17$ & $-4.18_{-0.04}^{+0.04}$ & $-3.82$ & $-3.89_{-0.10}^{+0.11}$ & $-4.05$ & $-3.99_{-0.08}^{+0.06}$ \\
        PA $(^\circ)$ & 98.478 & $98.464_{-0.033}^{+0.019}$ & 163.2 & $163.29_{-0.16}^{+0.19}$ & 128.72 & $128.73_{-0.02}^{+0.03}$ & 128.72 & $128.74_{-0.03}^{+0.05}$ \\
        $\Delta \alpha$ (mas) & 36 & $36_{-2}^{+3}$ & 12 & $11_{-6}^{+7}$ & $-19$ & $-20_{-7}^{+6}$ & $-17$ & $-22_{-7}^{+7}$ \\
        $\Delta \delta$ (mas) & $-14.0$ & $-14.0_{-0.09}^{+0.09}$ & $-20$ & $-29_{-5}^{+5}$ & $-9$ & $-9_{-6}^{+5}$ & $-10$ & $-8_{-6}^{+5}$ \\
        $i$ $(^\circ)$ & 86.753 & $86.735_{-0.041}^{+0.017}$ & 14.5 & $14.4_{-0.3}^{+1.1}$ & 88.4 & $88.42_{-0.06}^{+0.03}$ & 88.41 & $88.40_{-0.08}^{+0.04}$ \\
        $F_{*}$ $(\rm{\mu Jy})$ & 21 & $25_{-6}^{+6}$ & 23.1 &  $23.6_{-1.3}^{+1.7}$ & \multicolumn{4}{c}{See Table \ref{tab:HD197481-fluxes}} \\
        \hline
    \end{tabular}
    \tablefoot{The surface density normalisation $\Sigma_{\rm{c}}$ is parametrised as an exponent, with $10^{\Sigma_{\rm{c}}}$ in units of $\rm{g\;cm^{-2}}$. For HD\,197481, the significance of the gap centred at 36\,au is ambiguous, so we report results for two parametrisations here.}
\end{table*}

\begin{table*}
    \centering
    \caption{Stellar fluxes for HD\,197481. }
    \label{tab:HD197481-fluxes}
    \begin{tabular}{c cc cc cc}
        \hline \hline
        Date & \multicolumn{2}{c}{ARKS Fitting} & \multicolumn{2}{c}{Disc + Ring} & \multicolumn{2}{c}{Fiducial}\\
        & Best Fit & Median & Best Fit & Median & Best Fit & Median\\
        \hline
        March 26, 2014 & 327 & $343_{-19}^{+18}$ & 370 & $390_{20}^{+20}$ & 400 & $390_{-20}^{+20}$\\
        August 18, 2014 & 130 & $140_{-20}^{+30}$ & 140 & $150_{-30}^{+20}$ & 170 & $160_{-30}^{+20}$\\
        June 24, 2015 & 196 & $205_{-18}^{+18}$ & 210 & $220_{-20}^{+20}$ & 240 & $240_{-20}^{+20}$\\\hline
    \end{tabular}
    \tablefoot{ARKS fitting results are from the best parametric model for HD\,197481, a triple power law with one gap. Disc + ring and fiducial model results are from \cite{Daley2019}. All fluxes are in units of $\mu \rm{Jy}$.}
\end{table*}

\section{Model and residual maps from parametric modelling}
This appendix displays the model images and residual maps from parametric modelling. Details of the fitting methods are described in Sect.~\ref{sec:parametric_models}.

\begin{figure*}
    \centering
    \includegraphics[width=\textwidth]{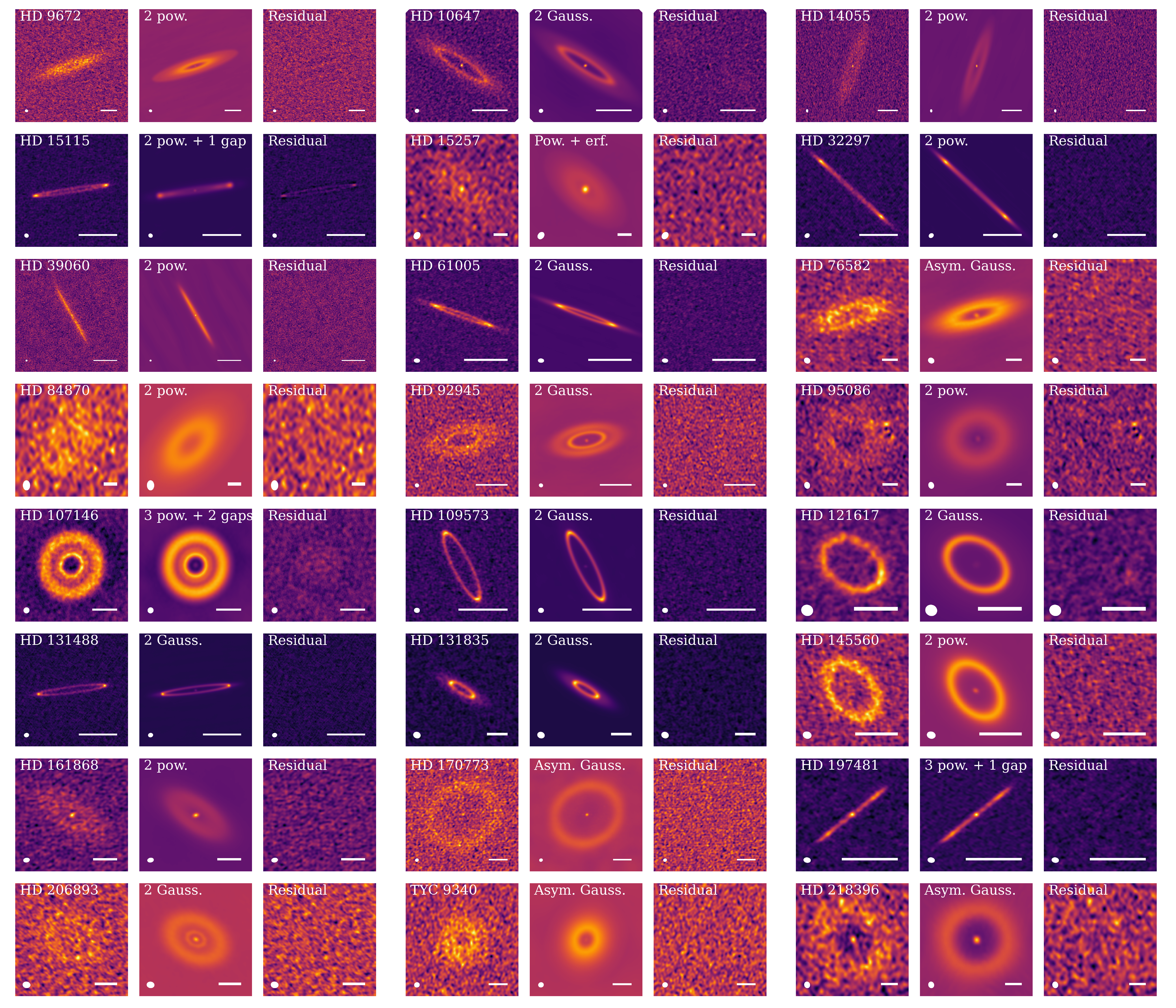}
    \caption{Data (left), model (middle), and residual (right) images for the best-fit parametric model for each disc. We compare the images in the visibility domain, and we apply a Briggs weighting with a robust parameter of 0.5 by default. Scale bars are 100\,au for all discs, except HD\,197481, which has a scale bar of 50\,au.}
    \label{fig:parametric_DMR}
\end{figure*}

\FloatBarrier
\section{AIC and BIC parametric model comparison}
\begin{figure*}
    \centering
    \includegraphics[width=0.95\linewidth]{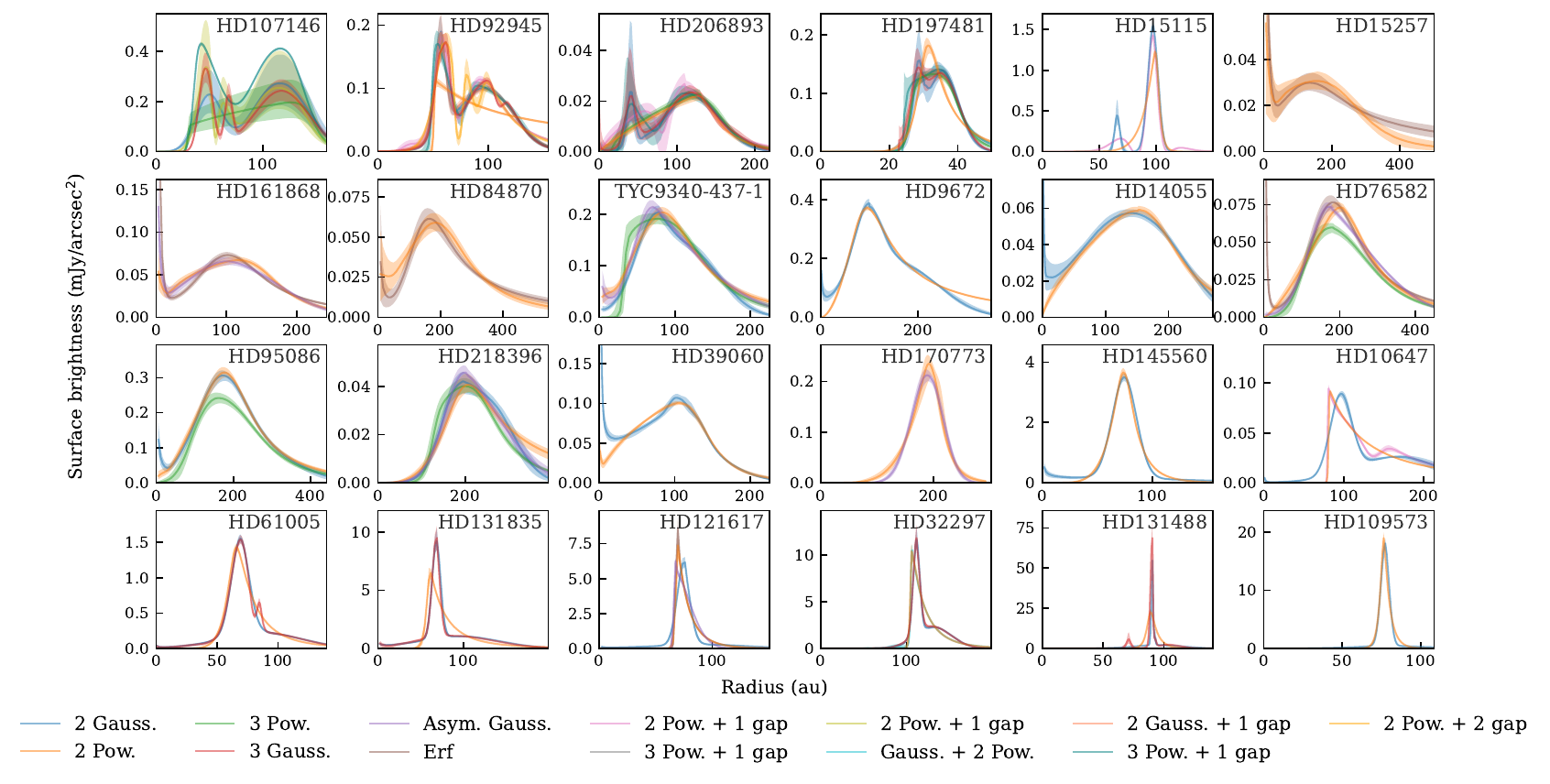}
    \caption{Deprojected radial profiles fitted under various parametric models attempted for each disc.}
    \label{fig:parametric_profiles}
\end{figure*}

This Appendix lists the results of model comparison among the parametric models fitted to each system in Table~\ref{tab:AICBIC} and displays the fitted parametric radial profiles in Fig.~\ref{fig:parametric_profiles}. Details of the fitting methods are described in Sect.~\ref{sec:parametric_models}.

\begin{table*}
    \centering
    \caption{AIC and BIC comparison of parametric models.}
    \label{tab:AICBIC}
    \begin{tabular}{c|l|cc}
        \hline
        \hline
        Disc & Functional Forms & AIC Confidence ($\sigma$) & $\Delta$BIC \\
        \hline
        \multirow{2}{*}{HD\,9672} & \textbf{Double power law} & --- & --- \\
                                 & Double Gaussian & 2.9 & $>10$ \\
        \hline
        \multirow{2}{*}{HD\,10647} & \textbf{Double Gaussian} & --- & 8.2 \\
                                  & Double power law & 3.6 & --- \\
                                  & Double power law + 1 gap & 2.7 & $>10$ \\
                                  
        \hline
        \multirow{2}{*}{HD\,14055} & \textbf{Double power law} & --- & --- \\
                                  & Double Gaussian & 0.77 & $>10$ \\
        \hline
        \multirow{2}{*}{HD\,15115} & \textbf{Double power law + 1 gap} & --- & ---\\
                                   & Double Gaussian & 4.9 & $>10$ \\
                                   & Double power law & $>5$ & $>10$ \\
        \hline
        \multirow{2}{*}{HD\,15257} & \textbf{Power law + error function} & --- & --- \\
                                   & Double power law & 0.97 & 0.16 \\
        \hline
        \multirow{3}{*}{HD\,32297} & \textbf{Double power law} & 1.4 & --- \\
                                  & Double Gaussian & 4.7 & $>10$ \\
                                  & Gaussian + double power law & --- & $>10$ \\
        \hline
        \multirow{2}{*}{HD\,39060} & \textbf{Double power law} & 1.8 & --- \\
                                  & Double Gaussian & --- & $>10$ \\
        \hline
        \multirow{3}{*}{HD\,61005} & \textbf{Double Gaussian} & --- & --- \\
                                   & Double power law & $>5$ & $>10$ \\
                                   & Triple Gaussian & 0.17 & $>10$ \\
        \hline
        \multirow{4}{*}{HD\,76582} & \textbf{Asymmetric Gaussian} & --- & --- \\
                                  & Double power law & 2.8 & $>10$ \\
                                  & Power law + error function & 4.1 & $>10$ \\
                                  & Triple power law & 1.1 & $>10$ \\
        \hline
        \multirow{2}{*}{HD\,84870} & \textbf{Double power law} & --- & --- \\
                                  & Power law + error function & 1.7 & 4.7 \\
        \hline
        \multirow{4}{*}{HD\,92945} & \textbf{Double Gaussian} & --- & --- \\
                                  & Double Gaussian + 1 gap & 1.9 & $>10$ \\
                                  & Triple Gaussian & 1.8 & $>10$ \\
                                  & Double power law & $>5$ & $>10$ \\
                                  & Double power law + 1 gap & 4.2 & $>10$ \\
                                  & Double power law + 2 gaps & 2.3 & $>10$ \\
                                  & Triple power law + 1 gap & 2.4 & $>10$ \\
        \hline
        \multirow{3}{*}{HD\,95086} & \textbf{Double power law} & 1.9 & --- \\
                                  & Double Gaussian & 2.2 & $>10$ \\
                                  & Triple power law & --- & $>10$ \\
        \hline
        \multirow{5}{*}{HD\,107146} & \textbf{Triple power law + 2 gaps} & --- & --- \\
                                   & Triple power law + 1 gap & $>5$ & $>10$ \\
                                   & Triple power law & $>5$ & $>10$ \\
                                   & Double Gaussian & $>5$ & $>10$ \\ 
                                   & Triple Gaussian & $>5$ & $>10$ \\
        \hline
        \multirow{2}{*}{HD\,109573} & \textbf{Double Gaussian} & --- & --- \\
                                   & Single Gaussian & $>5$ & $>10$ \\
                                   & Asymmetric Gaussian & $>5$ & $>10$ \\
                                   & Double power law & 5.0 & 4.7 \\
        \hline
        \multirow{2}{*}{HD\,121617} & \textbf{Double Gaussian} & --- & 7.3 \\
                                   & Single Gaussian & $>5$ & $>10$ \\
                                   & Power law + error function & 3.9 & --- \\
                                   & Double power law & 4.0 & 0.15 \\
                                   & Asymmetric Gaussian & $>5$ & $>10$ \\
        \hline
        \multirow{2}{*}{HD\,131488} & \textbf{Double Gaussian} & --- & --- \\
                                   & Triple Gaussian & $>5$ & $>10$ \\
                                   & Double power law & $>5$ & $>10$ \\
        \hline
        \multirow{2}{*}{HD\,131835} & \textbf{Double Gaussian} & --- & --- \\
                                   & Triple Gaussian & 1.4 &  $>10$ \\
                                   & Double power law & $>5$ & $>10$ \\
        \hline
        \multirow{3}{*}{HD\,145560} & \textbf{Double power law} & 3.9 & ---\\
                                   & Double Gaussian & --- & 7.1 \\
                                   & Single Gaussian & $>5$ & $>10$ \\
    \end{tabular}
\end{table*}
\begin{table*}

    \centering
\addtocounter{table}{-1}
    \caption{Continued.}
    \begin{tabular}{c|l|cc}
        \hline \hline
        Disc & Functional Forms & AIC Confidence ($\sigma$) & $\Delta$BIC \\
        \hline
        \multirow{3}{*}{HD\,161868} & \textbf{Double power law} & --- & --- \\
                                   & Power law + error function & 0.72 & 1.5 \\
                                   & Asymmetric Gaussian & 1.1 & 2.5 \\
        \hline
        \multirow{2}{*}{HD\,170773} & \textbf{Asymmetric Gaussian} & --- & --- \\
                                   & Double power law & 0.33 & 0.61 \\
        \hline
        \multirow{3}{*}{HD\,197481} & \textbf{Triple power law + 1 gap} & --- & $>10$ \\
                                   & Triple power law & 3.4 & --- \\
                                   & Double power law & 6.6 & $>10$ \\
                                   & Double Gaussian & 4.2 & 6.2 \\
                                   & Triple Gaussian & 2.8 & $>10$ \\
        \hline
        \multirow{5}{*}{HD\,206893} & \textbf{Double Gaussian} & --- & $>10$ \\
                                   & Triple Gaussian & 1.3 & $>10$ \\
                                   & Double power law & 2.5 & --- \\
                                   & Double power law + 1 gap & 1.1 & $>10$ \\
                                   & Triple power law & 3.3 & $>10$ \\
                                   & Triple power law + 1 gap & 1.3 & $>10$ \\
        \hline
        \multirow{4}{*}{TYC\,9340-437-1} & \textbf{Asymmetric Gaussian} & 2.2 & --- \\
                                         & Double power law & 2.3 & 0.85 \\
                                         & Triple power law & --- & $>10$ \\
                                         & Double Gaussian & 3.3 & $>10$ \\
        \hline
        \multirow{4}{*}{HD\,218396} & \textbf{Asymmetric Gaussian} & 0.18 & --- \\
                                    & Double Gaussian & --- & $>10$ \\
                                    & Double power law & 3.0 & $>10$ \\
                                    & Triple power law & 1.2 & $>10$ \\
        \hline
    \end{tabular}
    \tablefoot{We calculated AIC and BIC differences for each disc relative to the functional form with the lowest criterion value, indicated by a dash (---). The best-fit functional form is determined on a case-by-case basis for each disc, as described throughout the results section, and is indicated by bold text. We convert AIC differences to a confidence interval, while BIC differences $>10$ are considered to be a strong indicator of significance.}
\end{table*}


\FloatBarrier
\section{Model and residuals maps from \frank and \rave models} \label{appendix:residuals}
This appendix displays the model images and residual maps for the radial profile methods used in this study. These methods are described in Sect.~\ref{sec:modelling}.

\begin{figure*}
    \centering
    \includegraphics[width=0.7\linewidth]{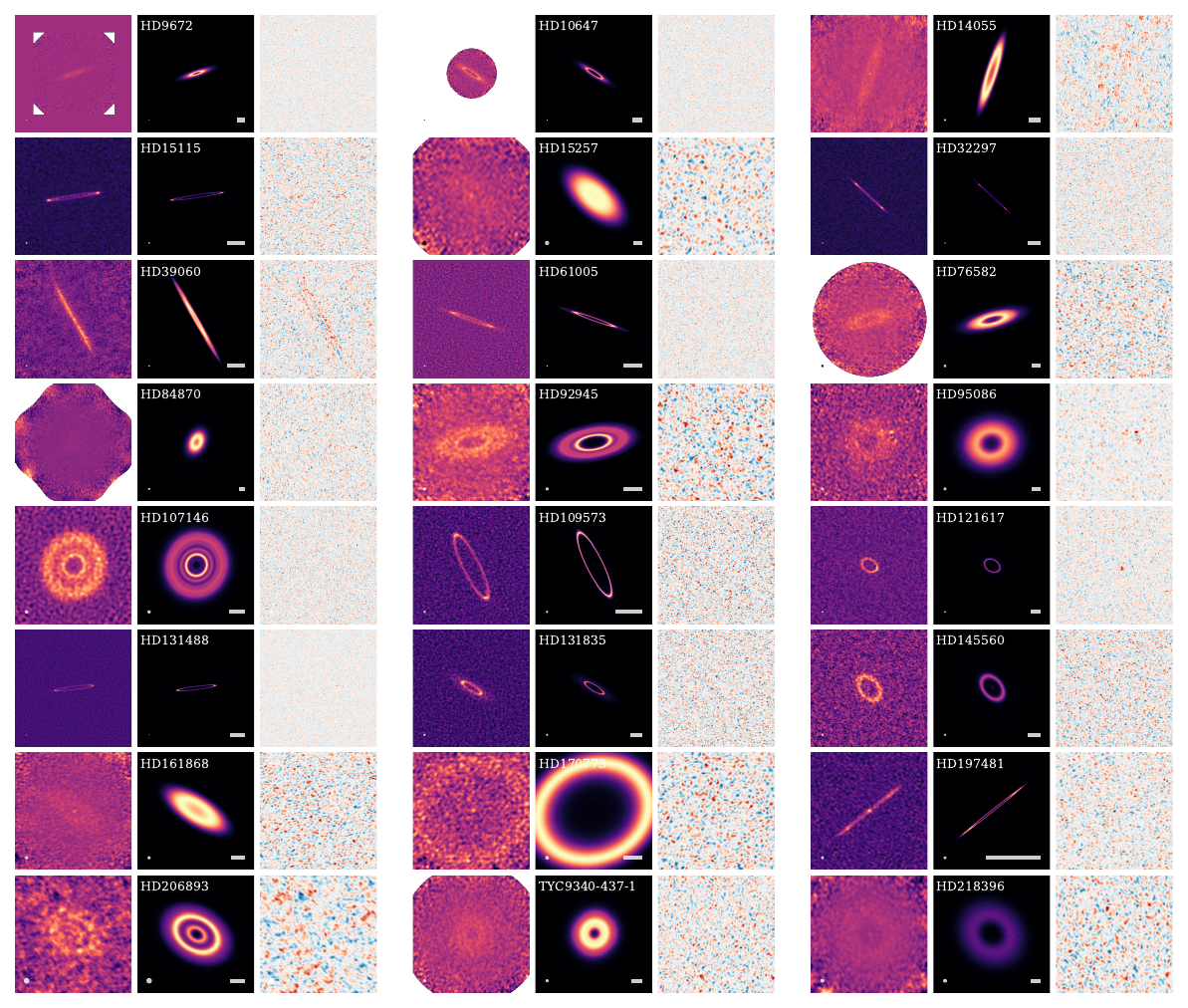}
    \caption{\clean image, \frank model (convolved), and \frank imaged residuals. The scale bars indicate 50~au and the ellipses at the bottom left of each panel indicate the beam size. }
    \label{fig:frank_residuals}
\end{figure*}

\begin{figure*}
    \centering
    \includegraphics[width=0.7\linewidth]{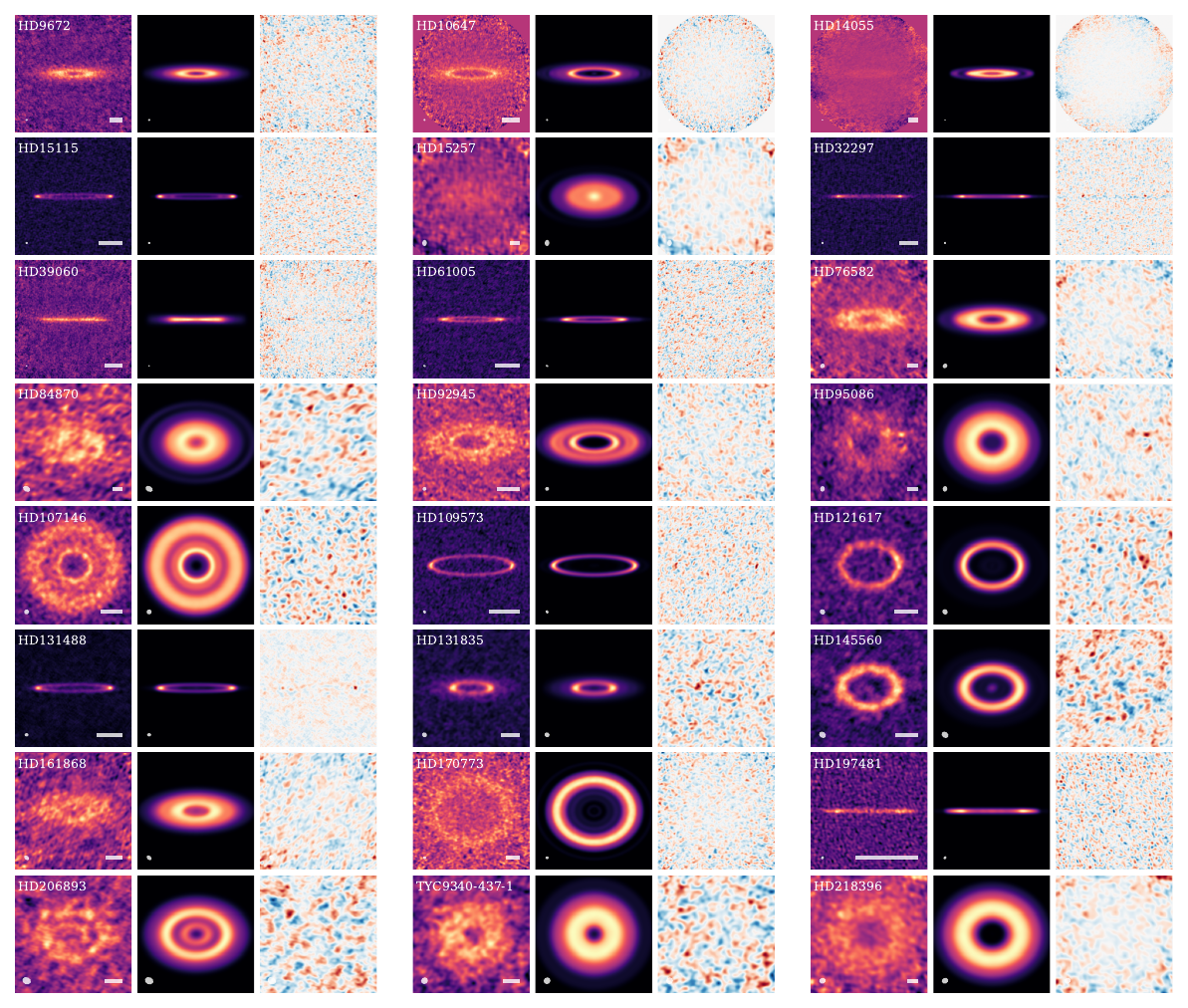}
    \caption{\clean image, \rave model (convolved), and \rave residuals with the major axis aligned horizontally. The scale bars indicate 50~au. }
    \label{fig:rave_residuals}
\end{figure*}

\begin{figure*}
    \centering
    \includegraphics[width=0.95\linewidth]{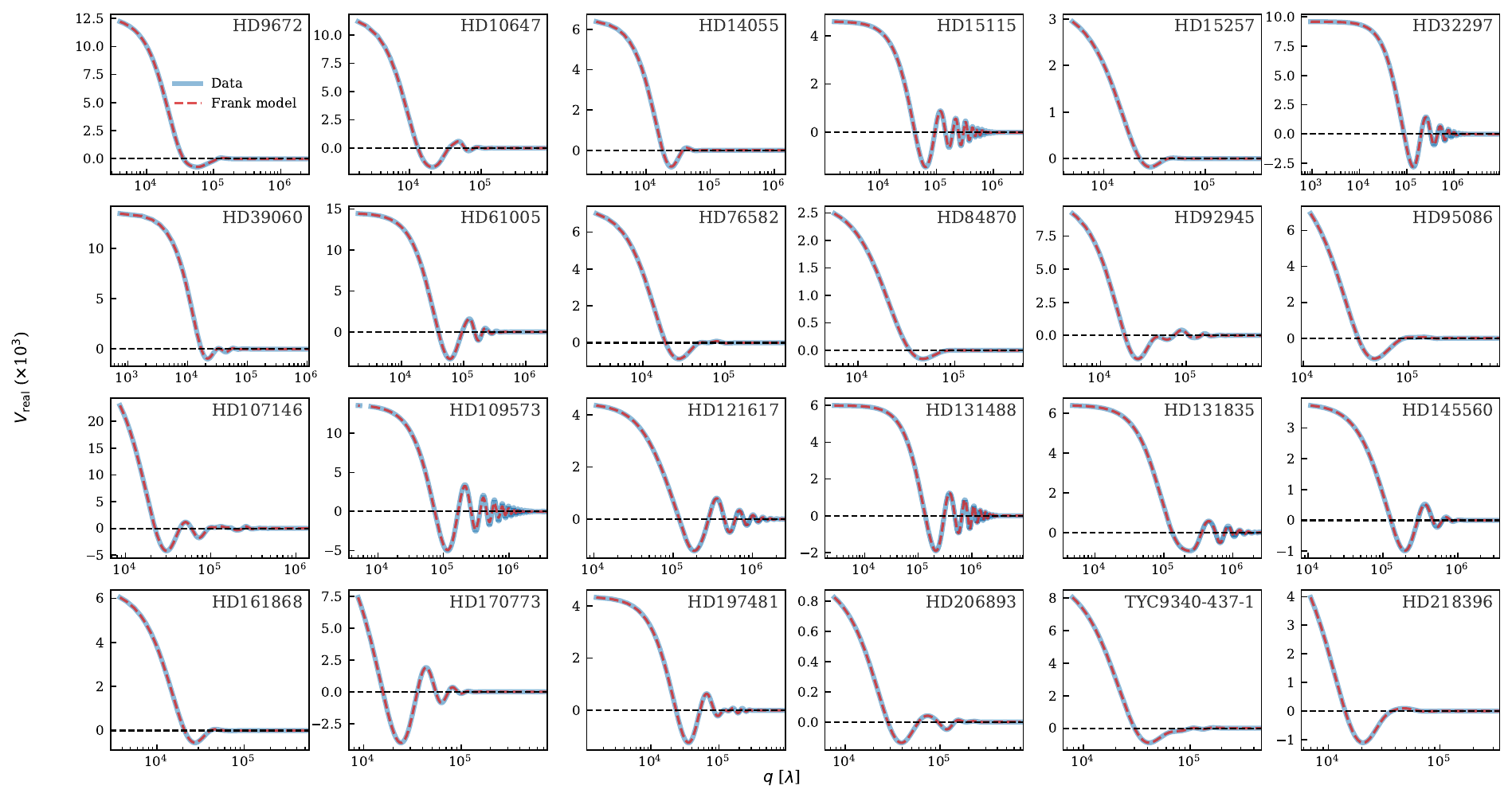}
    \caption{\frank visibility amplitudes of the observations (solid blue line) and \frank models (dotted orange line). }
    \label{fig:frank_1d}
\end{figure*}

\begin{figure*}
    \centering
    \includegraphics[width=0.95\linewidth]{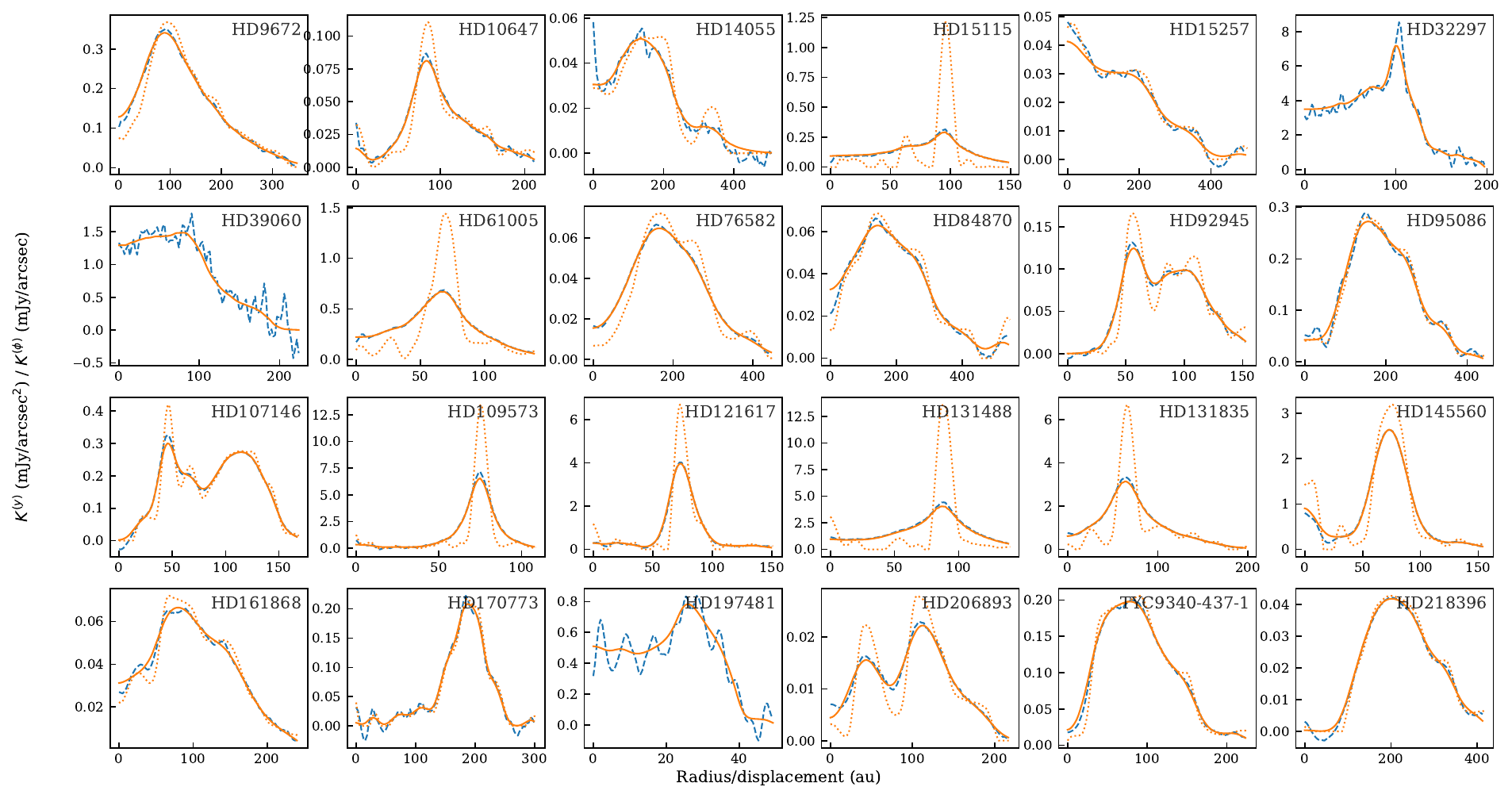}
    \caption{1D quantities that \rave fits to. This includes the vertically summed flux ($K^{(y)}$) of the \clean image (blue dashed line) and \rave model (solid orange line) for the edge-on discs (HD\,39060, HD\,197481, and HD\,32297), and the azimuthally averaged profiles ($K^{(\phi)}$) of the \clean image (blue dashed line) and (convolved) \rave model image (solid orange line), with the (deconvolved) \rave profile overplotted (dotted orange line), for all other discs. }
    \label{fig:rave_1d}
\end{figure*}

\FloatBarrier
\section{\frank and \rave hyperparameters}
This Appendix lists the hyperparameters used in the non-parametric fitting in Table~\ref{table:hyperparameters}. Details of the fitting methods are described in Sects.~\ref{sec:frank_methods} and \ref{sec:rave_methods}.

\begin{table*}
    \centering
    \caption{\frank and \rave hyperparameters used to fit the deconvolved and deprojected radial profiles. }
    \label{table:hyperparameters}
\begin{tabular}{l|llll|lll|ll}
\hline \hline
 & \frank &  &  &  & \rave &  &  & \clean &  \\
Target & $\alpha$ & $w_\text{smooth}$ & $N$ & $r_\text{out}$ [$^{\prime \prime}$] & $N$ & $r_\text{max}$ [$^{\prime \prime}$] & $y_\text{max}$ [$^{\prime \prime}$] & \rave robust & Display profile robust \\
\hline
HD\,9672 & 1.05 & 0.001 & 300 & 10 & 10 & 6.1 & 1.5 & 1.0 & 0.5 \\
HD\,10647 & 1.05 & 0.001 & 200 & 20 & 15 & 12.3 & 5.5 & 0.5 & 0.5 \\
HD\,14055 & 1.05 & 0.1 & 300 & 12 & 10 & 14.3 & 2.9 & 2.0 & 2.0 \\
HD\,15115 & 1.01 & 0.1 & 300 & 5 & 20 & 3.1 & 0.5 & 0.5 & 0.5 \\
HD\,15257 & 1.05 & 0.1 & 300 & 12 & 7 & 10.2 & 5.1 & 2.0 & 0.5 \\
HD\,32297 & 1.05 & 0.001 & 300 & 3 & 17 & 1.5 & 0.2 & 0.5 & 0.5 \\
HD\,39060 & 1.3 & 0.1 & 300 & 12 & 7 & 11.5 & 2.3 & 2.0 & 1.0 \\
HD\,61005 & 1.05 & 0.1 & 200 & 6 & 15 & 3.8 & 0.8 & 0.5 & 0.0 \\
HD\,76582 & 1.01 & 0.0001 & 300 & 15 & 10 & 9.2 & 3.7 & 2.0 & 0.5 \\
HD\,84870 & 1.01 & 0.1 & 300 & 12 & 10 & 6.1 & 3.7 & 2.0 & 0.5 \\
HD\,92945 & 1.01 & 0.1 & 200 & 10 & 15 & 7.2 & 3.6 & 1.0 & 1.0 \\
HD\,95086 & 1.01 & 0.1 & 200 & 8 & 10 & 5.1 & 4.1 & 2.0 & 0.5 \\
HD\,107146 & 1.03 & 0.01 & 300 & 10 & 20 & 6.1 & 6.1 & -0.5 & -0.5 \\
HD\,109573 & 1.01 & 0.1 & 300 & 2 & 20 & 1.5 & 0.4 & 0.5 & 0.0 \\
HD\,121617 & 1.01 & 0.1 & 200 & 3 & 18 & 1.3 & 0.8 & 0.5 & 0.5 \\
HD\,131488 & 1.1 & 0.1 & 200 & 2 & 15 & 0.9 & 0.2 & 1.0 & 0.0 \\
HD\,131835 & 1.03 & 0.0001 & 200 & 3 & 15 & 1.5 & 0.5 & 0.5 & 0.0 \\
HD\,145560 & 1.01 & 0.1 & 200 & 3 & 15 & 1.3 & 0.8 & 2.0 & 0.5 \\
HD\,161868 & 1.05 & 0.1 & 300 & 12 & 10 & 8.2 & 4.1 & 2.0 & 0.5 \\
HD\,170773 & 1.02 & 0.1 & 300 & 9 & 20 & 8.2 & 6.5 & 2.0 & 0.5 \\
HD\,197481 & 1.01 & 0.1 & 300 & 10 & 9 & 5.1 & 0.5 & 0.0 & 0.0 \\
HD\,206893 & 1.01 & 0.001 & 300 & 8 & 10 & 5.4 & 4.3 & 2.0 & 1.0 \\
TYC\,9340-437-1 & 1.01 & 0.0001 & 300 & 12 & 10 & 6.1 & 6.1 & 2.0 & 0.5 \\
HD\,218396 & 1.01 & 0.1 & 100 & 15 & 10 & 10.2 & 10.2 & 2.0 & 0.5 \\
\hline
\end{tabular}
\end{table*}

\FloatBarrier
\section{Mean-motion resonances} \label{app:app_mmr_l}
Undetected planets could exist in mean-motion resonances and shape the structure of the debris discs that we observe. Three discs in the ARKS sample exhibit more than one radial gap as suggested by the \frank profiles (see Table~\ref{table:gaps}). We find that the radial location of the two gaps in HD\,131835 and HD\,197481 (AU~Mic) could coincide with a 2:1 resonance, whereas the two gaps in HD\,107146 could coincide with the 3:2 resonance. While these commensurabilities between orbital periods at key locations in the discs could exist, it must be remembered that by chance, a fraction of radial features based on an arbitrary sample of radial profiles are expected to fall close to a resonance. A discussion on the probability and implications of these resonances is beyond the scope of this work, and we simply list these potential resonances as topics of investigation to encourage future work on this subject. 

\FloatBarrier
\section{Disc features measured from non-parametric profiles} \label{appendix:rp_labelled}
\begin{figure*}
    \centering
    \includegraphics[width=0.8\linewidth]{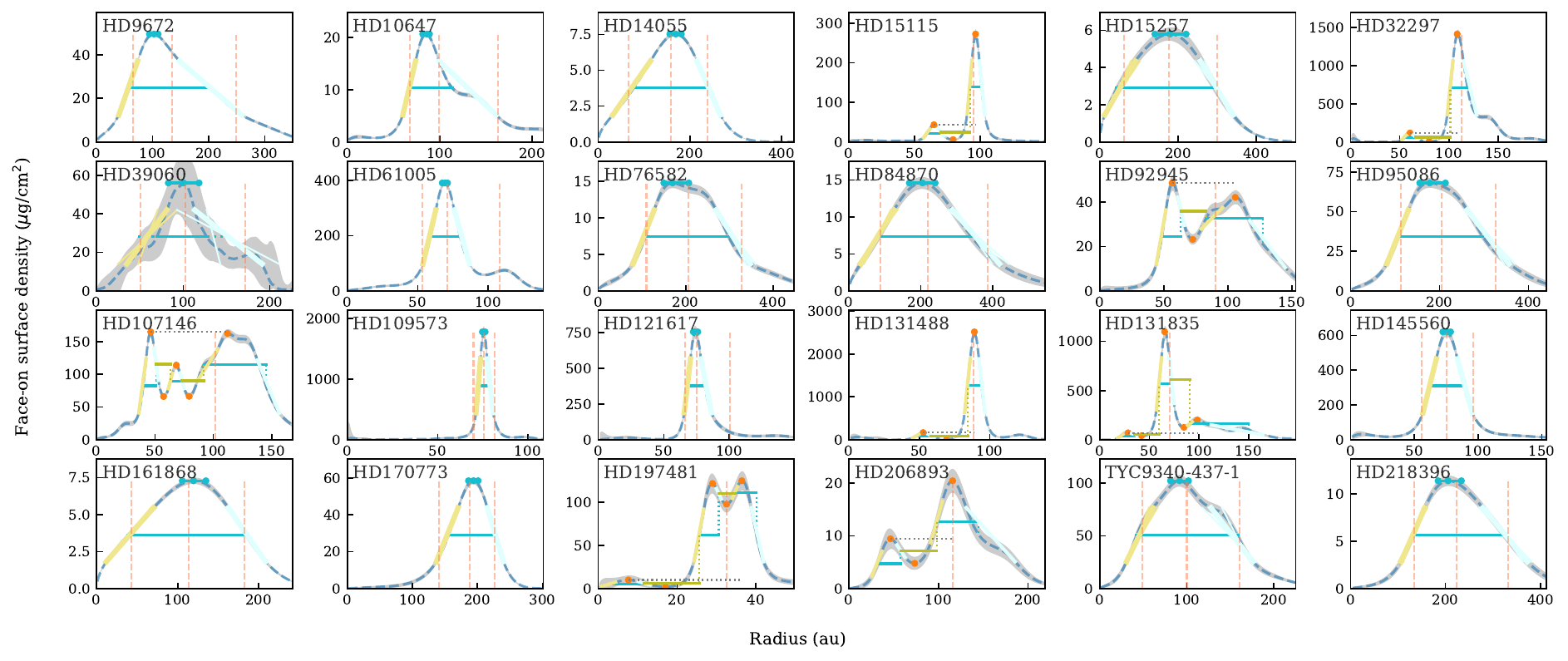}
    \caption{\frank surface density profiles labelled with measured disc features. Orange dots at peaks and troughs indicate the individual rings and gaps in multi-ring systems. The line segments labelled on each profile indicate the edges, fractional widths, and the uncertainties in the peak radius detected from the radial profiles (see Sect.~\ref{sec:measurements} for details). }
    \label{fig:rp_labelled}
\end{figure*}

This appendix lists the radial features measured from the \frank surface density profiles in Table~\ref{table:rings}, with a visualisation of these measurements shown in Fig.~\ref{fig:rp_labelled}. Details of these measurements are described in Sect.~\ref{sec:measurements}.

\begin{table*}
    \centering
    \caption{Ring and edge locations and widths in units of au measured from the \frank surface density profiles. }
    \label{table:rings}
\begin{tabular}{lllllllll}
\hline \hline
Target & Ring & $r_\text{peak}$ & $r_\text{centroid}$ & $\Delta r$ & $r_\text{inner}$ & $\Delta r_\text{inner}$ & $r_\text{outer}$ & $\Delta r_\text{outer}$ \\
\hline
HD\,9672 & 1 & 102.3$^{+7.4}_{-8.1}$ & 134.4 & 146.2$^{+1.7}_{-1.8}$ & 58.2$^{+0.7}_{-0.8}$ & 33.5$^{+1.2}_{-1.1}$ & 204.4$^{+2.7}_{-2.8}$ & 111.2$^{+3.4}_{-3.3}$ \\
HD\,10647 & 1 & 86.0$^{+2.9}_{-4.4}$ & 99.5 & 45.8$^{+1.4}_{-1.2}$ & 67.7$^{+0.5}_{-0.5}$ & 13.9$^{+0.7}_{-0.7}$ & 113.6$^{+2.2}_{-1.9}$ & 61.8$^{+1.5}_{-1.5}$ \\
HD\,14055 & 1 & 169.3$^{+12.1}_{-13.1}$ & 158.1 & 161.8$^{+2.2}_{-2.2}$ & 77.7$^{+2.6}_{-2.6}$ & 82.4$^{+2.4}_{-2.4}$ & 239.5$^{+1.8}_{-1.8}$ & 51.1$^{+1.9}_{-1.9}$ \\
HD\,15115 & 1 & 64.7$^{+2.5}_{-2.3}$ &  & 10.4$^{+1.0}_{-1.1}$ & 59.5$^{+1.0}_{-0.9}$ & 2.2$^{+1.3}_{-1.1}$ & 69.8$^{+1.0}_{-1.1}$ & 4.2$^{+1.4}_{-1.2}$ \\
 & 2 & 96.3$^{+1.1}_{-1.0}$ &  & 9.0$^{+0.2}_{-0.2}$ & 92.0$^{+0.2}_{-0.2}$ & 0.5$^{+0.3}_{-0.3}$ & 101.0$^{+0.3}_{-0.3}$ & 4.2$^{+0.3}_{-0.3}$ \\
HD\,15257 & 1 & 179.4$^{+41.1}_{-38.4}$ & 177.5 & 250.9$^{+7.7}_{-9.1}$ & 45.1$^{+7.8}_{-6.4}$ & 78.1$^{+7.0}_{-6.7}$ & 296.0$^{+9.0}_{-10.4}$ & 83.8$^{+10.5}_{-10.1}$ \\
HD\,32297 & 1 & 60.5$^{+3.4}_{-4.3}$ &  & 11.6$^{+2.0}_{-2.3}$ & 54.5$^{+2.2}_{-1.8}$ & 4.3$^{+2.2}_{-2.1}$ & 66.1$^{+2.1}_{-2.3}$ & 5.1$^{+2.6}_{-2.5}$ \\
 & 2 & 108.4$^{+2.7}_{-1.4}$ &  & 17.4$^{+0.4}_{-0.4}$ & 101.5$^{+0.3}_{-0.3}$ & 0.7$^{+0.4}_{-0.4}$ & 118.9$^{+0.5}_{-0.5}$ & 13.5$^{+6.4}_{-1.1}$ \\
HD\,39060 & 1 & 100.5$^{+18.7}_{-20.3}$ & 102.4 & 62.3$^{+16.3}_{-13.8}$ & 66.1$^{+12.0}_{-16.7}$ & 44.1$^{+12.8}_{-19.9}$ & 128.4$^{+15.9}_{-15.7}$ & 78.5$^{+17.2}_{-26.6}$ \\
HD\,61005 & 1 & 69.5$^{+1.6}_{-2.1}$ & 71.1 & 21.2$^{+0.3}_{-0.3}$ & 58.9$^{+0.3}_{-0.3}$ & 9.1$^{+0.4}_{-0.4}$ & 80.1$^{+0.3}_{-0.3}$ & 10.8$^{+0.5}_{-0.5}$ \\
HD\,76582 & 1 & 170.1$^{+37.1}_{-20.0}$ & 206.1 & 191.3$^{+4.1}_{-4.0}$ & 107.4$^{+3.4}_{-3.5}$ & 48.7$^{+4.6}_{-4.4}$ & 298.7$^{+4.7}_{-4.7}$ & 78.0$^{+9.1}_{-7.5}$ \\
HD\,84870 & 1 & 204.6$^{+35.0}_{-36.6}$ & 219.0 & 259.1$^{+10.8}_{-10.3}$ & 81.8$^{+8.7}_{-8.9}$ & 92.9$^{+8.7}_{-9.1}$ & 340.9$^{+12.6}_{-11.9}$ & 131.3$^{+18.4}_{-15.8}$ \\
HD\,92945 & 1 & 56.5$^{+3.3}_{-2.3}$ &  & 14.9$^{+0.9}_{-1.0}$ & 48.5$^{+0.8}_{-0.7}$ & 1.8$^{+1.0}_{-0.9}$ & 63.4$^{+1.1}_{-1.2}$ & 5.1$^{+1.5}_{-1.2}$ \\
 & 2 & 105.8$^{+5.7}_{-6.4}$ &  & 44.3$^{+2.8}_{-3.2}$ & 83.0$^{+3.4}_{-2.2}$ & 13.1$^{+4.0}_{-6.5}$ & 127.3$^{+3.5}_{-3.1}$ & 27.5$^{+2.8}_{-2.8}$ \\
HD\,95086 & 1 & 178.3$^{+35.1}_{-22.8}$ & 205.5 & 189.6$^{+4.8}_{-4.7}$ & 108.4$^{+3.5}_{-3.5}$ & 48.5$^{+4.2}_{-4.0}$ & 298.0$^{+6.1}_{-5.9}$ & 94.4$^{+9.7}_{-8.6}$ \\
HD\,107146 & 1 & 46.4$^{+1.4}_{-1.7}$ &  & 11.0$^{+0.4}_{-0.4}$ & 39.9$^{+0.4}_{-0.4}$ & 1.0$^{+0.6}_{-0.5}$ & 50.9$^{+0.5}_{-0.5}$ & 3.7$^{+0.6}_{-0.5}$ \\
 & 2 & 68.3$^{+2.0}_{-2.4}$ &  & 9.9$^{+0.8}_{-0.8}$ & 63.2$^{+0.8}_{-0.8}$ & 1.7$^{+1.0}_{-0.8}$ & 73.1$^{+0.8}_{-0.8}$ & 3.7$^{+1.1}_{-0.9}$ \\
 & 3 & 112.1$^{+6.9}_{-3.8}$ &  & 53.3$^{+1.5}_{-2.8}$ & 92.0$^{+4.6}_{-1.9}$ & 2.6$^{+1.3}_{-1.3}$ & 145.3$^{+1.0}_{-1.1}$ & 18.9$^{+1.4}_{-1.3}$ \\
HD\,109573 & 1 & 75.7$^{+0.7}_{-0.8}$ & 75.6 & 6.0$^{+0.2}_{-0.2}$ & 72.7$^{+0.2}_{-0.2}$ & 2.5$^{+0.2}_{-0.2}$ & 78.8$^{+0.2}_{-0.2}$ & 3.2$^{+0.3}_{-0.3}$ \\
HD\,121617 & 1 & 73.6$^{+2.3}_{-1.0}$ & 75.4 & 13.1$^{+0.5}_{-0.5}$ & 68.2$^{+0.4}_{-0.4}$ & 4.8$^{+0.5}_{-0.5}$ & 81.3$^{+0.6}_{-0.6}$ & 7.9$^{+0.8}_{-0.8}$ \\
HD\,131488 & 1 & 52.7$^{+3.5}_{-2.8}$ &  & 10.6$^{+20.3}_{-1.8}$ & 47.7$^{+1.7}_{-39.1}$ & 4.4$^{+2.1}_{-2.2}$ & 58.3$^{+1.5}_{-1.8}$ & 4.3$^{+2.1}_{-1.8}$ \\
 & 2 & 89.1$^{+0.8}_{-0.7}$ &  & 9.9$^{+0.2}_{-0.2}$ & 84.4$^{+0.2}_{-0.2}$ & 0.4$^{+0.2}_{-0.2}$ & 94.3$^{+0.2}_{-0.2}$ & 4.4$^{+0.2}_{-0.2}$ \\
HD\,131835 & 1 & 28.6$^{+7.4}_{-6.2}$ &  & 14.1$^{+4.6}_{-4.6}$ & 20.7$^{+3.0}_{-1.8}$ & 4.8$^{+3.5}_{-2.4}$ & 34.8$^{+7.4}_{-6.2}$ & 5.4$^{+4.1}_{-2.7}$ \\
 & 2 & 65.4$^{+2.4}_{-1.0}$ &  & 12.1$^{+0.4}_{-0.5}$ & 60.0$^{+0.4}_{-0.4}$ & 1.1$^{+0.5}_{-0.6}$ & 72.0$^{+0.5}_{-0.5}$ & 5.0$^{+0.6}_{-0.5}$ \\
 & 3 & 98.4$^{+6.5}_{-6.5}$ &  & 59.0$^{+3.6}_{-16.1}$ & 90.9$^{+4.1}_{-3.0}$ & 4.3$^{+4.6}_{-2.2}$ & 149.9$^{+4.1}_{-28.0}$ & 52.0$^{+7.0}_{-7.9}$ \\
HD\,145560 & 1 & 75.6$^{+2.8}_{-3.2}$ & 75.1 & 26.1$^{+0.6}_{-0.7}$ & 62.7$^{+0.6}_{-0.6}$ & 10.5$^{+0.7}_{-0.8}$ & 88.8$^{+0.7}_{-0.7}$ & 11.7$^{+0.8}_{-0.8}$ \\
HD\,161868 & 1 & 119.6$^{+15.3}_{-13.8}$ & 114.0 & 144.6$^{+2.5}_{-2.6}$ & 39.4$^{+2.9}_{-2.7}$ & 58.3$^{+2.2}_{-2.2}$ & 184.1$^{+2.3}_{-2.4}$ & 46.2$^{+2.6}_{-2.5}$ \\
HD\,170773 & 1 & 194.3$^{+6.9}_{-7.9}$ & 188.7 & 70.8$^{+1.6}_{-1.6}$ & 156.0$^{+1.6}_{-1.6}$ & 33.5$^{+2.1}_{-2.0}$ & 226.8$^{+1.5}_{-1.6}$ & 24.8$^{+1.6}_{-1.7}$ \\
HD\,197481 & 1 & 7.7$^{+4.1}_{-6.8}$ &  & 9.1$^{+2.5}_{-4.5}$ & 2.5$^{+5.2}_{-2.5}$ & 3.7$^{+2.0}_{-1.8}$ & 11.6$^{+2.5}_{-3.9}$ & 2.9$^{+3.1}_{-1.5}$ \\
 & 2 & 29.0$^{+1.4}_{-1.4}$ &  & 4.9$^{+0.8}_{-0.9}$ & 25.7$^{+0.3}_{-0.3}$ & 0.7$^{+0.4}_{-0.4}$ & 30.6$^{+1.2}_{-1.4}$ & 1.2$^{+1.2}_{-0.6}$ \\
 & 3 & 36.4$^{+1.4}_{-1.5}$ &  & 5.4$^{+0.7}_{-0.8}$ & 34.7$^{+1.2}_{-1.1}$ & 1.3$^{+1.3}_{-0.7}$ & 40.1$^{+0.4}_{-0.4}$ & 3.1$^{+0.5}_{-0.4}$ \\
HD\,206893 & 1 & 46.4$^{+8.5}_{-8.4}$ &  & 26.1$^{+4.1}_{-4.7}$ & 31.8$^{+3.3}_{-2.4}$ & 5.9$^{+4.5}_{-2.9}$ & 57.8$^{+5.8}_{-6.1}$ & 9.0$^{+9.1}_{-4.5}$ \\
 & 2 & 115.8$^{+8.8}_{-8.5}$ &  & 46.7$^{+5.6}_{-4.6}$ & 98.2$^{+3.2}_{-2.9}$ & 6.5$^{+3.8}_{-3.2}$ & 145.0$^{+8.3}_{-6.0}$ & 54.5$^{+5.1}_{-5.9}$ \\
TYC\,9340-437-1 & 1 & 91.9$^{+10.1}_{-10.5}$ & 100.1 & 114.5$^{+2.7}_{-2.9}$ & 42.8$^{+2.4}_{-2.2}$ & 30.4$^{+3.5}_{-3.3}$ & 157.3$^{+3.1}_{-3.4}$ & 48.3$^{+7.0}_{-8.2}$ \\
HD\,218396 & 1 & 205.4$^{+28.0}_{-21.0}$ & 224.1 & 190.8$^{+6.0}_{-6.0}$ & 133.1$^{+4.5}_{-4.4}$ & 50.3$^{+5.2}_{-5.1}$ & 323.9$^{+7.6}_{-7.6}$ & 93.5$^{+9.4}_{-8.8}$ \\
\hline
\end{tabular}
\end{table*}

\end{appendix}

\end{document}